\documentclass[onecolumn,showpacs,amsmath,amssymb,superscriptaddress,nofootinbib,floatfix,preprintnumbers]{revtex4-1}
 
\usepackage{graphicx}
\usepackage{multirow}
\usepackage{tabularx}
\usepackage{makecell}
\usepackage{booktabs}

\usepackage{paralist}
\usepackage{subfigure}

\usepackage{slashed,cases,comment,bm,here,colortbl}
\usepackage{framed}
\definecolor{shadecolor}{rgb}{0.95,0.95,0.95}
\newenvironment{claim}{\begin{shaded}\noindent\ignorespaces}{\end{shaded}}

\usepackage[colorlinks=true, linkcolor=BrickRed, citecolor=Blue, urlcolor=Blue, filecolor=Blue]{hyperref}
\usepackage[dvipsnames,table]{xcolor}

\usepackage{newtxtext,newtxmath}

\newcommand{\vev}[1]{ \left\langle {#1} \right\rangle }

\newcommand{\1}{\mbox{1}\hspace{-0.25em}\mbox{l}}

\newcommand{\der}{\partial}

\newcommand{\C}{\mathcal{C}}

\newcommand{\dd}{\mathrm{d}}

\newcommand{\abs}[1]{\left\vert {#1} \right\vert}

\newcommand{\Tr}{\text{Tr}}

\makeatletter
\@addtoreset{equation}{section}

\makeatother

\begin{document}

\title{Dark Matter Sommerfeld-enhanced annihilation\\ and Bound-state decay at finite temperature}

\author{Tobias Binder}
\email{tobias.binder@theorie.physik.uni-goettingen.de}
\affiliation{Institute for Theoretical Physics, Georg-August University G\"ottingen, Friedrich-Hund-Platz 1, G\"ottingen, D-37077 Germany
}
\author{Laura Covi}
\email{covi@theorie.physik.uni-goettingen.de}
\affiliation{Institute for Theoretical Physics, Georg-August University G\"ottingen, Friedrich-Hund-Platz 1, G\"ottingen, D-37077 Germany
}
\author{Kyohei Mukaida}
\email{kyohei.mukaida@desy.de}
\affiliation{Deutsches Elektronen-Synchrotron (DESY), Notkestra\ss e 85, Hamburg, D-22607 Germany
}

\date{\today}

\begin{abstract} 
\noindent

Traditional computations of the dark matter (DM) relic abundance, for models where attractive self-interactions are mediated by light force-carriers and bound states exist, rely on the solution of a coupled system of classical on-shell Boltzmann equations. This idealized description misses important thermal effects caused by the tight coupling among force-carriers and other charged relativistic species potentially present during the chemical decoupling process. We develop for the first time a comprehensive ab-initio derivation for the description of DM long-range interactions in the presence of a hot and dense plasma background directly from non-equilibrium quantum field theory. Our results clarify a few conceptional aspects of the derivation and show that under certain conditions the finite temperature effects can lead to sizable modifications of DM Sommerfeld-enhanced annihilation and bound-state decay rates. In particular, the scattering and bound states get strongly mixed in the thermal plasma environment, representing a characteristic difference from a pure vacuum theory computation. The main result of this work is a novel differential equation for the DM number density, written down in a form which is manifestly independent under the choice of what one would interpret as a bound or a scattering state at finite temperature. The collision term, unifying the description of annihilation and bound-state decay, turns out to have in general a non-quadratic dependence on the DM number density. This generalizes
the form of the conventional Lee-Weinberg equation which is typically adopted to describe the freeze-out process. We prove that our 
number density equation is consistent with previous literature results under certain limits. In the limit of vanishing finite temperature corrections our central equation is fully compatible with the classical on-shell Boltzmann equation treatment. So far, finite temperature corrected annihilation rates for long-range force systems have been estimated from a method relying on linear response theory. We prove consistency between the latter and our method in the linear regime close to chemical equilibrium.

\end{abstract}

\maketitle
\preprint{DESY 18-123}

\tableofcontents

\section{Introduction}
\label{sec:intro}
The cosmological standard model successfully describes the evolution of the large-scale structure of our Universe. It requires the existence of a cold and collisionless matter component, called dark matter (DM), which dominates over the baryon content in the matter dominated era of our Universe. The Planck satellite measurements of the Cosmic Microwave Background (CMB) temperature anisotropies have nowadays determined the amount of dark matter to an unprecedented precision, reaching the level of sub-percentage accuracy in the observational determination of the abundance when combining CMB and external data \cite{Ade:2015xua, Aghanim:2018eyx}, e.g.\ measurements of the baryon acoustic oscillation.

Interestingly, astrophysical observation and structure formation on sub-galactic scales might point towards the nature of dark matter as velocity-dependent self-interacting elementary particles. On the one hand, observations of galaxy cluster systems, where typical rotational velocities are of the order $v_0 \sim 1000 \; \text{km}/\text{s} $, set the most stringent bounds on the self-scattering cross section to be less than $\sigma/m_{\text{DM}} \lesssim 0.7 \;(0.1) \;\text{cm}^2/\text{g}$ in the bullet cluster \cite{Randall:2007ph} (in order to guarantee the production of elliptical halos \cite{MiraldaEscude:2000qt, Peter:2012jh}). On the other hand, a DM self-scattering cross section of the order $\sigma/m_{\text{DM}} \sim 1 \;\text{cm}^2/\text{g}$ on dwarf-galactic scales, where velocities are of the order $v_0 \sim 30-100 \; \text{km}/\text{s} $, would lead to a compelling solution of the cusp-core and diversity problem without strongly relying on uncertain assumptions of modelling the barionic feedbacks in simulations. This velocity-dependence of the self-scattering cross section can naturally be realized in models where a light mediator acts as a long-range force between the dark matter particles. For a recent review on self-interacting DM, see \cite{Tulin:2017ara}.

Generically, long-range forces can lead to sizable modifications of the DM tree-level annihilation cross section in the regime
where the annihilating particles are slow. For the most appealing DM candidates, known as WIMP Dark Matter~\cite{Lee:1977ua, Ellis:1983ew, Arcadi:2017kky}, such that the relic abundance in the Early Universe is set by the thermal freeze-out mechanism when
the DM is non-relativistic,  these effects can be sizable already at the time of chemical decoupling.
Then the inclusion of the long-range force modification in the computation of the relic abundance is necessary to reach the 
 required level of the accuracy set by the Planck precision measurement \cite{Ade:2015xua, Aghanim:2018eyx}. 
If the light mediators induce an attractive force between the annihilating DM particles, the total cross section is typically enhanced~\cite{Hisano:2003ec, Hisano:2006nn}
which is often referred as the \emph{Sommerfeld enhancement} \cite{doi:10.1002/andp.19314030302} or \emph{Sakharov enhancement} \cite{Sakharov:1948yq}. 
Additionally, such attractive forces can lead to the existence of DM \emph{bound-state} solutions~\cite{Pospelov:2008jd,MarchRussell:2008tu,Shepherd:2009sa}. This opens the possibility for conversion processes between scattering and bound states via radiative processes, influencing the evolution of the abundance of the stable scattering states during the DM thermal history. DM scenarios with Sommerfeld enhancement or with bound states have been widely studied in the literature~\cite{Belotsky:2004st, ArkaniHamed:2008qn, Slatyer:2009vg, Feng:2010zp, vonHarling:2014kha, Petraki:2015hla, An:2016gad,Asadi:2016ybp,Johnson:2016sjs, Hryczuk:2010zi, Beneke:2016ync,
Freitas:2007sa, Hryczuk:2011tq, Harigaya:2014dwa, Harz:2014gaa, Beneke:2014gja, Ellis:2015vna, Kim:2016kxt, ElHedri:2016onc, Liew:2016hqo, Mitridate:2017izz, Harz:2018csl,
Petraki:2016cnz, Harz:2017dlj, 
Cirelli:2007xd, MarchRussell:2008yu,
Hisano:2004ds, Cirelli:2008id,Cirelli:2016rnw,
Fan:2013faa, Bhattacherjee:2014dya, Ibe:2015tma, Beneke:2016jpw,
Berger:2008ti,
Blum:2016nrz,
Biondini:2017ufr, Biondini:2018pwp, Biondini:2018xor} and it has been shown that the main effect of such corrections is to shift in the parameter space the upper bounds on the DM mass, otherwise the theoretically predicted DM density would be too large (overclosure bound).

Classic WIMP candidates with large corrections via Sommerfeld enhancement or bound states are particles charged under the electroweak interactions, like the Wino 
neutralino in supersymmetric models~\cite{Jungman:1995df} or the first Kaluza-Klein excitation of the gauge boson in models with extra 
dimensions~\cite{Servant:2002aq}. 
For the supersymmetric case it was realized very early on by \cite{Hisano:2003ec, Hisano:2006nn} that the Sommerfeld effect 
reduces the Wino density up to 30~\% and pushes the mass of Wino Dark Matter candidate to few TeVs in order to obtain the correct 
relic density. These studies have later been extended to the case of general components of neutralino~\cite{Hryczuk:2010zi, Beneke:2016ync}.
Similar and even stronger effects from the Sommerfeld enhancement and bound states were found in the case of coannihilation of the WIMP particle with charged or colored 
states \cite{Freitas:2007sa, Hryczuk:2011tq, Harigaya:2014dwa, Harz:2014gaa, Beneke:2014gja, Ellis:2015vna, Kim:2016kxt, ElHedri:2016onc, Liew:2016hqo, Mitridate:2017izz, Harz:2018csl}.  
If the electroweakly charged Dark Matter is sufficiently heavy, the Sommerfeld enhancement or the presence of bound states due to the exchange of electroweak 
gauge or Higgs bosons, see e.g.~\cite{Petraki:2016cnz,Harz:2017dlj}, are very generic as it was shown for example in the minimal 
Dark Matter model \cite{Cirelli:2007xd,Mitridate:2017izz} and in Higgs portal models~\cite{MarchRussell:2008yu}.
In these cases, long-range force effects play an important role also for the indirect detection
limits~\cite{Hisano:2004ds, Cirelli:2008id, Pospelov:2008jd,MarchRussell:2008tu,Cirelli:2016rnw} and especially for the Wino the Sommerfeld enhancement has lead to the exclusion of most parameter 
space~\cite{Fan:2013faa, Bhattacherjee:2014dya, Ibe:2015tma, Beneke:2016jpw}.
Note that this effect can be important also when the Dark Matter is not itself a WIMP, but it is produced by WIMP 
decay out-of-equilibrium, like in the SuperWIMP mechanism \cite{Covi:1999ty, Feng:2003xh}. Indeed in such scenario, 
the DM inherits part of the energy density of the mother particle and so any change in the latter freeze-out density is directly 
transferred to the superweakly interacting DM and can relax the BBN constraints on the mother particle~\cite{Berger:2008ti}.

While a lot of effort has been made to compute quantitatively the effects of a long-range force on the DM relic density
employing the classical Boltzmann equation method, it is still unclear if that is a sufficient description. Indeed,
considering the presence of a thermal plasma on the long-range force leads on one side to a possible screening by
the presence of a thermal mass, on the other to the issue if the coherence in the (in principle infinite) ladder diagram
exchanges  between the two slowly moving annihilating particles is guaranteed.
Moreover, from a conceptional point of view, there is yet no consistent formulation in the existing literature of how to deal 
with long-range forces \emph{at finite temperature}, especially if the dark matter is, or, enters an out-of-equilibrium state 
(already the standard freeze-out scenario is a transition from chemical equilibrium to out-of-chemical-equilibrium). 
The main concern of our work will be to clarify conceptional aspects of the derivation and the solution of the number density equation for DM particles with attractive long-range force interactions in the presence of a hot and dense plasma background, starting from first principles. From the beginning, we work in the \emph{real-time formalism}, which 
has a smooth connection to generic out-of-equilibrium phenomena. 

The simplified DM system we would like to describe in the presence of a thermal environment is similar to heavy quarks in 
a hot quark gluon plasma. For this setup it has been shown that finite temperature effects can lead to a melting of the 
heavy-quark bound states which influences, e.g.\ the annihilation rate of the heavy quark pair into dilepton \cite{Burnier:2007qm}. 
For DM or heavy quark systems, the Sommerfeld enhancement at finite temperature has been discussed in the literature, 
where the chemical equilibration rate is {\it i)} estimated from \emph{linear response theory}~\cite{Bodeker:2012zm, Kim:2016zyy, Kim:2016kxt}  and {\it ii)} based on \emph{classical} rate arguments \cite{Bodeker:2012gs}, is then inserted into the 
\emph{non-linear} Boltzmann equation for the DM number density 'by hand'. 
Relying on those estimates, it has been shown that the overclosure bound of the DM mass can be strongly affected by the 
melting of bound states in a plasma environment \cite{Biondini:2017ufr, Biondini:2018pwp, Biondini:2018xor}. However, strictly speaking, the
linear response theory method is only applicable if the system is close to chemical equilibrium. Indeed the computation has been done in the \emph{imaginary-time formalism} so far, capturing the physics of thermal equilibrium. One of our goals is to obtain a more general result directly from the \emph{real-time} formalism, valid as well for systems way out of equilibrium. 

Most of the necessary basics of the real-time method we use are provided in Section \ref{sec:rtf} as a short review of out-of-equilibrium Quantum Field Theory. Within this mathematical framework, an exact expression for the DM number density equation of our system is derived in Section \ref{sec:eoms}, where the Sommerfeld-enhanced annihilation or decay rate at finite temperature can be computed from 
a certain component of a four-point correlation function. 
We derive the equation of motion for the four-point correlation function on the real-time contour in Section \ref{sec:closedequations} which becomes in 
its truncated form a Bethe-Salpeter type of equation. Since we close the correlation functions hierarchy by truncation, 
the system of coupled equations we have to solve contains only terms with DM two- and four-point functions. 
In Section \ref{sec:solutions}, we derive a simple semi-analytical solution of the four-point correlator under certain assumptions valid for WIMP-like freeze-outs. 
Our result does not rely on linear response theory and it is therefore quite general applying also in case of large deviations 
from chemical equilibrium. The limit of vanishing finite temperature corrections is taken in Section~\ref{sec:vacuum}, showing the consistency between our general results and the classical Boltzmann equation treatment. Here, we also compare to the linear response theory method and clarify the assumptions needed to reproduce those results. Our main numerical results for the finite temperature case are given in Section~\ref{sec:finitetemp}, both for a gauge theory and for a Yukawa potential, and
discussed in detail in Section \ref{sec:discussion}.  Finally, we conclude in Section \ref{sec:conclusion}.

\section{Real-time formalism prerequisites}
\label{sec:rtf}
\begin{figure}[h]
\includegraphics[scale=1.0]{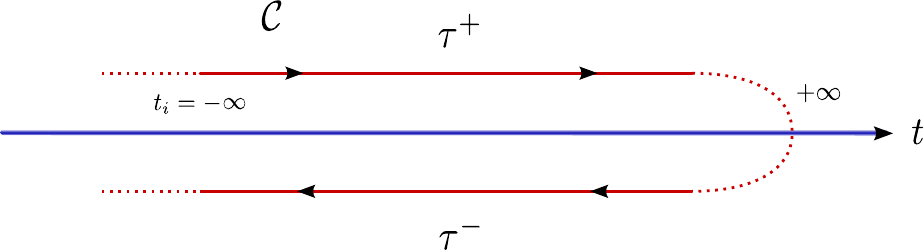}
\caption{Keldysh-Schwinger approximation of the closed-time-path contour $\mathcal{C}$, consisting of two time branches $\tau^+$ and $\tau^-$.}
\label{fig:contour}
\end{figure}
The generalization of Quantum field theory on the \emph{closed-time-path} (CTP) contour, or \emph{real-time formalism}, is a mathematical method which allows to describe the dynamics of quantum systems out-of-equilibrium. Prominent applications are systems on curved space-time and/or systems having a finite temperature. In this work, we assume that the equilibration of DM in the early Universe is a fast process, and consequently, the initial memory effects before the freeze-out process can be ignored. This leads to the fact that the adiabatic assumption for such a system is an excellent approximation, motivating us to  take the Keldysh-Schwinger prescription\footnote{In ordinary QFT the initial vacuum state $\Omega$ appearing in correlation functions $\langle \Omega |_{\text{in}} T[O(x,...)] |\Omega\rangle_{\text{in}}$ is equivalent up to a phase to the final vacuum state. For this special situation the operators are ordered along the 'flat' time axis ranging from $t_{\text{in}}=-\infty$ to $t_{\text{out}}=\infty$. By means of LSZ reduction formula it is then possible to relate correlation functions to the $S$ matrix and compute cross sections. This in-out formalism breaks down once, for example, the initial vacuum is not equivalent to the final state vacuum. An expanding background or external sources can introduce such a time dependence. In our work, there are mainly two sources of breaking the time translation invariance. First, since we have a thermal population, we consider traces of time-ordered operator products, where the trace is taken over all possible states. The many particle states are in general time dependent. Second, we have a density matrix next to the time ordering. The CTP, or, in-in formalism we adopt in this work can be, pragmatically speaking, seen as just a mathematical way of how to deal with such more general expectation values. The Keldysh description of the CTP contour applies if initial correlations can be neglected and we refer for a more detailed discussion and limitation to \cite{stefanucci_vanleeuwen_2013}.} of the CTP contour, as illustrated in Fig.~\ref{fig:contour}. The time contour $\mathcal{C}$ in the Keldysh-Schwinger prescription consists of two branches denoted by $\tau^+$ and $\tau^-$. The upper time contour $\tau^+$ ranges from the initial time $t_i=-\infty$ to $t=\infty$ while the lower contour $\tau^-$ is considered to go from $\infty$ back to $-\infty$. Therefore, times on the $\tau^-$ branch are said to be always later compared to the times on $\tau^+$. The time ordering of operator products on $\mathcal{C}$ can generically be written as
\begin{align}
T_{\mathcal C}[O_1(t_1)...O_n(t_n)] \equiv \sum_P \; (-1)^{F(p)} \theta_{\mathcal C}\left(t_{p(1)},t_{p(2)}\right) \;...\;\theta_{\mathcal C}\left(t_{p(n-1)},t_{p(n)}\right)\; O_{p(1)}(t_{p(1)})\;... \;O_{p(n)}(t_{p(n)}),
\end{align}
where the sum is over all the permutations $P$ of the set of operators $O_i$ and $F(p)$ is the number of permutations of fermionic operators. The unit step function and the delta distribution on the Keldysh-Schwinger contour is defined as
\begin{align}
\theta_{\mathcal C}(t_1,t_2) \equiv \left\{ \begin{array}{c} \theta(t_1-t_2) \; \text{ if } t_1,t_2 \in \tau^+ \\ \theta(t_2-t_1) \;  \text{ if } t_1,t_2 \in \tau^- \\1 \text{ if } t_1 \in \tau^-, t_2 \in \tau^+ \\ 0 \text{ if } t_1 \in \tau^+, t_2 \in \tau^-\end{array} \right. \!\hat{=} \left(\begin{array}{cc}
		\theta (t_1-t_2) & 0 \\
		1 & \theta (t_2-t_1)
	\end{array}\right),\quad \delta_{\mathcal C}(t_1,t_2)\hat{=} \left(\begin{array}{cc}
		\delta (t_1-t_2) & 0 \\
		0 & - \delta (t_1-t_2)
	\end{array}\right).
\end{align}
Correlation functions, i.e.\ contour $\cal C$-ordered operator products averaged over all states where the weight is the density matrix of the system denoted by $\hat \rho$, are defined by
\begin{align}
	\vev{ T_{\mathcal C} O(x_1,x_2,\dots,x_n)} \equiv \Tr \left[ \hat \rho \;T_{\mathcal C} O(x_1,x_2,\dots,x_n)  \right]. 
	\label{def:thermal_ev}
\end{align}
Let us introduce commonly used notations and properties of two-point correlation functions of fermionic or bosonic operator pairs relevant for this work. Because of the two-time structure, there are four possibilities to align the times $x^0$ and $y^0$ on $\mathcal{C}$ and hence four different components of a general two-point function denoted by $G(x,y)$, where in matrix notation it can be written as
\begin{align}
G(x,y) \equiv \vev{ T_{\mathcal C} \psi(x) \psi^{\dagger}(y)}= \theta_{\mathcal C} (x^0, y^0) \vev{ \psi (x) \psi^\dag (y)}
	\mp \theta_{\mathcal C} (y^0, x^0) \vev{ \psi^\dag (y) \psi (x) } \hat{=} \left(
		\begin{array}{cc}
		G^{++}(x,y) & G^{+-}(x,y) \\
		G^{-+}(x,y) & G^{--}(x,y)
		\end{array}
	\right).
	\label{eq:matrix}
\end{align}
Here, $G^{\sigma_x \sigma_y}$ means $x^0 \in \tau^{\sigma_x}$ and $y^0 \in \tau^{\sigma_y}$ with $\sigma_{i} = \pm$ for $i= x,y$ and the four different components of $G(x,y)$ are defined as:
\begin{align}
G^{-+}(x,y) &\equiv \vev{ \psi(x) \psi^{\dagger}(y)}, \\
G^{+-}(x,y) &\equiv \mp \vev{\psi^{\dagger}(y) \psi(x)}, \\
G^{++}(x,y) &\equiv \theta(x^0-y^0) G^{-+}(x,y) + \theta(y^0-x^0)G^{+-}(x,y),\\
G^{--}(x,y) &\equiv \theta(x^0-y^0) G^{+-}(x,y) + \theta(y^0-x^0)G^{-+}(x,y),
\end{align}
where $-$ $(+)$ on the r.h.s. of the second line applies for fermionic (bosonic) field operators. From these definitions one can recognize that not all components are independent, namely the following relation holds
\begin{align}
G^{++}(x,y)+G^{--}(x,y)=G^{+-}(x,y)+G^{-+}(x,y).
\end{align}
Furthermore, let us introduce \emph{retarded} and \emph{advanced} correlators defined by
\begin{align}
G^{R}(x,y)&\equiv \theta(x^0-y^0) \left[ G^{-+}(x,y) - G^{+-}(x,y)\right]= G^{++}(x,y) - G^{+-}(x,y) = -G^{--}(x,y) + G^{-+}(x,y), \\
G^{A}(x,y)&\equiv -\theta(y^0-x^0) \left[ G^{-+}(x,y) - G^{+-}(x,y)\right]= G^{++}(x,y) - G^{-+}(x,y) = -G^{--}(x,y) + G^{+-}(x,y),
\end{align}
as well as the \emph{spectral function} given by:
\begin{align}
G^{\rho}(x,y) \equiv G^{R}(x,y) - G^{A}(x,y) =  G^{-+}(x,y) - G^{+-}(x,y).\label{eq:sectral}
\end{align}
From these definitions we can derive further useful properties:
\begin{align}
G^A(x,y)&= - [G^R(y,x)]^{\dagger},\quad G^{+-}(x,y) = [G^{+-}(y,x) ]^{\dagger},\quad G^{-+}(x,y) = [G^{-+}(y,x) ]^{\dagger}.\label{eq:complexrelations}
\end{align}
In the case of free (unperturbed) propagators $G_0$, the following \emph{semigroup} property holds:
\begin{align}
G_0^{R}(x,y) &= \int\text{d}^3z \; G_0^{R}(x,z) G_0^{R}(z,y),\;\text{for }t_x>t_z>t_y.
\end{align}
This equality can be verified by noticing that for those time configurations the correlators are proportional to on-shell plane waves in Fourier space. Note that there is no time integration in the above equation. Together with the relations in Eq.~(\ref{eq:complexrelations}) further semigroup properties can be derived and all important ones are summarized for the use in subsequent sections in Appendix~\ref{app:semigr}.

As an example, the time integration over the Schwinger-Keldysh contour $\mathcal{C}$ of products of correlators can be written as: 
\begin{align}
C^{++}(x,y)&= \left[\int_{z^0 \in \mathcal{C}}\text{d}z^0\int \text{d}^3z\; A(x,z) B(z,y) \right]_{x^0,y^0 \in \tau^+ }
= \left[ \left(\int_{\tau^+}\text{d}z^0+\int_{\tau^-}\text{d}z^0 \right) \int \text{d}^3z\; A(x,z) B(z,y) \right]_{x^0,y^0 \in \tau^+ }\nonumber\\ &= \int_{-\infty}^{\infty}\text{d}z^0\int \text{d}^3z\; A^{++}(x,z) B^{++}(z,y) + \int_{\infty}^{-\infty}\text{d}z^0\int \text{d}^3z\; A^{+-}(x,z) B^{-+}(z,y)\nonumber\\
&=\int_{-\infty}^{\infty}\text{d}^4z\; \left( A^{++}(x,z) B^{++}(z,y)-A^{+-}(x,z) B^{-+}(z,y) \right). \label{eq:lagerethexample}
\end{align}
Eq.~(\ref{eq:lagerethexample}) is called a \emph{Lagereth rule} and it is straightforward to work out similar rules for, e.g.\ different components or double integrations of higher-order products of Keldysh-Schwinger correlators as they will appear later in this work.
Let us move on and define \emph{Wigner coordinates} according to
\begin{align}
T&=(t_x+t_y)/2,\; t=t_x-t_y,\\
\mathbf{R}&=(\mathbf{x}+\mathbf{y})/2,\; \mathbf{r}=\mathbf{x}-\mathbf{y}.
\end{align}
In the second line all variables are 3-vectors. The Wigner-transformed correlators are defined as
\begin{align}
\tilde{G}(t,\mathbf{r},\mathbf{R},T)\equiv G(T+t/2,\mathbf{R}+\mathbf{r}/2,T-t/2,\mathbf{R}-\mathbf{r}/2)=G(t_x \mathbf{x}, t_y \mathbf{y} ).
\end{align}
In all computations, the tilde will be dropped such that we can write for the Fourier transformation of $G(x,y)$:
\begin{align}
G(\omega,\mathbf{p},\mathbf{R},T)= \int \text{d}t \; \text{d}^3 r\; e^{i(\omega t-\mathbf{p}\cdot\mathbf{r})} G(t,\mathbf{r},\mathbf{R},T).
\end{align}
One of the great advantages of separating microscopic ($t,\mathbf{r}$) and macroscopic ($T,\mathbf{R}$) variables according to the Wigner transformation is that Fourier transformations of integral expressions can be considerably simplified by using the \emph{gradient expansion}.
For example, Eq.~(\ref{eq:lagerethexample}) in Fourier space can be written as
\begin{align}
C^{++}(\omega,\mathbf{p},\mathbf{R},T) &= A^{++}(\omega,\mathbf{p},\mathbf{R},T) \mathcal{G}_{AB} B^{++}(\omega,\mathbf{p},\mathbf{R},T)-A^{+-}(\omega,\mathbf{p},\mathbf{R},T) \mathcal{G}_{AB} B^{-+}(\omega,\mathbf{p},\mathbf{R},T) \\ &\simeq A^{++}(\omega,\mathbf{p},\mathbf{R},T) B^{++}(\omega,\mathbf{p},\mathbf{R},T) -A^{+-}(\omega,\mathbf{p},\mathbf{R},T) B^{-+}(\omega,\mathbf{p},\mathbf{R},T), \\
\mathcal{G}_{AB} &\equiv e^{-i(\partial^A_T\partial^B_\omega - \partial^A_\omega  \partial^B_T - \nabla^A_{\mathbf{R}}\cdot\nabla^B_{\mathbf{p}} +\nabla^A_{\mathbf{p }}\cdot\nabla^B_{\mathbf{R}})/2} \simeq 1.
\end{align}
Throughout this work, such exponentials containing derivatives are always approximated as in the last line, defining the leading order term of the gradient expansion. For homogeneous and isotropic systems the correlators do not depend on $R$ and thus for the spatial part the leading order term is exact. For a discussion of the validity of the leading order term of the temporal part we refer to \cite{Beneke:2014gla}. Let us emphasize here, that for typical DM scenarios the leading order term is always assumed to be a very good approximation.

Next, important properties of two-point correlators in thermal equilibrium are provided. Under this assumption, different components of correlators become related which are otherwise independent. For a system being in equilibrium (here we consider kinetic as well as chemical equilibrium), the density matrix takes the form
\begin{align}
\hat{\rho} \propto e^{-\beta H},
\end{align}
where $H$ is the Hamiltonian of the system and $\beta$ factor is the inverse temperature $T$ of the system. The density matrix in thermal equilibrium  can be formally seen as a time evolution operator, where the inverse temperature is regarded as an evolution in the imaginary time direction. Making use of the cyclic property of the trace it can be shown that under the assumption of equilibrium the components are related via
\begin{align}
G^{-+}(t_1- t_3) = \mp G^{+-}(t_1 - t_3 + i\beta).
\end{align}
This important property is called the \emph{Kubo-Martin-Schwinger} (KMS) relation, where $-$ ($+$) applies for two-point correlators of fermionic (bosonic) operators. Furthermore, in equilibrium the correlators should depend only on the difference of the time variables due to time translation invariance. Consequently, the KMS condition in Fourier space reads:
\begin{align}
G^{-+}(\omega,\mathbf{p})= \mp e^{\beta \omega} G^{+-}(\omega,\mathbf{p}).
\end{align}
From this condition, various equilibrium relations follow:
\begin{align}
G^{+-}(\omega,\mathbf{p})&= \mp\; n_{\text{F}/\text{B}}(\omega)G^{\rho}(\omega,\mathbf{p}),\quad G^{-+}(\omega,\mathbf{p})= \left[1\mp n_{\text{F}/\text{B}}(\omega)\right]G^{\rho}(\omega,\mathbf{p}),\\
G^{++}(\omega,p) & = \frac{G^{R}(\omega,\mathbf{p}) + G^{A}(\omega,\mathbf{p})}{2} + \! \left[\frac{1}{2} \mp  n_{\text{F}/\text{B}}(\omega)\right]\! G^{\rho}(\omega,\mathbf{p}).\label{eq:eqrel}
\end{align}
where the Fermi-Dirac or Bose-Einstein phase-space densities are given by $n_{\text{F}/\text{B}}(\omega) = 1/(e^{\beta\omega}\pm1)$. Thus in equilibrium, all correlator components can be calculated from the retarded/advanced components, where the spectral function $G^{\rho}$ is related to those via Eq.~(\ref{eq:sectral}).

General out-of-equilibrium observables, like the dynamic of the number density or spectral informations of the system, can be directly inferred from the equation of motions (EoM) of the corresponding correlators. Throughout this work, we derive the correlator EoM from the invariance principle of the path integral measure under infinitesimal perturbations of the fields. The equivalence of CTP-correlators and the path integral formulation is given by
\begin{align}
	\vev{ T_{\mathcal C} O(x_1,x_2,\dots,x_n)}  = \int \dd \mu_i \, \rho (\mu_i)\int_{\mu_i} [\dd \mu]\, O (x_1,x_2,\dots,x_n) e^{i \int_{x \in \mathcal{C}} \; \mathcal L(x)}. \label{eq:pathint}
\end{align}
and the action on the CPT contour is $S=\int_{x \in \mathcal{C}} \; \mathcal L(x)$. $\mu$ collectively represents the fields, and $ \rho$ stands for a state that could be either pure or mixed, as in Eq.~(\ref{def:thermal_ev}).
The second integral in Eq.~(\ref{eq:pathint}) is a path-integral with a boundary condition of $\mu_i$
at the initial time $t_i$ that we take to $- \infty$ in the Schwinger-Keldysh prescription of the contour, and the first one takes the average of $\mu_i$ with the weight of $\rho (\mu_i)$. Now, to derive the EoM for two-point correlators, let us consider an infinitesimal perturbation $O'{}^\dag = O^\dag + \epsilon$, satisfying $\epsilon (t_i) = 0$. By relying on the measure-invariance principle under this transformation, one obtains the EoM of the two-point correlators from
\begin{align}
	\vev{ O'{}^\dag (y)}
	&=
	\vev{ O^\dag (y)} + \epsilon(x) 
	+ i \int_{x \in \mathcal C} \epsilon (x) 
	\vev{ T_{\mathcal C} \frac{\delta S}{\delta_\text{L} O^\dag (x)} O^\dag (y)}
	+ \mathcal O (\epsilon^2) \\
	\Rightarrow \quad
	0 &= \delta_{\mathcal C} (x,y) + 
	i \vev{ T_{\mathcal C} \frac{\delta S}{\delta_\text{L} O^\dag (x)} O^\dag (y) },\label{eq:pathintegraleom}
\end{align}
where $\delta_\text{L}$ represents a derivative acting from the left. The same procedure can be applied to the case of $O $, as well as for deriving the EoM of higher correlation functions. The relation between the abstract EoM of correlators and differential equations for observables will be part of the next Section. In general, a correlator EoM depends on higher and lower correlators which is called the \emph{Martin-Schwinger Hierarchy}. For systems where the coupling expansion is appropriate, it might be sufficient to work in the one-particle self-energy approximation, where the EoM are closed in terms of two-point functions, and the kinetic equations can systematically be obtained by expanding the DM self-energy perturbatively in the coupling constant. The kinetic equations of two-point functions in the self-energy approximation are also known as the Kadanoff-Baym equations. For example, at NLO in the self-energy expansion of the two-point correlators the standard Boltzmann equation is recovered. 

Finite temperature corrections to non-perturbative systems, e.g.\ Sommerfeld-enhanced DM annihilations or bound-state decays, where a sub-class of higher-order diagrams becomes comparable to the LO in vacuum, are less understood in the CTP-formalism. The strategy in the next section will be to address this issue by going beyond the one-particle self-energy approximation \cite{Beneke:2014gla}. More precisely, we derive the exact Martin-Schwinger hierarchy of our particle setup in the CTP-formalism by using Eq.~(\ref{eq:pathintegraleom}) and truncate the hierarchy at the six-point function level. The system of equations is then closed with respect to two- \emph{and four-point} functions. This approach allows us to account for the resummation of the Coulomb diagrams and their finite temperature corrections at the same time. Furthermore, we show how to extract the DM number density equation from the EoM of two-point functions and that it depends on the four-point correlator. The complication is that the differential equation of the four-point correlator is coupled to the two-point correlator and in subsequent sections we solve this coupled systems of equations.

\section{Equations of motion in Real-time formalism for non-relativistic particles in a relativistic plasma environment}
\label{sec:eoms}
Throughout this paper, we consider the following minimal scenario capturing important effects to study long-range force enhanced DM annihilations and bound states under the influence of a hot and dense plasma environment:
\begin{align}
	\mathcal L = \bar \chi \left( i \slashed{\partial} - M \right) \chi + g_{\chi} \bar{\chi} \gamma^{\mu} \chi A_\mu + \bar \psi \left( i \slashed{\partial} - m \right) \psi + g_{\psi} \bar{\psi} \gamma^{\mu} \psi A_\mu - \frac{1}{4} F^{\mu \nu}F_{\mu \nu}.
\end{align}
The particles of the equilibrated plasma environment with temperature $T$ are the abelian mediators $A_\mu$ and the light fermionic particles $\psi$ with mass $m \ll T$. Fermionic DM $\chi$ is assumed to be nonrelativistic, i.e.\ $M \gg T$. All fermionic particles are considered to be of Dirac type. We assume the mediator to be massless, however, we provide the final results also for the case of a 
massive $A_\mu$ with mass $m_V \ll M$.

Let us illustrate how we can get the DM nonrelativistic effective action in the thermal medium of light particles. It is obtained in two steps. First, hard modes of $p \gtrsim M$ are integrated out. In this limit, 
the DM four component spinor $\chi$ splits into two parts, a term for the particle denoted by the two-component spinor $\eta$ and a term for the anti-particle denoted by $\xi$. Second, we assume that DM does not influence the plasma environment during the freeze-out process. This is typically the case since the DM number densities are smaller than that of light particles at this epoch. And thus, we may also integrate out the plasma fields by assuming they remain in thermal equilibrium. The resulting effective action on the CTP-contour for particle $\eta$ and anti-particle $\xi$ DM is given by:
\begin{align}
	\!S_\text{NR}[\eta,\xi] 
	\!= 
	\!\! \int\displaylimits_{x \in \mathcal C} \!
		\eta^\dag(x) \left[ i \der_{t} + \frac{\nabla^2}{2 M}  \right] \eta(x) +
		\xi^\dag(x) \left[i \der_{t} - \frac{\nabla^2}{2 M}  \right] \xi(x) 
	+ \!\!\!\! \int\displaylimits_{x,y \in \mathcal C}\!\!\! i \frac{g^2_{\chi}}{2} J(x) D(x,y) J (y) +
	i  O_s^\dag (x) \Gamma_s (x,y) O_s (y).\label{eq:nr_eft_u1}
\end{align}
Dark matter long-range force interactions are encoded in the first term of the interactions in Eq.~(\ref{eq:nr_eft_u1}). This term includes the current and the full two-point correlator of the electric potential on the CTP-contour which are defined by
\begin{align}
	J(x) \equiv \eta^\dag(x) \eta(x) + \xi^\dag(x) \xi(x), \quad D(x,y)\equiv \langle T_{\mathcal C} A_0(x) A_0(y) \rangle,\label{eq:currentandpotential}
\end{align}
respectively. The last term in Eq.~(\ref{eq:nr_eft_u1}) describes the annihilation part and we only keep the \textit{$s$-wave} contribution
\begin{align}
	\quad
	O_s(x) \equiv \xi^\dag(x) \eta(x), 
	\quad 
	\Gamma_s (x,y) \equiv 
	\frac{\pi (\alpha^2_{\chi} + \alpha_\chi \alpha_\psi)}{M^2}
	\left(
		\begin{array}{cc}
		\delta^4 (x - y) & 0 \\
		2 \delta^4 (x - y) & \delta^4 (x - y)
		\end{array}
	\right),
	\label{eq:ann_swave_u1}
\end{align}
with the fine structure constant being $ \alpha_i \equiv g^2_i / 4 \pi$, and summation over the spin indices are implicit. $\Gamma_s$ is shown in the matrix representation of the CTP formalism, see e.g.\ Eq.~\eqref{eq:matrix} in previous Section. Hence the delta functions on the right-hand-side are defined on the usual real-time axis. Similar to the vacuum theory, the annihilation part $\Gamma_s$ can be computed by cutting the box diagram (containing two $A_{\mu}$) and the vacuum polarization diagram (containing one $ A_{\mu} $
and a loop of light fermions $\psi$), where now all propagators are defined on the CPT-contour. Finite temperature corrections to these hard processes in $\Gamma_s$ are neglected\footnote{Assuming free plasma field correlators in the computation of $\Gamma_s$ is a good approximation since the energy scale of the hard process is $\sim M$ which is much larger for nonrelativistic particles than typical finite temperature corrections being of the order $\sim g_{\psi} T$. Consequently, the dominant thermal corrections should be in the modification of the long-range force correlator $D$, where the typical DM  momentum-exchange scale enters which is much lower compared to the annihilation scale.} and for a derivation we refer to Appendix \ref{app:annihilation}.

In our effective action Eq.~(\ref{eq:nr_eft_u1}), we have discarded higher order terms in $\nabla / M$ (like magnetic interactions)
and also interactions with ultra-soft gauge bosons,\footnote{
To fully study the dynamics of the bound state formation~\cite{Brambilla:2008cx, vonHarling:2014kha, Petraki:2015hla, Petraki:2016cnz} at late times of the freeze-out process (e.g. $T \lesssim |E_B|$ with $E_B$ being a typical binding energy), it is necessary to include emission and absorption via ultra-soft gauge bosons, e.g.\ via an electric dipole operator.
We drop for simplicity ultra-soft contributions and discuss in detail the limitation of our approach later in this work, see Section \ref{sec:ionizatineq}. Note that at high enough temperature $T \gtrsim |E_B|$ those processes are typically efficient, leading just to the ionization equilibrium among bounded and scattering states. To estimate the time when the ionization equilibrium is violated concretely, we have to take into account these processes in the thermal plasma, which will be presented elsewhere.
}
since we focus on threshold singularities of annihilations at the leading order in the coupling $g_{\chi}$
and the velocity $ \nabla / M$. Furthermore, our effective field theory is non-hermitian because we have integrated out (or traced out) hard and thermal degrees of freedom. The first source of non-Hermite nature is the annihilation term which originates from the integration of hard degrees of freedom. A similar term would also be present in vacuum \cite{Hisano:2003ec, Hisano:2004ds, Hisano:2006nn} and belongs to the ++ component of $\Gamma$. Thus, as a first result we have generalized the annihilation term towards the CTP contour. Another one stems from the gauge boson propagator $D$ that encodes interactions with the thermal plasma. While the annihilation term containing $\Gamma_s$ in our action breaks the number conservation of DM, the interaction term containing $D$ can not. From this observation, one may anticipate that the non-hermitian potential contributions of the gauge boson propagator never lead 
to a violation of the DM particle or anti-particle number conservation. Later, we will show this property directly from the EoM, respecting the global symmetries of our action.

In the next sections we proceed as in the following. First, we compute the finite temperature one-loop corrections contained in the potential term $D$ explicitly. Since the number density of DM becomes Boltzmann suppressed in the non-relativistic regime of the freeze-out process, the dominant thermal loop contributions arise from the relativistic species $\psi$. This implies that we can solve for $D$ independently of the DM system since we assume DM does not modify the property of the plasma. The correction terms for the DM self-interactions are screening effects on the electric potential, as well as imaginary contributions arising from soft DM-$\psi$ scatterings, derived and discussed in detail in Section~\ref{sec:potential}. Second, in Section~\ref{eq:dmeom} the kinetic equations for the DM correlators are derived. We show how to extract from these equations the number density equation, including finite temperature corrected processes for the negative energy spectrum (bound-state decays) as well as for the positive energy contribution (Sommerfeld-enhanced annihilation) in one single equation.

\subsection{Thermal corrections to potential term}
\label{sec:potential}
In this section, we briefly summarize how the electric component of the mediator propagator $D$ gets modified by the thermal presence of ultrarelativistic $\psi$ fields. The plasma environment is regarded to be perturbative and in the one-particle self-energy framework we can write down the Dyson equation on the Schwinger-Keldysh contour for the mediator:
\begin{align}
D_{\mu\nu}(x,y) &= D^0_{\mu\nu}(x,y) - i\int_{w,z \in \mathcal{C}} D^0_{\mu\alpha}(x,w)\Pi^{\alpha\beta}(w,z) D_{\beta\nu}(z,y),\label{eq:dysonmediator} \\
\Pi^{\mu\nu}(x,y)&=(-i)g_\psi^2 (-1)\Tr[\gamma^{\mu} S(x,y) \gamma^{\nu} S(y,x)],
\end{align}
where $S(x,y) \equiv \vev{T_{\mathcal{C}}\psi(x)\bar{\psi}(y)}_0 $ are unperturbed $\psi$ correlators.
$A_{\mu}$ and $\psi$ are assumed to be in equilibrium and thus, according to the discussion in Section \ref{sec:rtf}, we only need to compute the retarded/advanced propagators. From those, we can construct all other components by using KMS condition. The Dyson equation for retarded (advanced) mediator-correlator can be obtained by subtracting the $+-$ $(-+)$ component of Eq.~(\ref{eq:dysonmediator}) from the $++$ component of the same equation. In momentum space this results in:
\begin{align}
D_{\mu\nu}^{R/A} = (D^{R/A}_{\mu\nu})_0 -i (D^{R/A}_{\mu\alpha})_0 \Pi^{\alpha \beta}_{R/A}D_{\beta\nu}^{R/A},
\end{align}
where the mediator's retarded self-energy is defined as:
\begin{align}
\Pi^{\mu \nu}_R \left( P\right) = \Pi^{\mu \nu}_{++}  - \Pi^{\mu \nu}_{+-} =i g_\psi^2 \int \frac{\mathrm{d^4}K}{(2 \pi)^4} \left( \Tr[ \gamma^{\mu} S^{++}(K-P) \gamma^{\nu} S^{++}(K)] - \Tr[ \gamma^{\mu} S^{+-}(K-P) \gamma^{\nu} S^{-+}(K)] \right).\label{eq:thermaloneloop}
\end{align}
A sketch of efficiently calculating the thermal one-loop Eq.~(\ref{eq:thermaloneloop}) is provided in Appendix \ref{app:htl}. 
In the computation we utilize the \emph{Hard Thermal Loop} (HTL) approximation \cite{Braaten:1989mz} to extract leading thermal corrections.\footnote{
	Let us briefly summarize here the assumptions of the HTL approximation. First, we drop all vacuum contributions and only keep temperature dependent parts. Second, we assume the external energy $P^0$ and momentum $\mathbf{p}$ to be much smaller than typical loop momentum $\mathbf{k}$ which is of the order temperature (see Appendix \ref{app:htl}). The discussion of the validity of the HTL approximation depends on where the dressed mediator correlator is attached to. One can not naively argue for the case where one would attach it to the DM correlator that the external momentum $\mathbf{p}$ is the DM momentum which is of course much larger than temperature, thus invalidating the HTL approximation by this argumentation. For example, in our case the dressed mediator correlator enters the DM single-particle self-energy (see Fig.~\ref{fig:resummation} and Section~\ref{sec:htlstaticlimit}) and the dominant energy and momentum region in the loop diagram is where HTL effective theory is valid. 
} 
In the HTL approximation we are allowed to resum the self-energy contributions of the retarded/advanced component and the result is gauge-independent. We work in the non-covariant Coulomb gauge which is known to be fine at finite temperature since Lorentz invariance is anyway broken by the plasma temperature. We find for the dressed longitudinal component $\mu = \nu =0$ of the mediator propagator in the HTL resummed self-energy approximation and in the Coulomb gauge the following:
\begin{align}
D^{R,A}(\omega,\mathbf{p})&= \frac{-i}{\omega^2 -\mathbf{p}^2+ \Pi^{00}_{\text{R,A}}(\omega,\mathbf{p})\pm i\text{sign}(\omega)\epsilon},\\
\Pi^{00}_{\text{R,A}}(\omega,\mathbf{p}) &= -m^2_D \left[ 1- \frac{\omega}{2 |\mathbf{p}|} \ln{ \left( \left| \frac{\omega +|\mathbf{p}|}{\omega -|\mathbf{p}| } \right| \right)}  \pm i \frac{\omega}{|\mathbf{p}|} \frac{\pi}{2} \theta{\left(|\mathbf{p}|^2 -\omega^2\right)}\right],
\end{align}
where we introduced the Debye screening mass $m_D^2 = g_\psi^2 T^2 /3$. One can recognize that there is correction to the real part
of the mediator propagator as well as a branch cut for space-like exchange. Using the equilibrium relation Eq.~(\ref{eq:eqrel}), the $++$ component of $D$ in the static limit reads
\begin{align}
\lim_{\omega \rightarrow 0} D^{++}(\omega,\mathbf{p}) = \lim_{\omega \rightarrow 0} \left[ \frac{D^R(\omega,\mathbf{p})+D^A(\omega,\mathbf{p})}{2} + \left(\frac{1}{2}+n_B(\omega)\right)D^\rho(\omega,\mathbf{p})\right]= \frac{i}{\mathbf{p}^2+m_D^2} + \pi \frac{T}{|\mathbf{p}|} \frac{m_D^2}{\left(\mathbf{p}^2 + m_D^2 \right)^2}, \label{eq:staticHTL}
\end{align}
while for a massive mediator we have simply
\begin{align}
\lim_{\omega \rightarrow 0} D^{++}(\omega,\mathbf{p}) = \frac{i}{\mathbf{p}^2+m_V^2+m_D^2} + \pi \frac{T}{|\mathbf{p}|} \frac{m_D^2}{\left(\mathbf{p}^2 +m_V^2+ m_D^2 \right)^2}. \label{eq:staticHTLmassive}
\end{align}
The static $D^{++}$ component is of special importance for describing DM long-range interactions in a plasma environment as we will see later in this work. The first term in Eq.~(\ref{eq:staticHTL}) and Eq.~(\ref{eq:staticHTLmassive}) will result in a screened Yukawa potential after Fourier transformation while the second terms will lead to purely imaginary contributions. Physically, the latter part originates from the scattering of the photon with plasma fermions, leading to a damping rate~\cite{Bellac:2011kqa}.
Indeed in the quasi-particle picture, the mediator has a limited propagation time within the plasma, which limits as well the coherence of
the mediator exchange processes.
For what regards the DM particles, this term will later give rise to DM-$\psi $ scattering with zero energy transfer, leading also to a
thermal width for the DM states.
In following Sections, we try to keep generality and work in most of the computations with the unspecified form $D(x,y)$ and take just at the very end the static and HTL limit.
Let us finally remark that the simple form of Eq.~(\ref{eq:staticHTLmassive}) allows us to achieve semi-analytical results for the DM annihilation or decay rates in the presence of a thermal environment. 

\subsection{Exact DM number density equation from correlator equation of motion}
\label{eq:dmeom}
The main purpose of this section is to derive the equation for the DM number density directly from the exact EoM of our nonrelativistic action.
Defining the DM particle and anti-particle correlators as
\begin{align}
	G_\eta (x, y) & \equiv \vev{ T_{\mathcal C} \eta (x) \eta^\dag (y)}, \\
	G_\xi (x, y) & \equiv \vev{ T_{\mathcal C} \xi (x) \xi^\dag (y)},
\end{align}
we derive respective EoM from the path-integral formalism, as briefly explained at the end of Section~\ref{sec:rtf}, for the nonrelativistic effective action $S_\text{NR}$ given in Eq.~\eqref{eq:nr_eft_u1}:
\begin{align}
\left[i \der_{x^0} + \frac{\nabla^2_x}{2 M}\right]\! G_\eta (x,y)\! &=\! i \delta_\C (x,y)\!-\! i g^2_{\chi} \!\int\displaylimits_{z \in \mathcal{C}} D (x,z) \vev{T_\C \eta (x) J(z) \eta^\dag (y)} \!-\! i\! \int\displaylimits_{z \in \mathcal{C}}  \Gamma_s (x,z)  \vev{ T_\C \xi (x) \{ \xi^\dag (z) \eta (z) \} \eta^\dag (y) },\label{eq:Geta_x}\\
\left[- i \der_{y^0} + \frac{\nabla^2_y}{2 M}\right]\! G_\eta (x,y)\!&=\! i \delta_\C (x,y)\!-\! i g^2_{\chi}  \!\int\displaylimits_{z \in \mathcal{C}} D (y,z) \vev{T_\C \eta (x) J(z) \eta^\dag (y)}\! -\! i\! \int\displaylimits_{z \in \mathcal{C}} \Gamma_s (z,y)  \vev{ T_\C \eta (x) \{ \eta^\dag (z) \xi (z) \} \xi^\dag (y) },\label{eq:Geta_y}\\
\left[i \der_{x^0} - \frac{\nabla^2_x}{2 M}\right]\!G_\xi (x,y)\!	&=\! i \delta_\C (x,y)\!-\! i g^2_{\chi}  \!\int\displaylimits_{z \in \mathcal{C}} D (x,z) \vev{T_\C \xi (x) J(z) \xi^\dag (y)}\! -\! i\! \int\displaylimits_{z \in \mathcal{C}}  \Gamma_s (x,z)  \vev{ T_\C \eta (x) \{ \eta^\dag (z) \xi (z) \} \xi^\dag (y) },\label{eq:Gxi_x}\\
\left[- i \der_{y^0} - \frac{\nabla^2_y}{2 M}\right]\!G_\xi (x,y)\!&=\! i \delta_\C (x,y)\!-\! i g^2_{\chi} \!\int\displaylimits_{z \in \mathcal{C}} D (y,z) \vev{T_\C \xi (x) J(z) \xi^\dag (y)}\!- \!i\! \int\displaylimits_{z \in \mathcal{C}}  \Gamma_s (y,z)  \vev{ T_\C \xi (x) \{ \xi^\dag (z) \eta (z) \} \eta^\dag (y)  }.\label{eq:Gxi_y}
\end{align}
The anticipated structure in Eq.~(\ref{eq:Geta_x})-(\ref{eq:Gxi_y}) shows the dependence of the two-point correlators on higher correlation functions. Here, the curly brackets stand for the summation over the spin indices and $J$ is the current as already introduced in Eq.~(\ref{eq:currentandpotential}). It might be helpful to mention that we used a special property of two-point functions of Hermitian bosonic field operators: $D(x,y)=D(y,x)$. This exact property can be verified directly from the definition in Eq.~(\ref{eq:currentandpotential}).

In the following, the number density equation of particle and anti-particle DM is derived from this set of differential equations. First of all, we would like to clarify what is the number density in terms of fields appearing in $S_{\text{NR}}$ in Eq.~\eqref{eq:nr_eft_u1}. For this purpose, let us switch off the annihilation term $\Gamma_s\rightarrow0$ in $S_{\text{NR}}$ and seek for conserved quantities. In this limit, the theory has the following global symmetries: $\eta \mapsto e^{i \theta_\eta} \eta$  and $\xi \mapsto e^{- i \theta_\xi} \xi$.
The associated Noether currents for DM particle and anti-particle are:
\begin{align}
	J_\eta^\mu(x) = \left(
	\eta^\dag(x) \eta(x),\, \frac{1}{2 i M} \eta^\dag(x) \overleftrightarrow{\nabla} \eta(x)
	\right),
	\quad
	J_\xi^\mu(x) = \left(
	\xi(x) \xi^\dag(x),\, \frac{1}{2 i M} \xi(x) \overleftrightarrow{\nabla} \xi^\dag(x)
	\right).
\end{align}
The thermal-averaged zeroth component is the number density and the average over spatial component results in the current density.
We obtain the differential equation for the two DM currents directly from the two-point function EoM, by subtracting Eqs.~\eqref{eq:Geta_y} from Eqs.~\eqref{eq:Geta_x} and Eqs.~\eqref{eq:Gxi_y} from Eqs.~\eqref{eq:Gxi_x}, and by taking the spin-trace and the limit $y\rightarrow x$. For the particle DM, we obtain as an intermediate result after all these steps:
\begin{align}
	i \der_\mu \vev{J_\eta^\mu} 
	=& -
	\left(
		i \der_{x^0} + i \der_{y^0} + \frac{\nabla^2_x}{2 M} - \frac{\nabla^2_y}{2M}
	\right)
	\left. \Tr\, G_\eta (x,y) \right|_{y \to x} \nonumber\\
	= &
	+ i g^2_{\chi} \int_{ z \in \mathcal C} D (x,z) 
	\left[
		\vev{ T_{\mathcal C} \{ \eta (x) \eta^\dag (x) \} J (z)}
		-
		\vev{ T_{\mathcal C} \{ \eta (x) \eta^\dag (x) \} J (z)}
	\right] \nonumber\\
	&
	- i \int_{ z \in \mathcal C} 
	\left[ 
		\Gamma_s (x,z) \vev{ T_{\mathcal C} \{  \eta (z) \xi^\dag (z) \}  \{ \xi (x)  \eta^\dag (y)\}}
		- \Gamma_s (z,y) \vev{ T_{\mathcal C} \{\eta (x)\xi^\dag (y)\} \{  \xi (z) \eta^\dag (z) \} }
	\right]_{y \to x}.
	\label{eq:number_eta_ann}
\end{align}
The trace and the curly brackets indicate the summation over the spin indices.
It is important to note that the first line in the second equality cancels out, even in the case of a fully interacting correlator $D$ including finite temperature corrections. 
Thus, we confirm from the EoM that, e.g.\ non-Hermitian potential corrections arising from soft thermal DM-$\psi$ scatterings in the HTL approximation of $D$ [see Eq.~(\ref{eq:staticHTL})], never violate the current conservation of each individual DM species. For a homogeneous and isotropic system (vanishing divergence of current density) this would mean that the individual number densities of particles and antiparticles do not change by self-scattering processes, real physical DM-$\psi$ scatterings, soft DM-$\psi$ scatterings or other finite temperature corrections leading to potential contributions in $D$.  

It can be recognized, that the current conservation is only violated by the annihilation term $\Gamma_s$ in the last line in Eq.~\eqref{eq:number_eta_ann}, since this contribution does not cancel to zero. This term can be simplified by using Eq.~\eqref{eq:ann_swave_u1} and by fixing the time component $x^0$ to either $\tau^+$ or $\tau^-$. We have explicitly checked that both choices of $x^0$ lead to the same final result. With the definition of the four-point correlator on the closed-time-path contour
\begin{align}
	G_{\eta \xi,s} (x,y,z,w) \equiv 
	\vev{ T_{\mathcal C} \eta^i (x) \xi^\dag_i (y)  \xi^{j} (w) \eta^\dag_{j} (z) },
\end{align}
we obtain our final form of the current equations.
\begin{claim}
\textit{Current equations for particle $\eta$ and anti-particle $ \xi$}:
\begin{align}
	\der_\mu \langle J_\eta^\mu (x) \rangle &=  - 2 \frac{ \pi (\alpha^2_{\chi} + \alpha_\chi \alpha_\psi)}{M^2} G_{\eta \xi,s}^{++--} (x,x,x,x),\label{eq:Jeta}\\ 
	\der_\mu \langle J_\xi^\mu  (x) \rangle &= - 2 \frac{ \pi (\alpha^2_{\chi} + \alpha_\chi \alpha_\psi)}{M^2} G_{\eta \xi,s}^{++--} (x,x,x,x).\label{eq:Jxi}
\end{align}
\end{claim}
The current conservation is only violated by contributions coming from $\Gamma_s$. 
This is consistent with the expectations from the symmetry properties of the action when annihilation is turned on. Namely, only a linear combination of both global transformations leaves the action invariant which leads to the conservation of $\der_\mu(J_\eta^\mu - J_\xi^\mu)=0$, which is nothing but the DM asymmetry current conservation. The conservation of the total
DM number density, $\der_\mu(J_\eta^\mu + J_\xi^\mu)$, is violated by the annihilation term. 


Before we discuss the four-point correlator appearing in Eq.~(\ref{eq:Jeta}) and Eq.~(\ref{eq:Jxi}) in detail, let us now assume a Friedman-Robertson-Walker universe and make the connection to the Boltzmann equation for the number density that is typically adopted in the literature when calculating the thermal history of the dark matter particles. First, the spatial divergence on the left hand side of the current Eqs.~\eqref{eq:Jeta} and \eqref{eq:Jxi} vanishes due to homogeneity and isotropy. Second, the adiabatic expansion of the background introduces a Hubble expansion term $H$. Third, it can be seen from the sign of the right hand side of the current equations that only a DM loss term occurs. The production term is missing because we have assumed \emph{a priori}, when deriving the nonrelativistic action, that the DM mass is much larger than the thermal plasma temperature. Within this mass-to-infinity limit the DM production term is send to zero in the computation of the annihilation term $\Gamma_s$ and not expected to occur. Let us therefore add on the r.h.s of the current equations \emph{a posteriori} the production term of the DM via the assumption of detailed balance, resulting in the more familiar number density equations:
\begin{claim}
\begin{align}
\dot{n}_{\eta} + 3 H  n_{\eta} &= - 2(\sigma v_{\text{rel}}) \left[G_{\eta \xi,s}^{++--} (x,x,x,x) - G_{\eta \xi,s}^{++--} (x,x,x,x)\big|_{\text{eq}}\right],\label{eq:numberp}\\
\dot{n}_{\xi} + 3 H n_{\xi}  &= - 2(\sigma v_{\text{rel}}) \left[ G_{\eta \xi,s}^{++--} (x,x,x,x)-G_{\eta \xi,s}^{++--} (x,x,x,x)\big|_{\text{eq}}\right].\label{eq:numberantip}
\end{align}
\end{claim}

The tree-level s-wave annihilation cross section of our system was defined as $(\sigma v_{\text{rel}}) =  \pi (\alpha^2_{\chi} + \alpha_\chi \alpha_\psi)/M^2$ and $ |_{\text{eq}} $ in the last term means the evaluation at thermal equilibrium. Note that in the CTP formalism a cross section strictly speaking does not exist. The reason why this result is equal to the vacuum computation is because we computed the annihilation part $\Gamma_s$ at the leading order, where it is expected that zero and finite temperature results should coincide. The correlation function $G_{\eta \xi,s}^{++--}$ however is fully interacting. We summarize with two concluding remarks on our main results: 
\begin{itemize}
\item \emph{Sommerfeld-enhancement factor at finite temperature:} One of our findings is that the Sommerfeld factor is contained in a certain component of the interacting four-point correlation function, namely $G_{\eta \xi,s}^{++--}$. This result is valid for a generic out-of-equilibrium state of the dark matter system. The remaining task is to find a solution for this four-point correlator. As we will see later, the solution can be obtained from the Bethe-Salpeter equation on the CTP contour, derived in the next Section. For example, expanding the Bethe-Salpeter equation to zeroth order in the DM self-interactions $2 G_{\eta \xi,s}^{++--} (x,x,x,x) \simeq - 2 G_{\eta}^{+-}(x,x) G_{\xi}^{+-}(x,x)= n_{\eta} n_{\xi}$ and inserting this into Eq.~(\ref{eq:numberp}) and Eq.~(\ref{eq:numberantip}) would result in a well-known expression for the number density equation of the DM particles with velocity-independent annihilation. As we will see later, higher terms in the interaction or a fully non-perturbative solution contain the finite temperature corrected negative and positive energy spectrum. In other words, $G_{\eta \xi,s}^{++--}$ contains both, the bound state \emph{and} scattering state contributions at the same time and at finite temperature they turn out to be not separable as it is sometimes done in vacuum computations. Bound state contributions will automatically change the cross section into a decay width and thus, $n_{\eta}$ appearing on the l.h.s. of Eq.~(\ref{eq:numberp}) is the \emph{total} number of $\eta$ particles including the ones in the bound states and similar interpretation for the anti-particle $\xi$.
\item \emph{Particle number-conservation:} In Section \ref{sec:potential}, we have seen that the thermal corrections to the mediator propagator $D$ can contain, next to the real Debye mass, an imaginary contribution. It was shown that these non-hermitian corrections to the potential never violate the particle number-conservation due to the exact cancellation of the second line in Eq.~\eqref{eq:number_eta_ann}. This was expected from the beginning, since, when switching-off the annihilation $\Gamma_s\rightarrow0$, the nonrelativistic action in Eq.~(\ref{eq:nr_eft_u1}) has two global symmetries $	\eta \mapsto e^{i \theta_\eta} \eta$ and $\xi \mapsto e^{- i \theta_\xi} \xi$. The conserved quantities are the particle and anti-particle currents in Eq.~(\ref{eq:Jeta}) and (\ref{eq:Jxi}) in the limit $\Gamma_s\rightarrow0$ (vanishing r.h.s).
When annihilation is included, the nonrelativistic action is only invariant if both global transformations are performed at the same time, resulting in the conserved asymmetry current $J_\eta - J_\xi$. We conclude that thermal corrections can never violate these symmetries, even not at higher loop level. On the other hand, the solution of the Sommerfeld factor is contained in $G_{\eta \xi,s}^{++--}$ and hence the annihilation rate will depend on the thermal loop corrected longe-range mediator $D$, as we will see in the next section.
\end{itemize}

\section{Two-time Bethe-Salpeter equations}
\label{sec:closedequations}
The exact number density Eq.~(\ref{eq:numberp}) depends on the Keldysh four-point correlation function $G^{++--}_{\eta \xi,s}(x,x,x,x)$. In this Section, we derive the system of closed equation of motion needed in order to obtain a solution for this four-point function, including the full resummation of Coulomb divergent ladder diagrams. The result will be a coupled set of two-time Bethe-Salpeter equations on the Keldysh contour as given by the end of this section, Eq.~(\ref{eq:twotimeother})-(\ref{eq:twotimeret}). They apply in general for out-of-equilibrium situations and include in their non-perturbative form also the bound-state contributions if present. In order to arrive at those equations, a set of approximations and assumptions is needed. We therefore would like to start from the beginning in deriving those equations, which might lead to a better understanding of their limitation.

In the first simplification, we treat the annihilation term $\Gamma_s$ as a perturbation and ignore it in the following computations, since the leading order term in the annihilation part is already contained in  Eq.~(\ref{eq:numberp}). The exact set of EoMs for two- and four-point functions in the limit $\Gamma_s \rightarrow 0$ are given by:
\begin{align}
\left[i \der_{x^0} + \frac{\nabla^2_x}{2 M}\right]\! G_\eta (x,y)\! &=\! i \delta_\C (x,y)\!-\! i g^2_{\chi} \!\int_{z \in \mathcal{C}} D (x,z)[G_{\eta\xi}(x,z,y,z)-G_{\eta\eta}(x,z,y,z)],\label{eq:eta}\\
\left[i \der_{x^0} + \frac{\nabla^2_x}{2 M}\right]\!\bar{G}_\xi (x,y)\!	&=\! i \delta_\C (x,y)\!-\! i g^2_{\chi}  \!\int_{z \in \mathcal{C}} D (x,z) [G_{\eta\xi}(z,x,z,y)-G_{\xi\xi}(x,z,y,z)],\!\label{eq:xi}\\
\left[i \der_{x^0} + \frac{\nabla^2_x}{2 M} \right]G_{\eta \xi}(x,y,z,w) &= i \delta_\C (x,z) \bar{G}_{ \xi}(y,w)-i g^2_{\chi} \int_{\bar{x} \in \C} D(x,\bar{x} ) \vev{ T_\C \eta(x) J(\bar{x} ) \xi^{\dag}(y) \xi(w) \eta^\dag (z) }, \label{eq:eom4p6p}
\end{align}
where we will work for the rest of this work with the conjugate anti-particle correlator $\bar G_\xi$ and, here, the spin-uncontracted correlators are defined as
\begin{align}
	\bar G_\xi (x,y) &\equiv \vev{ T_\C \xi^\dag (x) \xi (y) } \label{eq:conjungate},\\
	G_{\eta \xi} (x,y,z,w) &\equiv \vev{T_\C \eta (x) \xi^\dag (y)  \xi (w) \eta^\dag (z)},\\
	G_{\eta \eta} (x,y,z,w) &\equiv \vev{T_\C \eta (x) \eta (y) \eta^\dag (w) \eta^\dag (z)},\\
	G_{\xi \xi} (x,y,z,w) &\equiv \vev{T_\C \xi^\dag  (x) \xi^\dag (y) \xi (w) \xi (z)}.\label{eq:etaetadef}
\end{align}
The EoMs for the two-point functions Eq.~(\ref{eq:eta}) and ~(\ref{eq:xi}) are equivalent to Eq.~(\ref{eq:Geta_x}) and (\ref{eq:Gxi_x}) in the limit $\Gamma_s \rightarrow 0$, and we have just rewritten them in terms of the four-point correlators and conjugate anti-particle propagator, defined in Eq.~(\ref{eq:conjungate})-(\ref{eq:etaetadef}). In our notation, the spinor indices of the operators having equal space-time arguments in the four-point correlators are summed and $J$ is the current as defined in Eq.~(\ref{eq:currentandpotential}).
The different conventions for the $\eta \eta$ and $\xi \xi$ four-point correlators are because $\eta^{\dagger}$ and $\xi$ are the creation operators. From the exact differential equation of the four-point correlator in Eq.~(\ref{eq:eom4p6p}), it can be seen that the correlator hierarchy is still not closed yet, since it depends on the six-point function.
We close this hierarchy of correlators by truncating the six-point function at the leading order: 
\begin{align}
\!\left[i \der_{y^0} + \frac{\nabla^2_y}{2 M}\right]\vev{ T_\C \eta(x)J(\bar{x}) \xi^{\dagger}(y) \xi(w) \eta^{\dagger}(z) } &\simeq  i  \delta_\C(y,w) [G_{\eta\xi}(x,\bar{x},z,\bar{x})-G_{\eta\eta}(x,\bar{x},z,\bar{x})]  \!- \!i \delta_\C(\bar{x},y) G_{\eta \xi} (x,y,z,w) , \label{eq:closerms}
\end{align}
i.e. \emph{only the integral kernel of the six point function containing the eight-point function was dropped}.
The set of correlator differential equations is now closed under this truncation procedure. A fully self-consistent solution requires in principle to solve the equations for the five correlators $G_{\eta}, G_{\xi},G_{\eta\xi},G_{\eta\eta},G_{\xi \xi}$ simultaneously. This is beyond the scope of this work and we have to further approximate the system in order to obtain at least a simple semi-analytical solution at the end. 

The solution of our target component $G_{\eta \xi}$ can formally be decoupled from the solution of $G_{\eta\eta}$ and $G_{\xi \xi} $ by approximating the latter two quantities as the leading order contribution (dropping the integral kernel, known as Hartree-Fock approximation). Then, one can recognize that
\begin{align}
[G_{\eta\xi}(x,z,y,z)-G_{\eta\eta}(x,z,y,z)] &\simeq [G_{\eta}(x,y) \bar{G}_{\xi}(z,z) - G_{\eta}(x,y) G_{\eta}(z,z) + G_{\eta}(x,z) G_{\eta}(z,y)]=   G_{\eta}(x,z) G_{\eta}(z,y),\label{eq:hartree1}\\
[G_{\eta\xi}(z,x,z,y) -G_{\xi\xi}(x,z,y,z) ] &\simeq [  \bar{G}_{\xi}(x,y) G_{\eta}(z,z) -\bar{G}_{\xi}(x,y) \bar{G}_{\xi}(z,z) + \bar{G}_{\xi}(x,z) \bar{G}_{\xi}(z,y)] =  \bar{G}_{\xi}(x,z) \bar{G}_{\xi}(z,y). \label{eq:hartree2}
\end{align}
In both equations, the last step is a strict equality only if the DM particle $G_{\eta}(z,z) = - n_{\eta}$  and anti-particle $\bar{G}_{\xi}(z,z)=-n_{\xi}$ number densities are equal. This is true if there is no DM asymmetry which we will assume throughout this work.

In the last approximation, we perform a coupling expansion of Eq.~(\ref{eq:eta})-(\ref{eq:eom4p6p}). After inserting the results of the two-point functions into the equation of $G_{\eta\xi}$, by using the relations Eq.~(\ref{eq:hartree1}) and (\ref{eq:hartree2}) in the free limit, we obtain for the four-point correlator to the leading order in $g_{\chi}$:
\begin{align}
G_{\eta \xi} (x,y,z,w)  \simeq G_{\eta,0} (x,z) \bar{G}_{\xi,0} (y,w) + g^2_{\chi}&\int_{\bar{x}, \bar{y} \in \mathcal{C}}  G_{\eta,0} (x,\bar{x}) \bar{G}_{\xi,0} (y,\bar{y}) D(\bar{x},\bar{y}) G_{\eta,0} (\bar{x},z) \bar{G}_{\xi,0} (\bar{y},w) \nonumber\\
+ &\int_{\bar{x}, \bar{y} \in \mathcal{C}} G_{\eta,0} (x,z) \bar{G}_{\xi,0} (y,\bar{y})(-i) \bar{\Sigma}_{\xi}(\bar{y} ,\bar{x})\bar{G}_{\xi,0} (\bar{x},w) \nonumber \\
+ &\int_{\bar{x}, \bar{y} \in \mathcal{C}}  G_{\eta,0} (x,\bar{x}) \bar{G}_{\xi,0} (y,w)(-i) \Sigma_{\eta}( \bar{x},\bar{y})G_{\eta,0} (\bar{y},z)\label{eq:couplingexp},
\end{align}
where in the last two terms, particle and anti-particle are disconnected and we have introduced the \emph{single-particle self-energies} according to 
\begin{align}
\Sigma_{\eta}(x,y) \equiv -i g^2_{\chi} D(x,y) G_{\eta,0}(x,y),\quad \bar{\Sigma}_{\xi}(x,y)\equiv -i g^2_{\chi}  D(x,y) \bar{G}_{\xi,0}(x,y).
\label{eq:self}
\end{align}
The first integral term in Eq.~(\ref{eq:couplingexp}) contains the ladder diagram exchange between particle and anti-particle. Similar equations for $G_{\eta \eta}$ and $G_{\xi \xi}$ can also be obtained by applying the same steps. In order to obtain the spectrum of bound state solutions as well as a fully non-perturbative treatment of the Sommerfeld-enhancement we have to resumm Eq.~(\ref{eq:couplingexp}) somehow.
\begin{claim}
\emph{We define our resummation scheme of the four-point correlator by resumming the ladder exchange, as well as the self-energy contributions in Eq.~(\ref{eq:couplingexp}) on an equal footing}.
\end{claim}
In other words, we are resumming the leading order terms in the coupling expansion of the four-point correlator $G_{\eta \xi}$ and similar for $G_{\eta \eta}$ and $G_{\xi \xi}$. On the one hand, this is a critical point, since this procedure is not exact and we cannot guarantee that other contributions do not play an important role as well, e.g.\ one of the limitations are systems with large coupling constants or large DM density where we can not decouple the solution of $G_{\eta \xi}$ from $G_{\eta\eta}$ and $G_{\xi \xi} $. The latter limitation might not be a problem since when focussing on the freeze-out the DM density becomes dilute.
On the other hand, as we will discuss in detail later in this work, the final result based on this resummation scheme behaves physically, fulfils KMS condition in equilibrium\footnote{Another resummation scheme for $G_{\eta \xi}$ we tested, can be obtained
directly after the steps of the truncation in Eq.~(\ref{eq:closerms}) and Hartree-Fock approximation in Eqs.~(\ref{eq:hartree1})-(\ref{eq:hartree2}). After some algebra, this would result in the Bethe-Salpeter equation $G_{\eta \xi}(x,y,z,w) = G_{\eta}(x,z) \bar{G}_{\xi}(y,w) + g^2_{\chi} \int_{\bar{x},\bar{y} \in \C} G_{\eta}(x,\bar{x}) \bar{G}_{\xi}(y,\bar{y}) D(\bar{x},\bar{y}) G_{\eta \xi}(\bar{x},\bar{y},z,w)$. Note that this equation would be equivalent to the original vacuum BS equation when naively extending the time integration in the latter equation towards the Keldysh contour $\mathcal{C}$. The main difference here compared to our resummation scheme is that the two-point functions are fully interacting. The l.h.s of the $++--$ or $--++$ component of this equation fulfils the KMS condition in equilibrium, when taking the the two-time limit. The right hand side depends in general on three times. It can be reduced to only two-times by assuming a static form for the mediator correlator $D(x,y)=\delta_{C}(t_x,t_y) V(\mathbf{x}-\mathbf{y})$. Integrating over Keldysh-contour delta function leads to the fact that also the r.h.s fulfils the KMS condition. However, this is a strong assumption on the form of the mediator correlation function, sending the off-diagonal terms $D^{+-}$ or $D^{-+}$ to zero. The simplest possibility to take into account the off-diagonal terms and at the same time obtain a two-time structure also of the r.h.s would be to assume that every component of $D$ is proportional to a time delta-function. A crucial observation we have made is that this type of approximation seems to violate the KMS condition through the off-diagonal terms, although the r.h.s has a two-time structure. The reason might be in the resummation of different orders in the coupling, as caused by the interacting DM two-point correlators in the BS kernel. The coupling expansion and the resummation of terms of equal order in the coupling parameter (our scheme), seems to be essential in order to obtain our final BS Eq.~(\ref{eq:twotimeother}), fulfilling the KMS condition.}, reproduces the correct vacuum limit, enables us to study bound states and DM Sommerfeld-enhanced annihilation at finite temperature, and in some other limits we recover the literature results based on linear response theory. 
Furthermore, a combined resummation of one-particle self-energies and ladder-diagram exchanges seems to be necessary at finite temperature, since similar combinations would also occur when calculating the effective potential from Wilson-loop lines \cite{Laine:2006ns} guaranteeing the gauge invariance.

Before we can resum Eq.~(\ref{eq:couplingexp}), we need some further exact rearrangements and manipulations of the four-point function components. Since we are interested in the solution of $G^{++--}_{\eta \xi}$ at equal times, it is sufficient to only consider two-time four-point functions, where we will adopt the short notation $G_{\eta \xi}(t,t^{\prime})= G_{\eta \xi}(t\mathbf{x}\mathbf{y},t^{\prime} \mathbf{z}\mathbf{w})$. Certain combinations of the components of Eq.~(\ref{eq:couplingexp}) turn out to be closed, e.g.\ let us define \footnote{This combination is not obvious at first place, but when subtracting the advanced component from the retarded it can be shown that the resulting spectral function has a similar completeness relation as the spectral function of two-point correlators \cite{1999physics...3039B}.}
\begin{align}
G_{\eta \xi}^{R}(t,t^{\prime}) &\equiv \theta(t-t^{\prime}) \bigg[ G_{\eta \xi}^{++--}(t,t^{\prime}) - G_{\eta \xi}^{+--+}(t,t^{\prime})-G_{\eta \xi}^{-++-}(t,t^{\prime})+G_{\eta \xi}^{--++}(t,t^{\prime}) \bigg],\label{eq:defret}\\
G_{\eta \xi}^{A}(t,t^{\prime}) &\equiv -\theta(t^{\prime}-t) \bigg[ G_{\eta \xi}^{++--}(t,t^{\prime}) - G_{\eta \xi}^{+--+}(t,t^{\prime})-G_{\eta \xi}^{-++-}(t,t^{\prime})+G_{\eta \xi}^{--++}(t,t^{\prime}) \bigg].\label{eq:deadv}
\end{align}
In the following we will show that, by using the semigroup properties of free correlators, the following structure of the retarded equations can be achieved
\begin{align}
G_{\eta \xi}^{R}(t,t^{\prime})=G_{\eta,0}^R(t,t^{\prime})G_{\xi,0}^R(t,t^{\prime}) + g^2_{\chi} \int \text{d}t_1G_{\eta,0}^R(t,t_1)G_{\xi,0}^R(t,t_1) \int\text{d}t_2 \Sigma_{\eta\xi}^R(t_1,t_2)G_{\eta,0}^R(t_2,t^{\prime})G_{\xi,0}^R(t_2,t^{\prime}), \label{eq:retcouplingexp}
\end{align}
and similar for the advanced. The precise terms contained in the two-particle self-energy $\Sigma_{\eta\xi}$ will be given later. From the form of Eq.~(\ref{eq:retcouplingexp}) it is clear that our resummation scheme as described above is just the replacement of the free two-point correlators at the end by $ G_{\eta,0}^R(t_2,t^{\prime})G_{\xi,0}^R(t_2,t^{\prime}) \rightarrow G_{\eta \xi}^{R}(t_2,t^{\prime}) $. 

Let us in the following sketch the way how to obtain this resummable structure.
The first term on the r.h.s in Eq.~(\ref{eq:retcouplingexp}) can be obtained by using simple relations
\begin{align}
G_{\eta,0}^{+-}G_{\xi,0}^{+-} - G_{\eta,0}^{+-}G_{\xi,0}^{-+} -  G_{\eta,0}^{-+}G_{\xi,0}^{+-} + G_{\eta,0}^{-+}G_{\xi,0}^{-+} =  G^{\rho}_{\eta,0}G^{\rho}_{\xi,0} =  G^{R}_{\eta,0}G^{R}_{\xi,0} + G^{A}_{\eta,0}G^{A}_{\xi,0}.
\end{align}
In the last step we used $G^RG^A=0$ for equal time products and when multiplying this with the unit step function as in the definition Eq.~(\ref{eq:defret}) and (\ref{eq:deadv}) this just projects out the respective product. The integral term is more complicated. Let us consider for simplicity only the last integral term in Eq.~(\ref{eq:couplingexp}) and perform the sum over the different contributions of the components for the retarded two-time four-point correlator, resulting in:
\begin{align}
I = \int \text{d}t_1 \text{d}t_2 G_{\eta,0}^R (t,t_1) \bar{G}_{\xi,0}^R (t,t^{\prime})(-i) \Sigma_{\eta}^R( t_1,t_2)G_{\eta,0}^R (t_2,t^{\prime}), \label{eq:retartedcontribution}
\end{align}
where the retarded one-particle self-energy is defined as $\Sigma^R = \Sigma^{++} -\Sigma^{+-} $.
Since all propagators are free, we can use the semigroup property (see also Appendix \ref{app:semigr})
\begin{align}
\bar{G}_{\xi,0}^R (t,t^{\prime}) = \bar{G}_{\xi,0}^R (t,t_1)\bar{G}_{\xi,0}^R (t_1,t_2)\bar{G}_{\xi,0}^R (t_2,t^{\prime}), \text{ for } t < t_1 < t_2 < t^{\prime}.
\end{align}
For brevity, we suppress the space integration here. This property can be used in Eq.~(\ref{eq:retartedcontribution}) since the times satisfy the inequality due to the product of retarded correlators, resulting in:
\begin{align}
I= \int\text{d}t_1 G_{\eta,0}^R (t,t_1) \bar{G}_{\xi,0}^R (t,t_1)\int\text{d}t_2 \big[(-i) \Sigma_{\eta}^R( t_1,t_2) \bar{G}_{\xi,0}^R (t_1,t_2) \big]G_{\eta,0}^R (t_2,t^{\prime})\bar{G}_{\xi,0}^R (t_2,t^{\prime}).
\end{align}
Comparing with Eq.~(\ref{eq:retcouplingexp}), we indeed find the anticipated structure. Applying similar steps to all integral terms in Eq.~(\ref{eq:couplingexp}), we find the following \emph{Two-time Bethe-Salpeter equations}\footnote{Similar equations where also obtained in \cite{1999physics...3039B}, although the derivation, particle content and potential is slightly different compared to our. Nevertheless, further helpful steps for bringing the integral terms into a resumable form by using semigroup properties can be found in the Appendix of \cite{1999physics...3039B}.}.

\begin{claim}
\textit{Two-Time Bethe-Salpeter equations}:
\begin{align}
\!G_{\eta \xi}^{\Phi}(t\mathbf{x}\mathbf{y},t^{\prime} \mathbf{z}\mathbf{w}) = \mathcal{G}_{\eta \xi}^{\Phi}(t\mathbf{x}\mathbf{y},t^{\prime}\mathbf{z}\mathbf{w})&+ \! \int\displaylimits_{t_1,\mathbf{x}_1,\mathbf{x}_2} \!\!\!\! \mathcal{G}_{\eta \xi}^{R}(t\mathbf{x}\mathbf{y},t_1 \mathbf{x}_1\mathbf{x}_2)\!\!\!\!\int\displaylimits_{t_2,\mathbf{x}_3,\mathbf{x}_4}\!\!\!\! \Sigma_{\eta \xi}^R(t_1 \mathbf{x}_1\mathbf{x}_2,t_2 \mathbf{x}_3\mathbf{x}_4)G_{\eta \xi}^{\Phi}(t_2 \mathbf{x}_3\mathbf{x}_4,t^{\prime} \mathbf{z}\mathbf{w}) \nonumber\\
&+ \!  \int\displaylimits_{t_1,\mathbf{x}_1,\mathbf{x}_2}\!\!\!\!  \mathcal{G}_{\eta \xi}^{R}(t\mathbf{x}\mathbf{y},t_1 \mathbf{x}_1\mathbf{x}_2)\!\!\!\!\int\displaylimits_{t_2,\mathbf{x}_3,\mathbf{x}_4}\!\!\!\! \sigma_{\eta \xi}^{\Phi}(t_1 \mathbf{x}_1\mathbf{x}_2,t_2 \mathbf{x}_3\mathbf{x}_4)G_{\eta \xi}^{A}(t_2 \mathbf{x}_3\mathbf{x}_4,t^{\prime} \mathbf{z}\mathbf{w})\nonumber\\
&+ \!  \int\displaylimits_{t_1,\mathbf{x}_1,\mathbf{x}_2}\!\!\!\!  \mathcal{G}_{\eta \xi}^{\Phi}(t\mathbf{x}\mathbf{y},t_1 \mathbf{x}_1\mathbf{x}_2)\!\!\!\!\int\displaylimits_{t_2,\mathbf{x}_3,\mathbf{x}_4} \!\!\!\!\Sigma_{\eta \xi}^{A}(t_1 \mathbf{x}_1\mathbf{x}_2,t_2 \mathbf{x}_3\mathbf{x}_4)G_{\eta \xi}^{A}(t_2 \mathbf{x}_3\mathbf{x}_4,t^{\prime} \mathbf{z}\mathbf{w})
\label{eq:twotimeother}
\end{align}
\textit{with $\Phi=\{ ++-- \};\{ +--+ \};\{ -++- \};\{ --++ \}$, and}
\begin{align}
\!G_{\eta \xi}^{R/A}(t\mathbf{x}\mathbf{y},t^{\prime} \mathbf{z}\mathbf{w}) = \mathcal{G}_{\eta \xi}^{R/A}(t\mathbf{x}\mathbf{y},t^{\prime}\mathbf{z}\mathbf{w})+ \!\!\!\! \int\displaylimits_{t_1,\mathbf{x}_1,\mathbf{x}_2} \!\!\!\! \mathcal{G}_{\eta \xi}^{R/A}(t\mathbf{x}\mathbf{y},t_1 \mathbf{x}_1\mathbf{x}_2) \!\!\!\!\int\displaylimits_{t_2,\mathbf{x}_3,\mathbf{x}_4} \!\!\!\! \Sigma_{\eta \xi}^{R/A}(t_1 \mathbf{x}_1\mathbf{x}_2,t_2 \mathbf{x}_3\mathbf{x}_4)G_{\eta \xi}^{R/A}(t_2 \mathbf{x}_3\mathbf{x}_4,t^{\prime} \mathbf{z}\mathbf{w})\label{eq:twotimeret}.
\end{align}
\end{claim}
The products of free correlators are defined as
\begin{align}
\mathcal{G}_{\eta \xi}^R(t\mathbf{x}\mathbf{y},t^{\prime}\mathbf{z}\mathbf{w}) \equiv  G_{\eta,0}^R(t\mathbf{x},t^{\prime}\mathbf{z}) \bar{G}_{\xi,0}^R(t \mathbf{y},t^{\prime} \mathbf{w})  ,\text{ and } \mathcal{G}_{\eta \xi}^A(t\mathbf{x}\mathbf{y},t^{\prime}\mathbf{z}\mathbf{w}) \equiv  -G_{\eta,0}^A(t\mathbf{x},t^{\prime}\mathbf{z}) \bar{G}_{\xi,0}^A(t \mathbf{y},t^{\prime} \mathbf{w}) ,
\end{align} 
and similar for the other components, e.g. $\mathcal{G}_{\eta \xi}^{++--} = G_{\eta,0}^{+-} G_{\eta,0}^{+-}$.
Furthermore, we introduced the \emph{two-particle self-energies} according to
\begin{align}
\Sigma_{\eta \xi}^{R}(t\mathbf{x}\mathbf{y},t^{\prime} \mathbf{w} \mathbf{z})\equiv & (-i)\Sigma_{\eta}^R(t \mathbf{x},t^{\prime} \mathbf{w}) \bar{G}_{\xi,0}^R(t \mathbf{y},t^{\prime} \mathbf{z}) +(-i)\Sigma_{\xi}^R(t \mathbf{y},t^{\prime} \mathbf{z}) G_{\eta,0}^R(t \mathbf{x},t^{\prime} \mathbf{w})\nonumber \\
&+ g^2_{\chi}\bar{G}_{\xi,0}^R(t \mathbf{y},t^{\prime} \mathbf{z}) D^{-+}(t \mathbf{x} , t^{\prime} \mathbf{z}) G_{\eta,0}^R(t \mathbf{x},t^{\prime} \mathbf{w}) +  g^2_{\chi}G_{\eta,0}^R(t \mathbf{x},t^{\prime} \mathbf{w})  D^{+-}(t^{\prime} \mathbf{w} , t \mathbf{y}) \bar{G}_{\xi,0}^R(t \mathbf{y},t^{\prime} \mathbf{z})\nonumber\\
&+ g^2_{\chi}\bar{G}_{\xi,0}^{+-}(t \mathbf{y},t^{\prime} \mathbf{z}) D^{R}(t \mathbf{x} , t^{\prime} \mathbf{z}) G_{\eta,0}^R(t \mathbf{x},t^{\prime} \mathbf{w}) +  g^2_{\chi}G_{\eta,0}^{+-}(t \mathbf{x},t^{\prime} \mathbf{w})  D^{A}(t^{\prime} \mathbf{w} , t \mathbf{y}) \bar{G}_{\xi,0}^R(t \mathbf{y},t^{\prime} \mathbf{z}),\label{eq:twotimeselfretarded}
\end{align}
where the retarded single-particle self-energy is defined in terms of the components of the definition Eq.~(\ref{eq:self}), namely $\Sigma^R_i \equiv \Sigma^{++}_i -  \Sigma^{+-}_i$. The advanced two-particle self-energy is given by
\begin{align}
-\Sigma_{\eta \xi}^{A}(t\mathbf{x}\mathbf{y},t^{\prime} \mathbf{w} \mathbf{z})\equiv & (-i)\Sigma_{\eta}^A(t \mathbf{x},t^{\prime} \mathbf{w}) \bar{G}_{\xi,0}^A(t \mathbf{y},t^{\prime} \mathbf{z}) +(-i)\Sigma_{\xi}^A(t \mathbf{y},t^{\prime} \mathbf{z}) G_{\eta,0}^A(t \mathbf{x},t^{\prime} \mathbf{w}) \nonumber\\
&+ g^2_{\chi}\bar{G}_{\xi,0}^A(t \mathbf{y},t^{\prime} \mathbf{z}) D^{-+}(t \mathbf{x} , t^{\prime} \mathbf{z}) G_{\eta,0}^A(t \mathbf{x},t^{\prime} \mathbf{w}) +  g^2_{\chi}G_{\eta,0}^A(t \mathbf{x},t^{\prime} \mathbf{w})  D^{+-}(t^{\prime} \mathbf{w} , t \mathbf{y}) \bar{G}_{\xi,0}^A(t \mathbf{y},t^{\prime} \mathbf{z})\nonumber\\
&+ g^2_{\chi}\bar{G}_{\xi,0}^{+-}(t \mathbf{y},t^{\prime} \mathbf{z}) D^{A}(t \mathbf{x} , t^{\prime} \mathbf{z}) G_{\eta,0}^A(t \mathbf{x},t^{\prime} \mathbf{w}) +  g^2_{\chi}G_{\eta,0}^{+-}(t \mathbf{x},t^{\prime} \mathbf{w})  D^{R}(t^{\prime} \mathbf{w} , t \mathbf{y}) \bar{G}_{\xi,0}^A(t \mathbf{y},t^{\prime} \mathbf{z}),
\label{eq:twotimeselfadvanced}
\end{align}
and the most important statistical components are defined by
\begin{align}
\sigma_{\eta \xi}^{++--}(t\mathbf{x}\mathbf{y},t^{\prime} \mathbf{w} \mathbf{z})  \equiv & (-i)\Sigma_{\eta}^{+-}(t \mathbf{x},t^{\prime} \mathbf{w}) \bar{G}_{\xi,0}^{+-}(t \mathbf{y},t^{\prime} \mathbf{z}) +(-i)\Sigma_{\xi}^{+-}(t \mathbf{y},t^{\prime} \mathbf{z}) G_{\eta,0}^{+-}(t \mathbf{x},t^{\prime} \mathbf{w}) \nonumber\\
&+ g^2_{\chi}G_{\eta,0}^{+-}(t \mathbf{x},t^{\prime} \mathbf{w})  D^{-+}(t^{\prime} \mathbf{w} , t \mathbf{y})\bar{G}_{\xi,0}^{+-} (t \mathbf{y},t^{\prime} \mathbf{z})  +  g^2_{\chi}G_{\eta,0}^{+-}(t \mathbf{x},t^{\prime} \mathbf{w})  D^{+-}(t \mathbf{x} , t^{\prime} \mathbf{z}) \bar{G}_{\xi,0}^{+-}(t \mathbf{y},t^{\prime} \mathbf{z}),
\label{eq:twotimeself++--}\\
\sigma_{\eta \xi}^{--++}(t\mathbf{x}\mathbf{y},t^{\prime} \mathbf{w} \mathbf{z})  \equiv & (-i)\Sigma_{\eta}^{-+}(t \mathbf{x},t^{\prime} \mathbf{w}) \bar{G}_{\xi,0}^{-+}(t \mathbf{y},t^{\prime} \mathbf{z}) +(-i)\Sigma_{\xi}^{-+}(t \mathbf{y},t^{\prime} \mathbf{z}) G_{\eta,0}^{-+}(t \mathbf{x},t^{\prime} \mathbf{w}) \nonumber\\
&+ g^2_{\chi}G_{\eta,0}^{-+}(t \mathbf{x},t^{\prime} \mathbf{w})  D^{+-}(t^{\prime} \mathbf{w} , t \mathbf{y})\bar{G}_{\xi,0}^{-+} (t \mathbf{y},t^{\prime} \mathbf{z})  +  g^2_{\chi}G_{\eta,0}^{-+}(t \mathbf{x},t^{\prime} \mathbf{w})  D^{+-}(t \mathbf{x} , t^{\prime} \mathbf{z}) \bar{G}_{\xi,0}^{-+}(t \mathbf{y},t^{\prime} \mathbf{z}).
\label{eq:twotimeself--++}
\end{align}
\begin{figure}[h]
\includegraphics[scale=1.3]{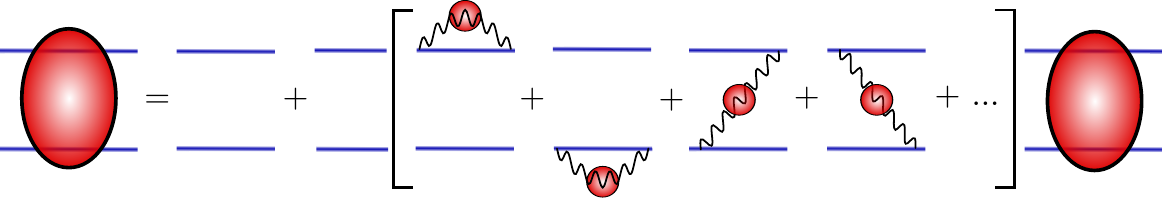}
\caption{Resummation scheme shown for the retarded four-point correlator Eq.~(\ref{eq:twotimeret}). The terms in the brackets belong to one-particle self-energy contributions as well as the mediator exchange between particle and antiparticle. Dots represent terms containing the DM distribution in the two-particle self-energy. Due to the Boltzmann suppression at the freeze-out, those contributions will be dropped later (see DM dilute limit in Section \ref{sec:dilutelimit}).}
\label{fig:resummation}
\end{figure}
A graphical illustration of the resummation scheme for the retarded or advanced component is shown in Fig.~\ref{fig:resummation}. Below, we summarize the essential properties of our two-time BS equations, based on our resummation of one-particle self-energies as well as the mediator exchanges on an equal footing.
\begin{itemize}
\item \emph{Two-time structure} The remarkable result of our resummation scheme is that we achieved a two-time dependence of the Bethe-Salpeter Eq.~(\ref{eq:twotimeother}) and (\ref{eq:twotimeret}) without assuming a static property of mediator correlator $D$ or, more general, without assuming \emph{any} particular form. 
\item \emph{KMS condition}
In general, the two-time correlators $G_{\eta \xi}^{++--}(t,t^{\prime})$ and $G_{\eta \xi}^{--++}(t,t^{\prime})$ are related in equilibrium via the KMS condition: $G^{--++}_{\eta \xi}(t \mathbf{x}\mathbf{y};t^{\prime}\mathbf{z}\mathbf{w}) = G^{++--}_{\eta \xi}(t+i\beta,\mathbf{x}\mathbf{y} ; t^{\prime}\mathbf{z}\mathbf{w}) \; \text{or} \; G^{--++}_{\eta \xi}(t\mathbf{x}\mathbf{y};t^{\prime}+i\beta,\mathbf{z}\mathbf{w}) = G^{++--}_{\eta \xi}(t\mathbf{x}\mathbf{y};t^{\prime}\mathbf{z}\mathbf{w})$. Now, one of the great features of our resummation scheme is that it respects this KMS condition which is non-trivial since the equations are not exact. This means that the \emph{left and right} hand side 
of BS Eq.~(\ref{eq:twotimeother}) for the components $G_{\eta \xi}^{++--}(t,t^{\prime})$ and $G_{\eta \xi}^{--++}(t,t^{\prime})$ transform respectively into each other in equilibrium.
This can be seen, by Fourier transforming the kernels, needed for proper analytic continuation of time, and by using the fact that in equilibrium all quantities only depend on the difference of the time variable. Indeed one finds that always the statistical part in the three kernels of Eq.~\eqref{eq:twotimeother} transform into their counterparts. The solution of our target component $G_{\eta \xi}^{++--}$ becomes a very simple expression when utilizing the power of the KMS condition as will see in the next section.

\item Similar to the two-point functions, two-time retarded and advanced components of the four-point correlator are related by complex conjugation $G_{\eta \xi}^{A}(t,t^{\prime}) = - \big[G_{\eta \xi}^{R}(t^{\prime},t)\big]^{\dagger}$, as can be shown directly from the definition Eq.~(\ref{eq:defret}) and Eq.~(\ref{eq:deadv}).
\item Other components of four-point correlators not listed above, can be constructed by the given ones \cite{1999physics...3039B}.
\item The full set of equations are able to describe Bose-Einstein condensation, e.g.\ relevant for fermionic systems where bound-state solutions exist and the density and chemical potential are in a critical regime. Since we focus on multi-TeV particles, the density will be always low enough to ignore those quantum-statistical effects.
\end{itemize}

\section{Two-particle spectrum at finite temperature}
\label{sec:solutions}
For general out of-equilibrium situations, the coupled system of two-time BS Eqs.~(\ref{eq:twotimeother})-(\ref{eq:twotimeret}) might require a fully numerical treatment.
However, when relying on some well-motivated assumptions which are guaranteed for WIMP like freeze-outs, we show in this chapter that the coupled equations can by drastically simplified. The result will be a formal solution of our target component $G^{++--}_{\eta \xi}$, appearing in our main number density Eqs.~(\ref{eq:numberp})-(\ref{eq:numberantip}), in terms of the DM \emph{two-particle spectral function}. Furthermore, we fully provide the details for finding the explicit solution of the DM two-particle spectral function from a  Schr\"odinger-like equation with an effective in-medium potential, including thermal corrections. For a better understanding of the limitation of this final solution we share, step by step, the approximations needed. All assumptions leading to this simple result will be made systematically and clearly separated section-wise in this chapter.

\subsection{Formal solution in grand canonical ensemble}
\label{sec:fs}
We assume the DM system to be in a grand canonical state where the density matrix takes the form:
\begin{align}
	\hat \rho = \frac{1}{Z} e^{- \beta \left( \hat H - \mu_\eta \hat N_\eta - \mu_\xi \hat N_\xi \right)};
	\qquad
	\hat H = \hat H_\text{NR} + M (\hat N_\eta + \hat N_\xi),\quad
	\hat N_\bullet \equiv \int_{\bm{x}} j^0_\bullet,
	\label{eq:grandcan}
\end{align}
where $\bullet = \eta,\xi$ and we assume symmetric DM resulting in equal chemical potentials $\mu = \mu_\eta = \mu_\xi$. Further, we assume the Hamiltonian to commute with the number operators by treating the annihilation $\Gamma_s$ as a perturbation. This is valid as long as the processes driving the system to a grand canonical state are much more efficient compared to annihilation or the decay of bound states. 
In this sense, the chemical potential is effectively time dependent. It is related to the total number density, appearing on the l.h.s in Eqs.~(\ref{eq:numberp})-(\ref{eq:numberantip}). The time dependence of the number density is set by the Hubble term and the production and loss terms appearing on the r.h.s..

Let us insert the grand canonical density matrix into the components $++--$ and $--++$ and show how they are related. Utilizing the commutation relations of $- [\hat N_\eta, \eta^{(\dag)}] = (-) \eta^{(\dag)}$, $[\hat N_\xi, \xi^{(\dag)}] = (-) \xi^{(\dag)}$,
one can derive the KMS relations for the two-time four-point correlators in the presence of the chemical potentials. Recalling that the Hamiltonian is a generator of the time evolution,
one finds the KMS condition for a grand canonical state:
\begin{align}
	G_{\eta \xi,s}^{++--} (t \bm{x} \bm{y}; t' \bm{z} \bm{w})
	= e^{- 2 \beta (M - \mu)} 
	G_{\eta \xi,s}^{--++} (t \bm{x} \bm{y}; t'+ i \beta, \bm{z} \bm{w}).
\end{align}
Its Fourier transform reads
\begin{align}
	G^{++--}_{\eta \xi,s} (\bm{r}, \bm{r'}; \omega, \bm{P})
	= e^{- \beta (\omega + 2 M - 2 \mu)} G^{--++}_{\eta \xi,s} (\bm{r},\bm{r'}; \omega, \bm{P}),
\end{align}
where we introduced the Wigner-transformed four-point correlators according to
\begin{align}
	G^{\bullet \bullet \bullet \bullet}_{\eta \xi,s} \left( t, \bm{X} + \frac{\bm{r}}{2},\bm{X} - \frac{\bm{r}}{2};
	t', \bm{X'} + \frac{\bm{r'}}{2}, \bm{X'} - \frac{\bm{r'}}{2}
	\right)
	\equiv
	\int_{\omega, \bm{P}} e^{- i \omega (t-t') + i \bm{P} \cdot (\bm{X} - \bm{X'})}
	G^{\bullet \bullet \bullet \bullet}_{\eta \xi,s} (\bm{r}, \bm{r'}; \omega, \bm{P}).
\end{align}
Here, we have used the fact that the operator $\hat \rho$ commutes with the Hamiltonian $\hat H$ and the translation operator $\hat{\bm{P}}$. Defining the \emph{two-particle spectral function} as
\begin{align}
G^\rho_{\eta \xi}(t \bm{x} \bm{y}; t' \bm{z} \bm{w}) \equiv G_{\eta \xi,s}^{--++} (t \bm{x} \bm{y}; t' \bm{z} \bm{w}) -G_{\eta \xi,s}^{++--} (t \bm{x} \bm{y}; t' \bm{z} \bm{w}),
\end{align}
our target component is formally solved in terms of the spectral function and chemical potential by utilizing the KMS relation for a grand canonical state.
\begin{claim}
\textit{Collision term for grand canonical state}:
\begin{align}
	G^{++--}_{\eta \xi,s} (x,x,x,x) & = \int_{\omega, \bm{P}} f_B (\omega + 2 M - 2 \mu) G^\rho_{\eta \xi} (\mathbf{0}, \mathbf{0}; \omega, \bm{P})\\
	&\simeq
	\int_{\omega, \bm{P}}
	e^{- \beta (\omega + 2 M - 2\mu)} G^\rho_{\eta \xi} (\mathbf{0},\mathbf{0};\omega, \bm{P}) \\
	&= 
	e^{- 2 \beta (M - \mu)}
	\int_{-\infty}^{\infty} \frac{\text{d}^3\bm{P}}{(2 \pi)^3} e^{- \beta \bm{P}^2 / 4 M}
	\int_{-\infty}^{\infty} \frac{\text{d}E}{(2\pi)} e^{- \beta E} G^\rho_{\eta \xi} (\mathbf{0},\mathbf{0}; E).
	\label{eq:coll_th}
\end{align}
\end{claim}
\noindent
In the second line we approximated the Bose-Einstein distribution $f_B$ as Maxwell-Boltzmann, assuming the DM system to be dilute $T \ll (M - \mu)$. The same limit should also be taken in the explicit solution of the spectral function, as it is done in the next Section. In the last equality \eqref{eq:coll_th}, we have used that the spectral function only depends on $E \equiv \omega - \bm{P}^2 / 4M$ (as we will explicitly see later)
and adopted the loose notation 
$G_{\eta \xi}^\rho (\mathbf{0}, \mathbf{0}; \omega, \bm{P})
=G_{\eta \xi}^\rho (\mathbf{0}, \mathbf{0}; E)$. As a consequence of Fourier transformation, the energy integration in Eq.~\eqref{eq:coll_th} ranges from minus infinity to plus infinity.

If the theory has bound-state solutions in the spectrum, the two-particle spectral function has strong contributions at particular negative values of $E$ (binding energy).
These contributions are further enhanced by the Boltzmann factor $e^{-\beta E}$ compared to the scattering solutions with positive energy, as can be directly seen from Eq.~\eqref{eq:coll_th}. This is expected from the assumption of a grand canonical system. All DM states of energy E must be populated with the Boltzmann factor $e^{- \beta E}$ for a given DM number density. As a result, the bound states are preferred compared to the scattering states because their energies are smaller.

Under the key assumption of the grand canonical ensemble, our main number density Eqs.~(\ref{eq:numberp})-(\ref{eq:numberantip}) are formally closed. This is because the effective chemical potential appearing in Eq.~\eqref{eq:coll_th} is related to the total number density by Legendre transformation.
On the one hand side, the validity of adopting a grand canonical state requires scattering processes to be efficient in order to keep DM 
in kinetic equilibrium with the plasma particles $A_{\mu}$ and $\psi$.
For light mediators this is indeed guaranteed for times much later than the freeze-out.
On the other hand, if the theory has bound-state solutions and is described by a grand canonical ensemble with a single chemical potential as we have introduced, there appears a hidden assumption on the internal chemical relation between scattering and bound state contributions. As we will see later, it automatically implies the Saha condition for ionization equilibrium. Ionization equilibrium can only be achieved by efficient radiative processes like the ultra-soft emissions of mediators. In the description of our theory we have traced out from the beginning these contributions but can now formally include them by assuming ionization equilibrium.
Thus, a grand canonical description of systems where bound states exists is only appropriate if the ionization equilibrium can be  guaranteed. In Section \ref{sec:ionizatineq}, we come back to this issue in detail, provide an explicit expression for the chemical potential and prove the implication of ionization equilibrium.

We would like to finally remark that once a grand canonical picture is appropriate, all finite temperature corrections to the annihilation or decay rate are contained in the solution of the two-particle spectral function in Eq.~\eqref{eq:coll_th} through Eqs.~\eqref{eq:numberp} and \eqref{eq:numberantip}. Indeed as we will see, the negative and positive energy solutions of the spectral function will merge continuously together if finite temperature effects are strong. In this case it turns out to be impossible to distinguish bound from scattering state solutions. Due to the form of Eq.~\eqref{eq:coll_th} it is, however, not required to distinguish between these two contributions. Integrating the spectral function over the whole energy range automatically takes into account \emph{all} contributions. In summary, for a grand canonical ensemble, the finite temperature corrections to Sommerfeld-enhanced annihilation and bound-state decay processes are contained in the two-particle spectral function and an explicit solution of the latter quantity is derived in following sections.

\subsection{Two-particle spectral function in DM dilute limit}
\label{sec:dilutelimit}
A key observation is that we have factored out in \eqref{eq:coll_th} the leading order contribution in the DM phase-space density and the remaining task is to find the two-particle spectral properties. For the computation of a two-particle spectral function we can now approximate the DM system to be \emph{dilute} which is the limit $T \ll (M - \mu)$:
\begin{align}
G_{\eta/\xi}^{+-}(x,y) \ll G_{\eta/\xi}^{-+}(x,y) \Rightarrow G_{\eta/\xi}^{-+}(x,y) \simeq G^\rho_{\eta/\xi}(x,y) = G_{\eta/\xi}^{R}(x,y)- G_{\eta/\xi}^{A}(x,y), \label{eq:dmdl}
\end{align}
where $G^\rho$ is the single particle spectral function as introduced in Section \ref{sec:rtf}.
In the DM dilute limit, the two-particle spectral function is related to the the imaginary part of the dilute solution of the retarded four-point correlator, where the result is given below.
\begin{claim}
\textit{Two-particle spectral function in the DM dilute limit}:
\begin{align}
\label{eq:KK_rel}
G^\rho_{\eta \xi} (\mathbf{0},\mathbf{0}; E) & \simeq G^{--++}_{\eta \xi} (\mathbf{0},\mathbf{0}; E) = G^{\text{ret}}_{\eta \xi} (\mathbf{0},\mathbf{0}; E) - G^{\text{adv}}_{\eta \xi} (\mathbf{0},\mathbf{0}; E) 
= 2 \Im \left[ i G^{\text{ret}}_{\eta \xi} (\mathbf{0},\mathbf{0}; E) \right].
\end{align}
\end{claim}
\noindent
Here, we defined $G^{\text{ret}}_{\eta \xi}$ and $G^{\text{adv}}_{\eta \xi}$ as the DM dilute limit of the Eq.~(\ref{eq:twotimeret}) for $G^{\text{R}}_{\eta \xi}$ and $G^{\text{A}}_{\eta \xi}$, respectively. For the rest of this section, the computation of the dilute limit of these equations is given, proving the claim $G^{--++}_{\eta \xi} (\bm{0},\bm{0}; E)= 2 \Im [i G^{\text{ret}}_{\eta \xi} (\bm{0},\bm{0}; E)] $.

Applying the DM dilute limit Eq.~(\ref{eq:dmdl}) to the two-time BS Eq.~(\ref{eq:twotimeret}) for $G^{R}_{\eta \xi}$, one finds:
\begin{align}
G_{\eta \xi}^{\text{ret}} &= G_{\eta,0}^R \bar{G}_{\xi,0}^R + \int G_{\eta,0}^R \bar{G}_{\xi,0}^R  \Sigma_{\eta \xi}^{\text{ret}}G_{\eta \xi}^{\text{ret}},\label{eq:retardeddil}\\
\Sigma_{\eta \xi}^{\text{ret}}(tt^{\prime}\mathbf{x}\mathbf{y}\mathbf{w}\mathbf{z}) &= g_{\chi}^2 G_{\eta,0}^R(t \mathbf{x},t^{\prime} \mathbf{w})\bar{G}_{\xi,0}^R(t \mathbf{y},t^{\prime} \mathbf{z})\left[-D^{-+}(t \mathbf{x},t^{\prime} \mathbf{w})- D^{-+}(t \mathbf{y},t^{\prime} \mathbf{z})+ D^{-+}(t \mathbf{x} , t^{\prime} \mathbf{z}) + D^{+-}(t^{\prime} \mathbf{w} , t \mathbf{y}) \right],
\end{align}
where $\Sigma_{\eta \xi}^{\text{ret}}$ is the dilute limit of the two-particle self-energy $\Sigma_{\eta \xi}^{R}$ in Eq.~(\ref{eq:twotimeselfretarded}). All space-time dependences remain the same as in Eq.~(\ref{eq:twotimeret}) but we suppress them hereafter for simplicity. Similar limit can be taken for the advanced component. Important to note is that due to the DM dilute limit, it can be recognized that the retarded Eq.~(\ref{eq:retardeddil}) and hence the two-particle spectral function are \emph{independent} of the DM number density. For the freeze-out process the DM dilute limit is an excellent approximation.
Now to continue with the proof of $G^{--++}_{\eta \xi} (\bm{0},\bm{0}; E)= 2 \Im [i G^{\text{ret}}_{\eta \xi} (\bm{0},\bm{0}; E)] $, the equation for the $G^{--++}_{\eta \xi} $ [see Eq.~(\ref{eq:twotimeother})] in the DM dilute limit is
\begin{align}
&G_{\eta \xi}^{--++}=\nonumber\\ 
&G_{\eta,0}^{R} \bar{G}_{\xi,0}^{R} + G_{\eta,0}^{A} \bar{G}_{\xi,0}^{A} + \int G_{\eta,0}^R \bar{G}_{\xi,0}^R  \Sigma_{\eta \xi}^{\text{ret}}\left[G_{\eta \xi}^{\text{ret}}-G_{\eta \xi}^{\text{adv}}\right] + G_{\eta,0}^R \bar{G}_{\xi,0}^R  \sigma_{\eta \xi,\text{dil.}}^{--++}G_{\eta \xi}^{\text{adv}} + \left[G_{\eta,0}^{R} \bar{G}_{\xi,0}^{R} + G_{\eta,0}^{A} \bar{G}_{\xi,0}^{A} \right] \Sigma_{\eta \xi}^{\text{adv}}G_{\eta \xi}^{\text{adv}}\\
&=G_{\eta \xi}^{\text{ret}}-G_{\eta \xi}^{\text{adv}} + \int G_{\eta,0}^R \bar{G}_{\xi,0}^R \left[ - \Sigma_{\eta \xi}^{\text{ret}} + \Sigma_{\eta \xi}^{\text{adv}} +  \sigma_{\eta \xi,\text{dil.}}^{--++}\right]G_{\eta \xi}^{\text{adv}},\label{eq:relation}
\end{align}
where in the step to the last equality we used two-time BS 
Eq.~\eqref{eq:twotimeret} backward. The statistical two-particle self-energy in the dilute limit is given by
\begin{align}
\sigma_{\eta \xi,\text{dil.}}^{--++}(tt^{\prime}\mathbf{x}\mathbf{y}\mathbf{w}\mathbf{z}) = g_{\chi}^2 &\left[G_{\eta,0}^R(t \mathbf{x},t^{\prime} \mathbf{w})\bar{G}_{\xi,0}^R(t \mathbf{y},t^{\prime} \mathbf{z}) + G_{\eta,0}^A(t \mathbf{x},t^{\prime} \mathbf{w})\bar{G}_{\xi,0}^A(t \mathbf{y},t^{\prime} \mathbf{z}) \right]\nonumber\\
&\times\left[-D^{-+}(t \mathbf{x},t^{\prime} \mathbf{w})- D^{-+}(t \mathbf{y},t^{\prime} \mathbf{z})+ D^{-+}(t \mathbf{x} , t^{\prime} \mathbf{z}) + D^{+-}(t^{\prime} \mathbf{w} , t \mathbf{y}) \right],
\end{align}
where we have used:
\begin{align}
G_{\eta,0}^{-+}(t,t^{\prime}) \bar{G}_{\xi,0}^{-+}(t,t^{\prime}) \simeq G_{\eta,0}^{\rho}(t,t^{\prime}) \bar{G}_{\xi,0}^{\rho}(t,t^{\prime})  = G_{\eta,0}^{R}(t,t^{\prime}) \bar{G}_{\xi,0}^{R}(t,t^{\prime}) + G_{\eta,0}^{A}(t,t^{\prime}) \bar{G}_{\xi,0}^{A}(t,t^{\prime}).
\end{align}
The first similarity is a consequence of the dilute limit. In the last equality we used $G^R G^A=0$ for equal time products. The integral term in Eq.~(\ref{eq:relation}) vanishes by noting that in the dilute limit we have indeed $- \Sigma_{\eta \xi}^{\text{ret}} + \Sigma_{\eta \xi}^{\text{adv}} +  \sigma_{\eta \xi,\text{dil.}}^{--++} =0$ which finally proves the claim $G^{--++}_{\eta \xi} (\bm{0},\bm{0}; E)= 2 \Im i G^{\text{ret}}_{\eta \xi} (\bm{0},\bm{0}; E)$.
\subsection{Retarded equation in static potential limit}
\label{sec:htlstaticlimit}
In the previous section, we related the two-particle spectrum to the solution of the retarded equation in the DM dilute limit: $G^\rho_{\eta \xi} (\mathbf{0},\mathbf{0}; E) = 2 \Im \left[ i G^{\text{ret}}_{\eta \xi} (\mathbf{0},\mathbf{0}; E) \right]$. As a final step, we further simplify the two-time Bethe-Salpeter Eq.~(\ref{eq:retardeddil}) for $G^{\text{ret}}_{\eta \xi}$ by taking the proper static limit of the mediator correlator $D$, resulting finally in a simple Schr\"odinger-like equation. We start by acting with the inverse two-particle propagator from the left on Eq.~(\ref{eq:retardeddil}), arriving at
\begin{align}
[\mathcal{G}_{\eta \xi}^{R}]^{-1}G_{\eta \xi}^{\text{ret}}(t\mathbf{x}\mathbf{y},t^{\prime} \mathbf{z}\mathbf{w}) = i \delta(t-t^{\prime})\delta^3(\mathbf{x}-\mathbf{z})\delta^{3}(\mathbf{y}-\mathbf{w})+ i\int\displaylimits_{t_2,\mathbf{x}_3,\mathbf{x}_4} \!\!\!\! \Sigma_{\eta \xi}^{\text{ret}}(t \mathbf{x}\mathbf{y},t_2 \mathbf{x}_3\mathbf{x}_4)G_{\eta \xi}^{\text{ret}}(t_2 \mathbf{x}_3\mathbf{x}_4,t^{\prime} \mathbf{z}\mathbf{w}).
\end{align}
Here, we suppress spin-indices for simplicity but quote the final result in full form later.
Let us simplify the interaction kernel in Fourier space, where we take Wigner transform in time $\tau\equiv t-t^{\prime}$:
\begin{align}
G_{\eta \xi}^{\text{ret}}(\mathbf{p}_1,\mathbf{p}_2,\mathbf{p}^\prime_1,\mathbf{p}^\prime_2,\omega) =\int \text{d}^3x\; \text{d}^3y\; \text{d}^3z\; \text{d}^3w\; \text{d}\tau\; e^{i(\omega \tau -\mathbf{p}_1\cdot \mathbf{x} -\mathbf{p}_2\cdot\mathbf{y}+ \mathbf{p}_1^{\prime}\cdot\mathbf{z}+\mathbf{p}_2^{\prime}\cdot\mathbf{w})}G_{\eta \xi}^{\text{ret}}(\mathbf{x} \mathbf{y} \mathbf{z} \mathbf{w}, \tau).
\end{align}
Taking this Fourier transform of the kernel leads to two distinct parts:
\begin{align}
\!\widehat{\int\Sigma_{\eta \xi}^{\text{ret}}G_{\eta \xi}^{\text{ret}}} \!=\! g_{\chi}^2 \!\int \!\frac{\text{d}^3q}{(2\pi)^3}\left[I_1(\mathbf{p}_1,\mathbf{p}_2,\mathbf{q},\omega)G_{\eta \xi}^{\text{ret}}(\mathbf{p}_1,\mathbf{p}_2,\mathbf{p}^\prime_1,\mathbf{p}^\prime_2,\omega) +I_2(\mathbf{p}_1,\mathbf{p}_2,\mathbf{q},\omega)G_{\eta \xi}^{\text{ret}}(\mathbf{p}_1-\mathbf{q},\mathbf{p}_2+\mathbf{q},\mathbf{p}^\prime_1,\mathbf{p}^\prime_2,\omega)\right],
\end{align}
where $I_1$ results from the sum of the two one-particle self-energy contributions, whereas $I_2$ originates from the exchange term between particle and antiparticle (see Fig.~\ref{fig:resummation}):
\begin{align}
I_1 &= (-i)\int \frac{\text{d}\omega_1\text{d}\omega_2\text{d}\omega_3}{(2\pi)^3} \frac{G_{\eta,0}^{\rho}(\omega_1,\mathbf{p}_1-\mathbf{q}) D^{-+}(\omega_3,\mathbf{q}) \bar{G}_{\xi,0}^{\rho}(\omega_2,\mathbf{p}_2) +G_{\eta,0}^{\rho}(\omega_1,\mathbf{p}_1) D^{-+}(\omega_3,\mathbf{q}) \bar{G}_{\xi,0}^{\rho}(\omega_2,\mathbf{p}_2-\mathbf{q}) }{\omega-\omega_1-\omega_2-\omega_3+i\epsilon}, \label{eq:self}\\
I_2 &= i \int \frac{\text{d}\omega_1\text{d}\omega_2\text{d}\omega_3}{(2\pi)^3} \frac{G_{\eta,0}^{\rho}(\omega_1,\mathbf{p}_1-\mathbf{q}) D^{-+}(\omega_3,\mathbf{q}) \bar{G}_{\xi,0}^{\rho}(\omega_2,\mathbf{p}_2) + G_{\eta,0}^{\rho}(\omega_1,\mathbf{p}_1) D^{+-}(-\omega_3,\mathbf{q}) \bar{G}_{\xi,0}^{\rho}(\omega_2,\mathbf{p}_2+\mathbf{q}) }{\omega-\omega_1-\omega_2-\omega_3+i\epsilon}.\label{eq:exch}
\end{align}
We can perform the two $\omega$ integrations over one-particle spectral function, where in the free limit they are given by:
\begin{align}
G_{\eta,0}^{\rho}(\omega,\mathbf{p}) = (2 \pi) \delta\left(\omega-\frac{\mathbf{p}^2}{2M}\right),\quad \bar G_{\xi,0}^{\rho}(\omega,\mathbf{p}) = (2 \pi) \delta\left(\omega-\frac{\mathbf{p}^2}{2M}\right).
\end{align}
Now the integrals are reduced to
\begin{align}
I_1 &= (-i)\int \frac{\text{d}\bar{\omega}}{(2\pi)} \frac{D^{-+}(\bar{\omega},\mathbf{q}) +  D^{+-}(-\bar{\omega},\mathbf{q})}{\Omega_1-\bar{\omega}+i\epsilon}, \\
I_2 &= i \int \frac{\text{d}\bar{\omega}}{(2\pi)} \frac{D^{-+}(\bar{\omega},\mathbf{q}) +  D^{+-}(-\bar{\omega},\mathbf{q})}{\Omega_2-\bar{\omega}+i\epsilon}.
\end{align}
where $\Omega_i$ contain the respective on-shell energies. Noticing that the Fourier transform of $D^{++}(t,r) = \theta(t)D^{-+}(t,r) + \theta(-t)D^{+-}(t,r)$ is given by
\begin{align}
D^{++}(\Omega_i,\mathbf{q}) = i \int \frac{\text{d}\bar{\omega}}{(2\pi)} \frac{D^{-+}(\bar{\omega},\mathbf{q})+ D^{+-}(-\bar{\omega},\mathbf{q}) }{\Omega_i-\bar{\omega} + i \epsilon},
\end{align}
we can now take the proper static limit $\Omega_i\rightarrow0$ which results in
\begin{align}
i\widehat{\int\Sigma_{\eta \xi}^{\text{ret}}G_{\eta \xi}^{\text{ret}}} = (-i) g_{\chi}^2\int \frac{\text{d}^3q}{(2\pi)^3} \left[D^{++}(0,\mathbf{q})G_{\eta \xi}^{\text{ret}}(\mathbf{p}_1,\mathbf{p}_2,\mathbf{p}^\prime_1,\mathbf{p}^\prime_2,\omega) -D^{++}(0,\mathbf{q})G_{\eta \xi}^{\text{ret}}(\mathbf{p}_1-\mathbf{q},\mathbf{p}_2+\mathbf{q},\mathbf{p}^\prime_1,\mathbf{p}^\prime_2,\omega)\right].
\end{align}
Introducing Wigner-momenta and Fourier transforming back with respect to the difference variables
\begin{align}
\mathbf{P}= (\mathbf{p}_1+\mathbf{p}_2)-(\mathbf{p}_1^{\prime}+\mathbf{p}_2^{\prime}),\; \mathbf{p}=(\mathbf{p}_1-\mathbf{p}_2)/2,\;\mathbf{p}^{\prime}=(\mathbf{p}_1^{\prime}-\mathbf{p}_2^{\prime})/2,\\
G_{\eta \xi}^{\text{ret}}(\mathbf{r},\mathbf{r}^{\prime},\mathbf{P},\omega) =\int \frac{\text{d}^3p}{(2\pi)^3} \frac{\text{d}^3p^{\prime}}{(2\pi)^3} e^{i(\mathbf{p}\cdot \mathbf{r} -\mathbf{p}^{\prime} \cdot \mathbf{r}^{\prime})}G_{\eta \xi}^{\text{ret}}(\mathbf{p},\mathbf{p}^{\prime},\mathbf{P}, \omega),
\end{align}
we finally end-up with the Schr\"odinger-like equation for the retarded four-point correlator in the static limit of the potential:
\begin{align}
\label{eq:BSeq_st_dil_omega}
\left[  \omega-\frac{\mathbf{P}^2}{4M}+  \frac{\bm{\nabla}^2_{\mathbf{r}}}{M}  +i\epsilon - V_{\text{eff}}(\mathbf{r}) \right]G_{\eta\xi}^{\text{ret}}(\mathbf{r},\mathbf{r}^{\prime};\mathbf{P}, \omega) = \Tr[\mathbf{1}_{2\times2}]i\delta^3(\mathbf{r} - \mathbf{r}^{\prime}).
\end{align}
Now we see that the retarded and hence also the spectral function only depends on $E\equiv \omega-\mathbf{P}^2/(4M)$.
\begin{claim}
Retarded two-time BS equation in static and dilute limit:
\begin{align}
\label{eq:BSeq_st_dil}
\left[  \frac{\bm{\nabla}^2_{\mathbf{r}}}{M} + E+i\epsilon - V_{\text{eff}}(\mathbf{r}) \right]G_{\eta\xi}^{\text{ret}}(\mathbf{r},\mathbf{r}^{\prime};E) = \Tr[\mathbf{1}_{2\times2}]i\delta^3(\mathbf{r} - \mathbf{r}^{\prime}),
\end{align}
where the effective in-medium potential is defined as
\begin{align}
V_{\text{eff}}(\mathbf{r}) \equiv -i g_{\chi}^2 \int\displaylimits_{-\infty}^{+\infty} \frac{\text{d}^3q}{(2\pi)^3} \left(1- e^{i\mathbf{q}\cdot \mathbf{r}} \right) D^{++}(0,\mathbf{q}).\label{eq:effectivepot}
\end{align}
\end{claim}
\noindent
The spectral function, we would actually like to compute, is obtained from the solution of this equation according to the relation 
$G^\rho_{\eta \xi} (\mathbf{0},\mathbf{0}; E) = 2 \Im \left[ i G^{\text{ret}}_{\eta \xi} (\mathbf{0},\mathbf{0}; E) \right]$~, as proven in the previous section. In the next Section, we will derive the explicit solution of the retarded equation where we will further approximate the static mediator correlator $D^{++}$ in the hard thermal loop limit, as already given in Eq.~(\ref{eq:staticHTL}). The first term in the effective potential in Eq.~(\ref{eq:effectivepot}) originates from the sum of the two single-particle self-energies, while the second term accounts for the mediator exchange between particle and anti-particle. The trace in the Schr\"odinger-like Eq.~(\ref{eq:BSeq_st_dil}) takes the correct spin summation into account, which we have suppressed in this Section for simplicity.

\subsection{Explicit solution in static HTL approximation}
\label{sec:explsol}
Taking the static HTL approximation of the massless mediator as derived in Eq.~(\ref{eq:staticHTL}), the effective potential according to Eq.~(\ref{eq:effectivepot}) results in
\begin{align}
V_{\text{eff}}(\mathbf{r}) &= - \alpha_{\chi} m_D - \frac{\alpha_{\chi}}{r}e^{-m_Dr} - i \alpha_{\chi} T \phi(m_D r)\label{eq:inmediumpot},\\
\phi(x)&= 2 \int_0^{\infty}  \text{d}z \frac{z}{(z+1)^2} \left(1- \frac{\sin(zx)}{zx} \right),
\end{align}
and $\phi(0)=0$ and $\phi(\infty)=1$. 
One can recognize the real part of the potential is corrected by the Debye mass as expected. It shifts the energy by $\alpha_{\chi} m_D$ (twice the Salpeter correction of single particle self-energies) and screens the Coulomb potential. At the same time, the effective potential contains an imaginary part. The physical meaning of this term is scatterings of DM with light particles in the thermal plasma. If particle and anti-particle are far away, the imaginary part must be solely determined by scattering with the thermal plasma without the Yukawa force. One can also see that this is actually the case, since the imaginary part becomes twice the thermal width of single particles $- i \alpha_\chi T$ for $r \to \infty$ (See single particle corrections in Appendix \ref{app:hf}, as well as Salpeter correction in Appendix \ref{app:salpeter}). This property follows from our resummation scheme, treating the DM self-energy on an equal footing with the ladder diagram exchange. Since the finite temperature corrections introduce a constant imaginary term for large distances, we can drop the $i\epsilon$ term in the following derivation of the explicit solution of Eq.~\eqref{eq:BSeq_st_dil}. This solution will be general and contains also the correct vacuum limit, where $i\epsilon$ has to be carefully taken into account. The effective potential in Eq.~(\ref{eq:inmediumpot}) was first obtained in \cite{Laine:2006ns} and reproduced subsequently by other methods \cite{Beraudo:2007ky,Brambilla:2008cx}. Let us remark that we derived it independently, i.e.\ for the first time starting from a set of two-time Bethe-Salpeter equations on the Keldysh contour.

To derive the solution, we expand $G_{\eta \xi}^\text{ret}$ in terms of partial waves
\begin{align}
G_{\eta\xi}^{\text{ret}}(\bm{r},\bm{r}^{\prime};E)=  \sum_l \frac{2l+1}{4 \pi} P_l(\cos \theta_{\bm{r},\bm{r}^{\prime}}) (-i) G^{\text{ret}}_{\eta \xi,l} (r, r'; E),
\end{align}
leading to the $l=0$ (s-wave) equation:
\begin{align}
\label{eq:BSeq_swave}
	\left[
	 -  \frac{1}{M}\frac{1}{r^2} \partial_r ( r^2 \partial_r) - E + V_{\text{eff}} (r) 
	\right]
	G^{\text{ret}}_{\eta \xi,s} (r, r'; E) = \Tr[\mathbf{1}_{2\times2}] \frac{1}{r r'}\delta (r - r').
\end{align}
The physically relevant boundary conditions we impose on $G^{\text{ret}}_{\eta \xi,s}$ are listed below.
\begin{itemize}
	\item $G_{\eta \xi,s}^{\text{ret}} (r, r'; E)$ is finite $ \forall $ $r $, $r'$.
	\item For $|r - r'| \to \infty$, $G_{\eta \xi,s}^{\text{ret}}$ decays exponentially (as a consequence of constant imaginary potential).
\end{itemize}
Note here that, since we are working in the dilute limit, the Feynman propagator and the retarded function are the same.
These requirements set the form of the solution uniquely (see also~\cite{Strassler:1990nw, Hisano:2004ds}):
\begin{align}
	G_{\eta \xi,s}^{\text{ret}} (r,r'; E)= 
	 \Tr[\mathbf{1}_{2\times2}]  \frac{M}{r r'} \left[
		\theta (r- r') g_> (r) g_< (r') + \theta (r' - r) g_< (r) g_> (r') 
	\right],
	\label{eq:general_solution}
\end{align}
where $g_{>/<}$ are the solutions of the homogeneous differential equation
\begin{align}
\left[-  \frac{1}{M}\partial_r^2  - E  + V_\text{eff} (r)\right] g_{>/<}(r) = 0,
\end{align}
whose boundary conditions are given as follows: 
\begin{itemize}
	\item $g_< (0) = 0$ and $g_<'(0) =1$.
	\item $g_> (0) = 1$ and $g_>(r)$ decays exponentially for $r \to \infty$.
\end{itemize}
One can explicitly check that the solution of the form \eqref{eq:general_solution} with the boundary conditions for $g_{>/<}$ fulfils the requirements on $G^{\text{ret}}_{\eta \xi,s}$ as listed above.

The two-particle spectral function, needed to compute the annihilation or decay rate according to Eqs.~\eqref{eq:coll_th} and \eqref{eq:KK_rel}, can be written in terms of the homogeneous solution
as:
\begin{align}
G_{\eta \xi}^{\rho}(\mathbf{0},\mathbf{0};E)|_{l=0} = 2 \Im \left[ i G_{\eta\xi}^{\text{ret}}(\mathbf{0},\mathbf{0};E) |_{l=0} \right]
= \frac{1}{2\pi} \Im \left[ \lim_{r,r^{\prime}\rightarrow0} G^{\text{ret}}_{\eta \xi,s} (r, r^{\prime}; E) \right] =
  \frac{1}{2\pi} \Tr[\mathbf{1}_{2\times2}]  M \Im [ g_>' (0) ]\; .
	\label{eq:rel_origin}
\end{align}
In the last term, prime stands for the derivative with respect to $r$.
Formally this solution is correct but might be be troublesome in the numerical evaluation, since the real part of $g_>' (0)$ has a singularity due to the $1/r$ behavior of the effective potential.
To resolve this issue, we rewrite the imaginary part of $g_>'(0)$ by means of a different solution.
We closely follow the discussion given in Ref.~\cite{Strassler:1990nw}. Let us define another singular solution $g_s$ whose boundary conditions are given by
$g_s (0) = 1$ and $\Im [ g_s' (0)] = 0$.
The outgoing solution for $r \to \infty$ can be expressed by a linear combination of
\begin{align}
	g_> = g_s + B g_<.
\end{align}
By definition, we have $\Im [g_>'(0)] = \Im [B]$.
The fact that $g_>$ is outgoing forces it to decay for $r \to \infty$, which determines $B$. And thus, one finds
\begin{align}
	\Im [g_>' (0) ] = \Im [B ]= - \Im \left[ \lim_{r \to \infty} \left[ 
		\frac{g_s(r)}{g_< (r)}
	 \right]\right]\; .
	 \label{eq:g>viag<}
\end{align}
The physical quantity $\Im [ B]$ we would like to compute does not depend on the choice of $\Re [g_s' (0)]$.
This is why we can get the correct result without handling the divergence of $\Re [g_s' (0)] $.
The final step is to rewrite the singular solution as
\begin{align}
	g_s (r) = - g_< (r) \int^r_0 \dd r' \frac{1}{g_<^2(r')}.
	\label{eq:g_s}
\end{align}
One may also write down the expression for $g_>$ by using Eq.~\eqref{eq:g_s}:
\begin{align}
	g_> (r) = g_< (r) \int_r^\infty \dd r' \frac{1}{g^2_< (r')}.
\end{align}
One can check that it fulfills the boundary conditions $g_s (0) = g_> (0) = \lim_{r \to 0} r (1/r) = 1$ and 
$\Im [g_s'(0)] = \Im [g_> (0)] - \Im [B] = 0$.
Plugging Eq.~\eqref{eq:g_s} into Eq.~\eqref{eq:g>viag<}
and recalling the relation Eq.~\eqref{eq:rel_origin},
we finally arrive at the convenient form to evaluate the imaginary part (see also \cite{Burnier:2007qm}):
\begin{align}
\Im \left[ G_{\eta \xi,s}^{\text{ret}}(0,0;E) \right] = \Tr[\mathbf{1}_{2\times2}] M \Im [g^{\prime}_>(0)]
= \Tr[\mathbf{1}_{2\times2}] M \lim_{\delta \rightarrow 0} \int_{\delta}^{\infty} \text{d}r \Im \left( \frac{1}{g_<(r)} \right)^2.
\end{align}
The great advantage of this general solution is that it applies for the whole two-particle energy spectrum of our theory, i.e.\ for negative and positive $E$.
Here, we have introduced the tolerance $\delta \ll 1$ as initial value for numerical studies.  
Let us finally introduce dimensionless variables, expressing distances in terms of the Bohr radius $x \equiv \alpha_{\chi} M r$, and summarize the equations in terms of these units below.
\begin{claim}
S-wave part of two-particle spectral function:
\begin{align}
G_{\eta \xi}^{\rho}(\mathbf{0},\mathbf{0};E)|_{l=0}= \frac{1}{2 \pi} \Tr[\mathbf{1}_{2\times2}] \alpha_{\chi} M^2 \lim_{x^{\prime}\rightarrow0} \int_{x^{\prime}}^\infty \text{d}x\; \Im{\bigg[\frac{1}{g_<(x) g_<(x)} \bigg]}. \label{eq:solutionnumerical}
\end{align}
Homogeneous equations for massless and massive mediator:
\begin{align}
&g_<^{\prime\prime}(x)+ \bigg[\frac{E}{\alpha_{\chi}^2M} + \frac{m_D}{\alpha_{\chi} M} + \frac{1}{x} e^{-\frac{m_D}{\alpha_{\chi} M} x} + i \frac{T}{\alpha_{\chi} M} \phi\left(\frac{m_D}{\alpha_{\chi} M} x\right) \bigg]g_<(x) =0 \;,\label{eq:homcoul}\\
&g_<^{\prime\prime}(x)+ \bigg[ \frac{E}{\alpha_{\chi}^2M} + \frac{\sqrt{m_V^2+m_D^2}-m_V}{\alpha_{\chi} M} + \frac{1}{x} e^{-\frac{\sqrt{m_V^2+m_D^2}}{\alpha_{\chi} M} x} + i \frac{T}{\alpha_{\chi} M} \frac{1}{\sqrt{1+ m_V^2/m_D^2}} \phi\bigg(\frac{\sqrt{m_V^2+m_D^2}}{\alpha_{\chi} M} x\bigg) \bigg]g_<(x)=0 \;.\label{eq:homyuk}
\end{align}
\end{claim}
\noindent
In practice, we use the following initial conditions which can be obtained by power series approach:
\begin{align}
g_<(x)&= x - x^2/2 +i \gamma_i x^5,\\
\gamma_C &=- \frac{1}{40} \frac{T}{\alpha_{\chi} M} \bigg[\frac{m_D}{\alpha_{\chi} M} \bigg]^2,\\
\gamma_Y &=- \frac{1}{40} \frac{T}{\alpha_{\chi} M} \frac{1}{\sqrt{1 + (m_V/m_D)^2}} \bigg[\frac{\sqrt{
     m_D^2 + m_V^2}}{\alpha_{\chi} M} \bigg]^2.
\end{align}
where $\gamma_C$ applies for the Coulomb case Eq.~(\ref{eq:homcoul}) while $\gamma_Y$ can be taken for the Yukawa case Eq.~(\ref{eq:homyuk}). To efficiently deal with the sometimes highly oscillatory integrand in Eq.~(\ref{eq:solutionnumerical}), local adaptive integrating methods are useful. In finding the initial power in $x$ for the imaginary part we have assumed that $\phi(x)\sim \frac{1}{2} x^2$ for small $x$ \cite{Burnier:2007qm} which is only approximately true. In general, one has to carefully check if the correct value of the potential is close enough to this approximation at the initial value which we have done for the numerical results presented in subsequent Sections.

\section{DM number density equation in grand canonical ensemble}
\label{sec:vacuum}
In the previous Section, we have obtained a formal solution of the out-of-equilibrium term $G_{\eta \xi,s}^{++--}$, entering our main number density Eq.~(\ref{eq:numberp}). This was achieved by assuming the DM system is in a grand canonical state, formally solving $G_{\eta \xi,s}^{++--}$ in terms of the chemical potential and two-particle spectral function by KMS relation. Inserting this formal solution given in Eq.~(\ref{eq:coll_th}) into the main number density  Eq.~(\ref{eq:numberp}) results in our
\begin{claim}
Master formula for the DM number density equation in a grand canonical ensemble:
\begin{align}
\dot{n}_{\eta} + 3 H  n_{\eta} &= - 2(\sigma v_{\text{rel}}) G_{\eta \xi,s}^{++--} (x,x,x,x)\big|_{\text{eq}}\left(e^{\beta 2 \mu_{\eta}[n_{\eta}] } - 1 \right),\label{eq:numbergrandcanonicalgeneral}
\end{align}
where a symmetric plasma $2\mu_{\eta}=\mu_{\eta}+\mu_{\xi}$ is assumed and $(\sigma v_{\text{rel}})$ is the s-wave tree level annihilation cross section. The latter quantity is averaged over initial internal degrees of freedom (spin) and summed over final.
\end{claim}
The chemical potential $\mu_{\eta}[n_{\eta}]$ is a function of the total number density $n_{\eta}$ as it appears on the left hand side of our master formula.
The term $G_{\eta \xi,s}^{++--}\big|_{\text{eq}}$ is the chemical equilibrium limit $\mu \rightarrow 0 $ of Eq.~(\ref{eq:coll_th}), given by:
\begin{align}
G^{++--}_{\eta \xi,s} (x,x,x,x)\big|_{\text{eq}} =  e^{-  \beta 2 M } \int_{-\infty}^{\infty} \frac{\text{d}^3\bm{P}}{(2 \pi)^3} e^{- \beta \bm{P}^2 / 4 M}
	\int_{-\infty}^{\infty} \frac{\text{d}E}{(2\pi)} e^{- \beta E} G^\rho_{\eta \xi} (\mathbf{0},\mathbf{0}; E)|_{l=0}.
\end{align}
We presented a general method in the previous Section of how to compute the in-medium two-particle spectral function $G^\rho_{\eta \xi}$ explicitly. It contains finite temperature corrections to the Sommerfeld enhancement and bound-state decay. The only parameter left in Eq.~(\ref{eq:numbergrandcanonicalgeneral}) is the chemical potential, which has not yet been explicitly solved. The chemical potential  $\mu_{\eta}[n_{\eta}]$ can be obtained in two steps as demonstrated in the following. First, the total number density as a function of the chemical potential is computed. For a grand canonical ensemble this follows from basic relations of quantum statistical mechanics and is given by:
\begin{align}
n_{\eta}[\mu_{\eta}] = \left. \frac{\partial p}{\partial \mu_{\eta}} \right|_{T,\Omega}, \label{eq:thermonumber}
\end{align}
where the total pressure is
\begin{align}
p \Omega = T \ln Z_{\text{gr}}(\Omega,T,\mu) = T \ln \Tr[e^{-\beta(H-\mu_{\eta} N_\eta-\mu_{\xi} N_\xi)} ].\label{eq:thermototalpressure}
\end{align}
Here, $\Omega$ is the volume and  $Z_{\text{gr}}$ is the grand canonical partition function. Second, by inversion of Eq.~(\ref{eq:thermonumber}) one obtains the chemical potential as a function of the total number density. The functional dependence of $\mu[n_{\eta}]$ on the total number density can be non-trivial especially for the case if bound-state solutions exist as we will see later. The $n_{\eta}$ on the l.h.s of Eq.~(\ref{eq:thermonumber}) is equivalent to $n_{\eta}$ appearing on the l.h.s of our Master Eq.~(\ref{eq:numbergrandcanonicalgeneral}). 

In subsequent sections of this chapter we demonstrate how powerful our Master Eq.~(\ref{eq:numbergrandcanonicalgeneral}) is. We self-consistently compute the component $G^{++--}_{\eta \xi,s} (x,x,x,x)\big|_{\text{eq}}$ and the chemical potential $\mu$. This means in both terms the same approximations should be made to obtain a well behaved number density equation.

For a better understanding, we would like to start in the next Section \ref{sec:zeroselfinteraction} with the simplest case of our theory by taking the zero self-interaction and zero finite temperature correction limit.
This means we take $g_{\chi} \rightarrow 0$ and $g_{\psi} \rightarrow 0$ in the effective in-medium potential Eq.~(\ref{eq:inmediumpot}) and compute the spectral function. Same limits are applied to the Hamiltonian entering in Eq.~(\ref{eq:thermototalpressure}) to compute the total pressure.
Under these limits, our Master Eq.~(\ref{eq:numbergrandcanonicalgeneral}) reduces to the conventional Lee-Weinberg equation, describing constant s-wave annihilation of DM.

As a next step, we allow for long-range self-interactions but neglect finite temperature corrections.
This corresponds to the limit $g_{\psi} \rightarrow 0$, leading to the fact that the remaining term in the effective in-medium potential is the standard Coulomb or unscreened Yukawa potential. In Section \ref{sec:relations}, the two-particle spectrum for this simple case of our theory is shown. We stress the point that only in this limit there is a direct relation between spectral function and standard expressions for the Sommerfeld-enhancement factor or the decay width of the bound states. Section~\ref{sec:ionizationeq_ideal} completes the results by computing the chemical potential for the same limit. Combining the analytic expressions for the spectral function and chemical potential, we prove our Master formula Eq.~(\ref{eq:numbergrandcanonicalgeneral}) to be
consistent with the classical on-shell Boltzmann equation treatment for vanishing thermal corrections. We also point out that adopting a grand canonical ensemble with one single time dependent chemical potential as in our master formula implies ionization equilibrium between the scattering and bound states. A detailed discussion is given on the validity of ionization equilibrium during the freeze-out process. If no bound-state solutions exist, the only limitation of our master formula is essentially kinetic equilibrium \cite{vandenAarssen:2012ag, Binder:2017rgn}.

We relax the assumption of zero finite temperature corrections in Section~\ref{sec:ionizatineq}. This brings us to another central result of this work: a DM number density equation, generalizing the conventional Lee-Weinberg equation and classical on-shell Boltzmann equation treatment as a consequence of accounting simultaneously for DM annihilation and bound-state decay at finite temperature. However, it should be noted that this equation in Section~\ref{sec:ionizatineq} strictly speaking only applies to the narrow thermal width case and is therefore less general compared to our Master Eq.~(\ref{eq:numbergrandcanonicalgeneral}). This means we have neglected in Section~\ref{sec:ionizatineq} imaginary-part corrections to the effective in-medium potential for the computation of the chemical potential. While we can fully account for these non-hermite corrections in the computation of $G^{++--}_{\eta \xi,s} (x,x,x,x)\big|_{\text{eq}}$, it remains an open question of this work of how to consistently compute the chemical potential for the broad thermal width case. The broad thermal width case for $G^{++--}_{\eta \xi,s} (x,x,x,x)\big|_{\text{eq}}$ we compute numerically later in this work (see Section \ref{sec:finitetemp}).
Nevertheless, we demonstrate that the chemical potential and the two-particle spectral function entering the number density equation in Section~\ref{sec:ionizatineq} can be evaluated self-consistently in the narrow thermal width limit. This approach, taking leading finite temperature real-part corrections into account, is already more general of what has been computed so far in the literature. In principle, it is possible to take a non-consistent approach and compute $G^{++--}_{\eta \xi,s} (x,x,x,x)\big|_{\text{eq}}$ including imaginary parts in the potential while only including real-part corrections to the chemical potential. However, some care must be taken when doing so. This is because the chemical potential corrects the functional form of the number density dependence in our master equation.
We discuss in more detail the possibility of taking a non-self-consistent approach by the end of Section~\ref{sec:ionizatineq}.

Finally in Section~\ref{sec:com2linear}, we compare our Master Eq.~(\ref{eq:numbergrandcanonicalgeneral}) to the previous literature, relying on the method of linear response theory. Consistency is proven in the linear regime close to chemical equilibrium.

\subsection{Recovering the Lee-Weinberg equation}
\label{sec:zeroselfinteraction}
We take the limit of zero self-interactions $\alpha_{\chi} \rightarrow 0$ while keeping the annihilation term $\Gamma_s$ as a perturbation. It should be emphasized again that we have to approximate the spectral function and the chemical potential both in the same limit in order to obtain a self-consistent solution.
The free spectral function without self-interactions and the ideal pressure are given by:
\begin{align}
G_{\eta \xi}^{\rho}(\mathbf{0},\mathbf{0};E)|_{l=0}  &=  \theta(E) \frac{1}{2\pi} \Tr[\mathbf{1}_{2\times2}] M^{3/2} E^{1/2},\; \\
p_0 \Omega &= T \ln \Tr[e^{-\beta(H_0-\mu_{\eta} N_\eta-\mu_{\xi} N_\xi)} ].
\end{align}
Here, $H_0$ is the free Hamiltonian and for a derivation of this result for the two-particle spectral function directly starting from the general expression Eq.~(\ref{eq:solutionnumerical}) can be found in Appendix \ref{app:spec_at_vacuum}.
The number density can be obtained from Eq.~(\ref{eq:thermonumber}) by using the ideal pressure:
\begin{align}
n_{\eta}[\mu_{\eta}] &= \left. \frac{\partial p_0}{\partial \mu_{\eta}} \right|_{T,\Omega} = G_{\eta,0}^{+-}(x,x) = e^{\beta \mu }  2 \int \frac{\text{d}^3 \mathbf{p}}{(2\pi)^3} e^{-\beta (M+\mathbf{p}^2/2M)} = e^{\beta \mu } n_{\eta,0}^{\text{eq}},\, \\
n_{\eta,0}^{\text{eq}} & = 2 \left( \frac{M T}{2 \pi} \right)^{3/2} e^{- \beta M} \label{eq:idealscatteringeq}.
\end{align}
In the second equality of the first line, we find the relation between ideal number density and the non-interacting correlator $G_{\eta,0}^{+-}(x,x)$. The latter quantity has to be evaluated in a grand canonical ensemble, which we have done in the third equality by using KMS condition and the DM dilute limit (see Appendix \ref{app:idealgas}). The DM dilute limit should be taken in the computation of $G_{\eta,0}^{+-}(x,x)$ to be consistent with the computation of the spectral function. For the latter quantity we have seen in Section \ref{sec:dilutelimit} only in the DM dilute limit it is independent of the DM number density and our general solution Eq.~(\ref{eq:solutionnumerical}) relies on this assumption. In the last equality of the first line we defined the conventional chemical equilibrium number density of ideal particles.
Finally, we obtain from the last equality the non-interacting (ideal) chemical potential by inversion: $\beta \mu = \ln[n_{\eta}/n_{\eta,0}^{\text{eq}}]$. Note that this inversion can only be done analytically if one approximates the Fermi-Dirac distribution as Maxwell-Boltzmann (which is our DM dilute limit). Entering these results of the spectral function and the chemical potential into our master formula for the DM number density Eq.~(\ref{eq:numbergrandcanonicalgeneral}), leads to the conventional Lee-Weinberg equation for DM particles with zero self-interactions:
\begin{align}
\dot{n}_{\eta} + 3 H  n_{\eta} &= - \langle\sigma v_{\text{rel}}\rangle\left[n_{\eta}^2 - \left(n_{\eta,0}^{\text{eq}}\right)^2 \right].
\end{align}
Here, we have recovered the standard thermal averaged cross section by using the simple substitution $E=M v_{\text{rel}}^2/4$ for the positive energy spectrum:
\begin{align}
\langle \sigma v_{\text{rel}} \rangle &= \frac{(M/T)^{3/2}}{2 \sqrt{\pi}} \int_{0}^{\infty} \text{d} v_{\text{rel}} \; e^{- \frac{v_{\text{rel}}^2 M}{4T}}   v_{\text{rel}}^2 (\sigma v_{\text{rel}})\\
&=(\sigma v_{\text{rel}}).
\end{align}
The last equality holds for constant s-wave annihilation cross section $(\sigma v_{\text{rel}})$ as it is the case for our model.

\subsection{Spectral function, Sommerfeld enhancement factor and decay width for vanishing thermal corrections}
\label{sec:relations}
\begin{figure}[h]
\includegraphics[scale=0.42]{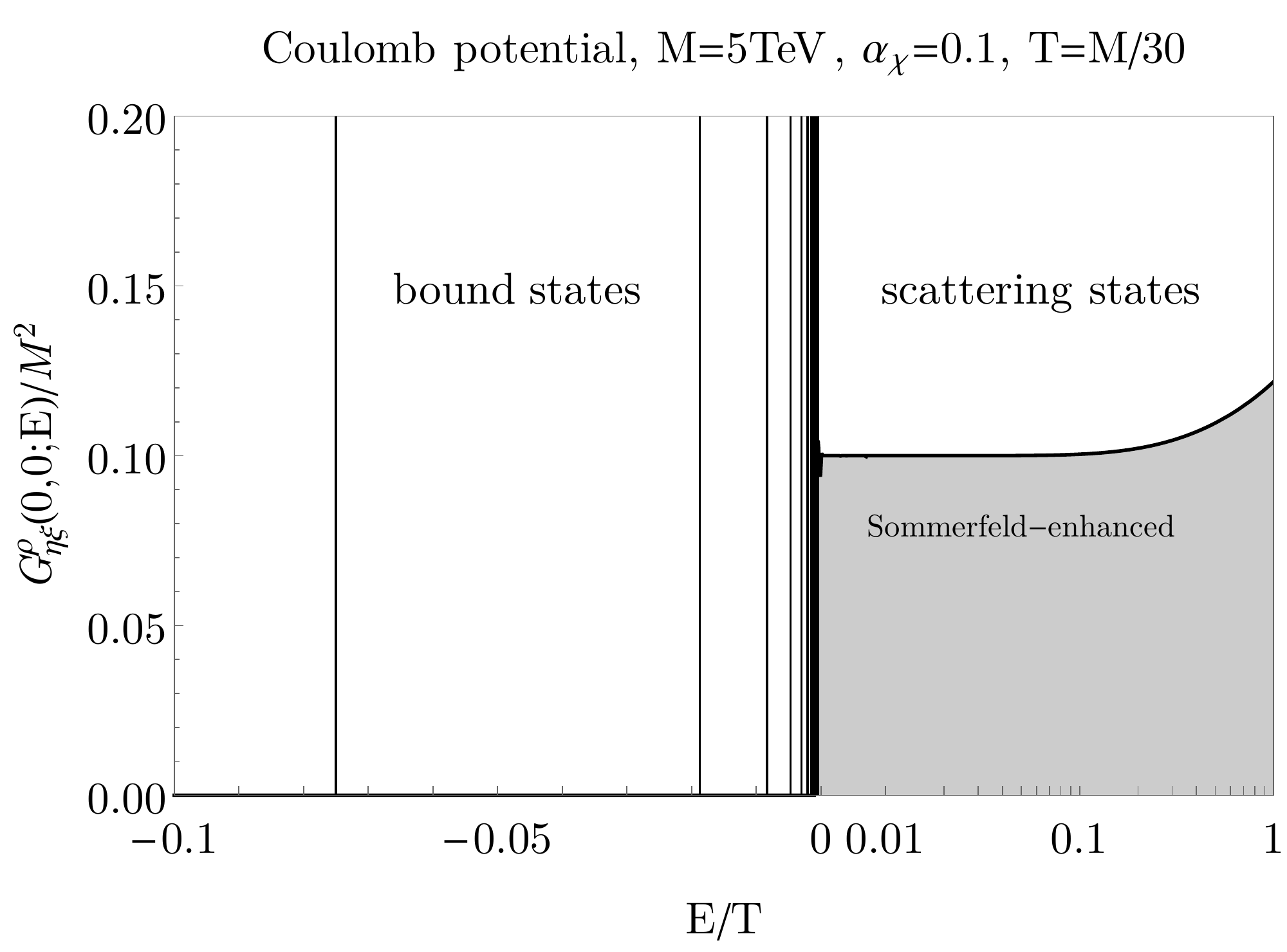}
\includegraphics[scale=0.42]{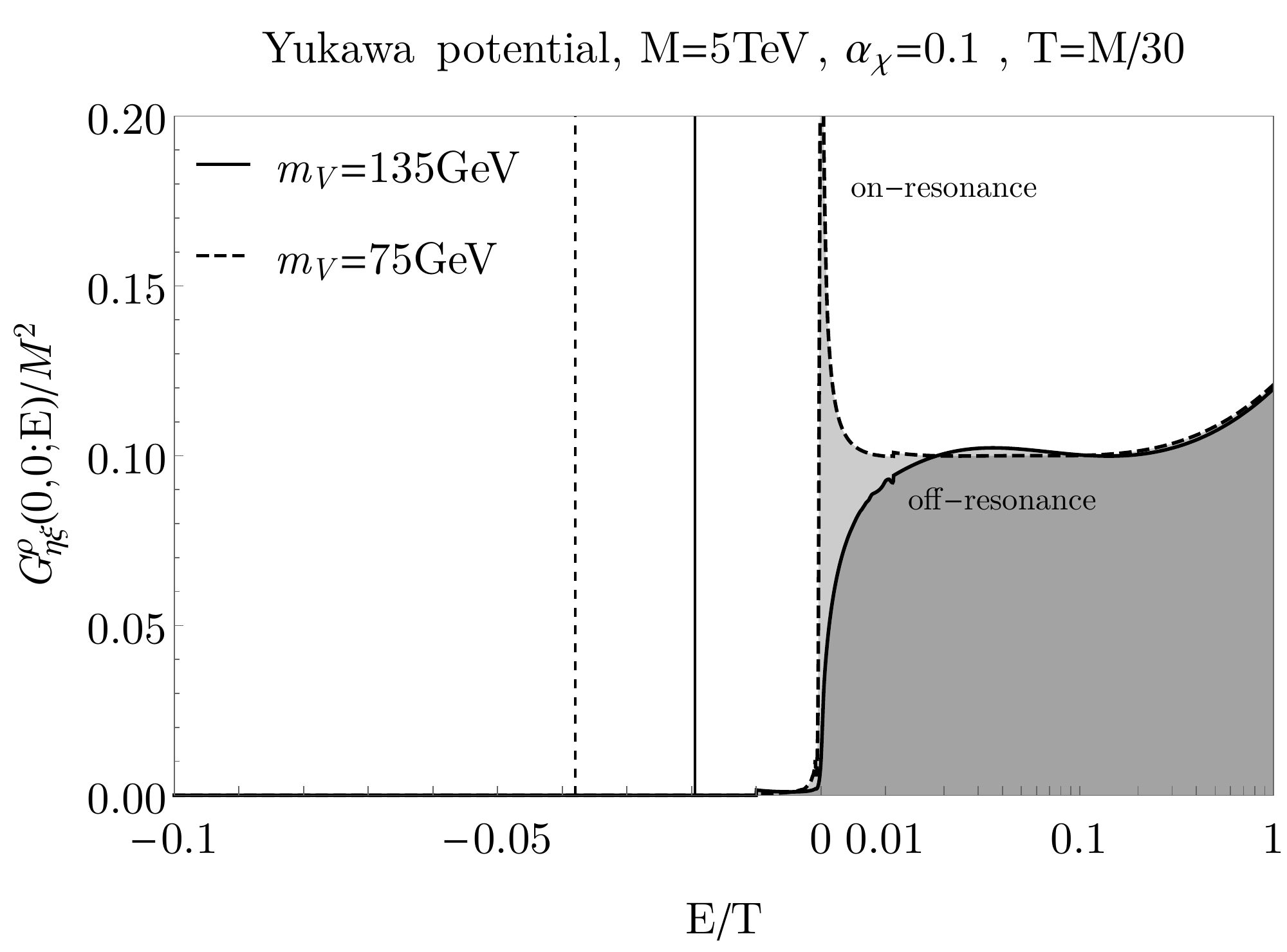}
\caption{S-wave two-particle spectral function vs.\ the energy $E$ in units of typical freeze-out temperature shown for a standard Coulomb (left) and Yukawa (right) potential. The two-particle spectral function enters directly our master formula and is weighted by the Boltzmann factor for all the energy range.}
\label{fig:spectralvacuum}
\end{figure}

We turn now to the interacting case $\alpha_{\chi} \neq 0$ and compute the two-particle spectral function. The s-wave two-particle spectral function is numerically solved according to Eq.~(\ref{eq:solutionnumerical}) in the limit of vanishing finite temperature corrections and the results are shown in Fig.~\ref{fig:spectralvacuum}. Poles in the negative energy spectrum represent the bound states, while the spectrum is continuous for the scattering states at positive energy. In the vacuum limit one clearly sees that the scattering states can be separated from the bound state contribution at $E=0$. Due to this separation, the solution of the spectral function is directly related to the Sommerfeld enhancement factor $S(v_\text{rel})$ and bound-state decay width $\Gamma_n$. These relations are given below and now it becomes clear that the two-particle spectral function as shown in Fig.~\ref{fig:spectralvacuum} is, for the vacuum case, just a convenient way of presenting all contributions simultaneously.
\begin{claim}
Relation between two-particle spectral function $G_{\eta \xi}^{\rho}$, Sommerfeld enhancement factor $S$ and decay width $\Gamma_n$ in the limit of vanishing finite temperature corrections:
\begin{align}
	(\sigma v_{\text{rel}}) G_{\eta \xi}^{\rho}(\mathbf{0},\mathbf{0};E)|_{E>0,l=0}	
	&=   
	\frac{1}{4 \pi} \Tr[\mathbf{1}_{2\times2}] M^2 v_\text{rel} (\sigma v_\text{rel}) S(v_\text{rel}),\label{eq:possol}\\
	(\sigma v_{\text{rel}}) G_{\eta \xi}^{\rho}(\mathbf{0},\mathbf{0};E)|_{E<0,l=0}
	&= 
	\frac{\pi}{2} \Tr[\mathbf{1}_{2\times2}] \sum_n \delta(E - E_{B_n}) \Gamma_n,\label{eq:negsol}
\end{align}
where $E=M v_{\text{rel}}^2/4$ for the scattering states, $ E_{B_n}$ is the (negative) binding energy for the bound states, and $(\sigma v_{\text{rel}})$ is the tree-level s-wave annihilation cross section. $\Gamma_n$ is decay width of the bound state (not to be confused with our annihilation term $\Gamma_s$ at the beginning of this work).
\end{claim}
These relations can be proven directly from our general solution Eq.~(\ref{eq:solutionnumerical}), see App.~\ref{app:spec_at_vacuum} for a derivation. On a first look, the spectral function in the vacuum case seems just a nice way of presentation. Instead one should emphasize that
the notion of spectral function is more general and unifies the picture of scattering state annihilation and bound-state decay. This observation becomes important for the finite temperature case discussed in Section~\ref{sec:finitetemp}, where it is impossible to separate or distinguish between annihilation and decay. The spectrum includes both. Only in the absolute vacuum case a clear distinction between annihilation and decay can be made.

Coming back to Fig.~\ref{fig:spectralvacuum} and now keeping in mind the relations Eqs.~(\ref{eq:possol})-(\ref{eq:negsol}). There is an infinite number of exited $S$-bound states for the Coulomb case (left plot) with binding energy $ E_{B_n}= -\alpha_{\chi}^2M/(4 n^2)$, where $n$ is the number of the exited states and $n=1$ is the ground state with lowest binding energy , shown as the pole most to the left. At small positive energies, where $v_{\text{rel}}\lesssim \alpha_{\chi}$, the spectral function is constant, resulting in the familiar scaling $S(v_{\text{rel}})\propto v_{\text{rel}}^{-1}$ according to Eq.~(\ref{eq:possol}).

In the Yukawa potential case, shown in the right plot of Fig.~\ref{fig:spectralvacuum}, there is a finite number of bound-state solutions. For certain ratios of $\epsilon_{\phi} \equiv m_V/(\alpha_{\chi} M)$ there exist a bound-state solution with zero binding energy ($E=0$). For those special cases the Sommerfeld enhancement factor scales as $S(v_{\text{rel}})\propto v_{\text{rel}}^{-2}$ for $v_{\text{rel}}\lesssim m_V/M \ll \alpha_{\chi}$, called on-resonance regime, leading to an interesting observational impact on cosmology at very late times \cite{Binder:2017lkj, Bringmann:2018jpr}. Roughly, those poles where the spectral function would diverge are at the on-resonance condition $\epsilon_{\phi}= 6/(m^2 \pi^2) $ where $m$ is integer\footnote{This is exactly true for the Hulth\'en potential case which is analytically solvable, while for the Yukawa case this resonance condition slightly deviates. }. The on-resonance divergences give rise to partial wave unitarity violation of the total cross section, as can be seen in the right plot of Fig.~\ref{fig:spectralvacuum} at $E=0$. It has been pointed out in \cite{Blum:2016nrz} that once the imaginary contribution of the annihilation part (proportional to our $\Gamma_s$) is included self-consistently in the solution of the Schr\"odinger-like Eq.~(\ref{eq:BSeq_st_dil}), then the Sommerfeld enhancement starts to saturate below the unitarity limit\footnote{Since we have treated the annihilation term $\Gamma_s$ as a perturbation, where the leading order contribution was found to contribute to the change in the number density, the contribution of $\Gamma_s$ does not occur in Eq.~(\ref{eq:BSeq_st_dil}). We will see later that at finite temperature the imaginary parts in the potential will dominate over the imaginary contribution from the annihilation term and thus at sufficiently finite temperature we will always get a saturation below the unitarity bound.}. This means that for some small velocity there is a transition from the divergent scaling $S(v_{\text{rel}})\propto v_{\text{rel}}^{-2}$ to $S(v_{\text{rel}})\propto \text{const.}$ which results in zero spectrum at zero energy. The saturation is always present if the on-resonance condition is not exactly fulfilled. The other extreme case is if $\epsilon_{\phi}$ is taken exactly in between neighbouring on-resonance values, called off-resonance. Then, $S(v_{\text{rel}})$ never scales stronger than  $S(v_{\text{rel}})\propto v_{\text{rel}}^{-1}$ and at some small velocity of the order $v_{\text{rel}}\lesssim m_V/M$ the Sommerfeld enhancement factor starts to saturate and the spectral function approaches zero.

\subsection{DM number density equation for vanishing thermal corrections}
\label{sec:ionizationeq_ideal}
In the previous section we have proven in the limit of vanishing finite temperature corrections a relation between spectral function, standard expression of Sommerfeld enhancement factor and the bound-state decay width. Inserting these relations Eqs.~(\ref{eq:possol})-(\ref{eq:negsol}) into our Master Eq.~(\ref{eq:numbergrandcanonicalgeneral}) leads to the following differential equation for the total number density $n_\eta $:
\begin{align}
	\dot n_\eta  + 3 H n_\eta 
	= - \langle \sigma v_{\text{rel}} \rangle \left[\left(n_{\eta,0}^\text{eq}\right)^2e^{\beta 2 \mu} - \left(n_{\eta,0}^\text{eq}\right)^2 \right]
	- \sum_i \Gamma_i \left[ n_{B_i,0}^\text{eq}e^{\beta 2 \mu} - n_{B_i,0}^\text{eq} \right]. \label{eq:numberidealzerotemp}
\end{align}
In the limit of zero chemical potential, defining chemical equilibrium, the r.h.s. vanishes as expected. Here, we recovered the thermal averaged Somerfeld enhancement factor:
\begin{align}
\langle \sigma v_{\text{rel}} \rangle &= \frac{(M/T)^{3/2}}{2 \sqrt{\pi}} \int_{0}^{\infty} \text{d} v_{\text{rel}} \; e^{- \frac{v_{\text{rel}}^2 M}{4T}}   v_{\text{rel}}^2 (\sigma v_{\text{rel}}) S(v_{\text{rel}}).
\end{align}
The chemical equilibrium number density for the scattering states $n_{\eta,0}^\text{eq}$ was coming out as already defined in Eq.~(\ref{eq:idealscatteringeq}). This outcome is fully consistent with the result one would get from integrating the Maxwell-Boltzmann equilibrium phase-space density (multiplied by spin factor 2) of non-relativistic particles. The chemical equilibrium number density of the bound-states was defined as
\begin{align}
n_{B_i,0}^{\text{eq} } 
	= \left( \frac{2 M T}{2 \pi} \right)^{3/2} e^{- \beta M_{B_i}}\label{eq:eqbound},
\end{align}
where the mass of bound state $i$ is $M_{B_i} = 2 M - |E_B^i|$ and the subscript '0' stands for ideal bound states, respectively. The term in front of the exponential in the bound state number density Eq.~(\ref{eq:eqbound}) needs some further explanation. Since we have only considered s-wave contributions to the spectral function, $n_{B_i,0}^{\text{eq} }$ is the equilibrium number density of the i-th exited Para-WIMPonium. The decay, as well as the annihilation, of Ortho-WIMPonium into three $A_{\mu}$ would be a p-wave process. To form Para-WIMPonium there is only one spin option while for Ortho-WIMPonium there are 3, consistent with the picture of having in total 4 spin degrees of freedom. Therefore, the spin factor 1 in Eq.~(\ref{eq:eqbound}) comes out correctly. When carefully looking at the term in front of the exponential in Eq.~(\ref{eq:eqbound}), it can be seen that the normalization of the distribution came out as like integrating the phase space density:
\begin{align}
n_{B_i,0}^{\text{eq} } = \int \frac{\text{d}^3 \mathbf{P}}{(2\pi)^3} e^{-\beta(M_{B_i} + \mathbf{P}/4M)}.
\end{align}
The kinetic term $\mathbf{P}/4M$ of the bound state misses the correction coming from the binding energy, because the conventional normalisation would give $(n_{B_i,0}^{\text{eq} })_{(c)} = (M_{B_i}/2\pi)^{3/2} e^{- \beta M_{B_i}}$. The reason why this correction of the order $\mathcal{O}(E_{B}^i/M)$ does not come out as in the conventional case can be explained by how we have approximated the Martin-Schwinger hierarchy. In our paper, we expand the equation for the four-point correlator around the product of two free propagators. This is why the spectral function only depends on $E=\omega-\mathbf{P}/4M$ as we have shown in Eq.~(\ref{eq:BSeq_st_dil_omega}). If we would iterate the solution, e.g.\ correcting the free correlators and inserting them again into the solution of the four-point correlator, we could obtain the conventional result.
However, note that the correction is small for perturbative systems, since typically $\mathcal{O}(E_{B}^i/M) \sim \mathcal{O}(\alpha^2)$. 

We are turning to the discussion of the chemical potential $\mu$ in Eq.~(\ref{eq:numberidealzerotemp}).
The chemical potential can be obtained by inverting the number density as a function of the chemical potential. For an \emph{ideal gas} description it is known that the total number density of $\eta$-particles $n_{\eta}$, as it appears on the l.h.s. of Eq.~(\ref{eq:numberidealzerotemp}), is in general just given by the sum of scattering and bound state contributions:
\begin{align}
n_{\eta} &= n_{\eta,0}+ \sum_i n_{B_i,0}\label{eq:totalnumberideal}\\
n_{\eta,0} &= n_{\eta,0}^{\text{eq}} e^{ \beta \mu_{\eta}},\\
n_{\xi,0} &= n_{\xi,0}^{\text{eq}} e^{ \beta \mu_{\xi}},\\
	n_{B_i,0} &= n_{B_i,0}^{\text{eq} }  e^{-\beta \mu_{B_i}}.
\end{align}
Since we have imposed a grand canonical state with only one single chemical potential, the chemical potentials for the scattering and bound states are related and therefore the number densities are not independent quantities.
Assuming a grand canonical ensemble with only one time dependent chemical potential $\mu$ implies ~$2 \mu = 2\mu_{\eta}= \mu_{B_i}$, which leads to the relation
\begin{align}
	n_{B_i,0} = 2 \left( \frac{\pi}{MT} \right)^{3/2} n_{\eta,0}^2 e^{-\beta |E_{B}^i|}.
	\label{eq:ionization_eq_ideal}
\end{align}
This is nothing but the \emph{Saha ionization equilibrium condition}. To see it explicitly, let us insert Eq.~(\ref{eq:ionization_eq_ideal}) into Eq.~(\ref{eq:totalnumberideal}), leading to a quadratic equation for the number density of free scattering states:
\begin{align}
n_{\eta} = n_{\eta,0}+ K^{\text{id}}(T)n_{\eta,0}n_{\eta,0},\quad K^{\text{id}}(T) = \frac{\sum_i n_{B_i,0}}{n_{\eta,0}n_{\eta,0}}. \label{eq:sahaideal}
\end{align}
$K^{\text{id}}(T)$ is according to Eq.~(\ref{eq:ionization_eq_ideal}) independent of the chemical potential.
This quadratic Eq.~(\ref{eq:sahaideal}) can be solved, leading to the degree of ionization $\alpha^{\text{id}}$ for ideal gases:
\begin{align}
\frac{n_{\eta,0}}{n_{\eta}} = \alpha^{\text{id}}(n_{\eta}K^{\text{id}}(T)),\quad \alpha^{\text{id}}(x) =  \frac{1}{2x} \left( \sqrt{1+4x} -1 \right).\label{eq:iodegreeideal}
\end{align}
The chemical potential can now be obtained from this equation by using $n_{\eta,0} = n_{\eta,0}^{\text{eq}} e^{ \beta \mu}$ resulting in:
\begin{align}
\beta \mu = \ln\left[ \frac{\alpha^{\text{id}}n_{\eta}}{n_{\eta,0}^{\text{eq}} }\right].
\end{align}
Inserting this chemical potential into Eq.~(\ref{eq:numberidealzerotemp}) we finally end up with the 
\begin{claim}
Boltzmann equation for vanishing finite temperature corrections, ideal gas approximation, and the system in a grand canonical state:
\begin{align}
	\dot n_\eta  + 3 H n_\eta 
	= - \langle \sigma v_{\text{rel}} \rangle \left[ \left(\alpha^{\text{id}} n_{\eta}\right)^2 - \left(n_{\eta,0}^\text{eq}\right)^2 \right]
	- \sum_i  \Gamma_i  n_{B_i,0}^\text{eq} \left[   \left(\frac{\alpha^{\text{id}} n_{\eta}}{n_{\eta,0}^{\text{eq}}}\right)^2  - 1 \right].\label{eq:sahaidealzerot}
\end{align}
\end{claim}
\noindent
The equation is closed in terms of the total number density $n_\eta $.
Before we discuss this result in detail, let us consider the case where we would have treated all bound and scattering states to be independent. This could have been realised in Eq.~(\ref{eq:numberidealzerotemp}) by assigning different chemical potentials to scattering and bound states. Then we would have ended up with decoupled equations:
\begin{align}
\dot{n}_{\eta,0}+ 3 H n_{\eta,0} = - \langle \sigma v_{\text{rel}} \rangle \left[  n_{\eta,0}^2 - \left(n_{\eta,0}^\text{eq}\right)^2 \right],\quad \sum_i \dot{n}_{B_i,0}+ 3 H n_{B_i,0} = - \sum_i  \Gamma_i   \left[   n_{B_i,0} - n_{B_i,0}^\text{eq} \right].\label{eq:decoupled}
\end{align}
These are the standard equations if bound and scattering states are decoupled.
They might be helpful to understand Eq.~(\ref{eq:sahaidealzerot}) better. 
Namely, when adding the well-known Boltzmann Eqs.~(\ref{eq:decoupled}) and \emph{imposing} ionization equilibrium, one would end up with Eq.~(\ref{eq:sahaidealzerot}).
We summarize and discuss the main findings of this Section below.
\begin{itemize}
\item The differential Eq.~(\ref{eq:sahaidealzerot}) describes the out-of-chemical equilibrium evolution of the total number density, including the reactions $ \eta\xi \rightleftharpoons \{ A A, \psi \psi \} $ (Sommerfeld-enhanced annihilation and production) and $ (\eta\xi)_{\text{B}} \rightleftharpoons \{ A A, \psi \psi \} $ (bound-state decay and production) under the constraint of ionization equilibrium for all times. The total number density $n_{\eta}$ counts both: free particles as well as particles in the bound state. Eq.~(\ref{eq:sahaidealzerot}) is equivalent to the coupled set of Boltzmann equations including soft emissions and absorptions \cite{vonHarling:2014kha} in the limit of ionization equilibrium. The equations are independent of the bound state formation or ionization cross section since the rates are, by assumption, balanced.
One can also see it from a different perspective. Via Eq.~(\ref{eq:sahaidealzerot}) it is very elegant to include bound state formation or dissociation processes without calculating the cross sections or solving a coupled system of differential equations.
Note that when comparing our single equation to the coupled set of Boltzmann equations in  Refs.~\cite{vonHarling:2014kha}, the last term in Eq.~(\ref{eq:sahaidealzerot}) accounting for the direct production of bound states from two photon annihilation (inverse decay) was dropped. It might have little impact since bound state formation from radiative processes can be much more efficient around the freeze-out.

\item Let us discuss some asymptotic regimes of Eq.~(\ref{eq:sahaidealzerot}), assuming the system has bound state contributions. For this case, the ionization degree $\alpha^{\text{id}}$ has a non-trivial dependence on the total number density $n_{\eta}$, as can be seen from Eq.~(\ref{eq:iodegreeideal}). This leads to the fact that the collision term of the number density Eq.~(\ref{eq:sahaidealzerot}) has in general neither a linear nor a quadratic dependence on $n_{\eta}$, as it is for the decoupled conventional Eqs.~(\ref{eq:decoupled}).
However, we recover correctly the quadratic and linear form in some regimes discussed now.
Close to the freeze-out, the temperature is much larger than the binding energy of bound states. As a consequence, the bound state contribution in the ionization degree can be neglected. This can be seen directly from Eq.~(\ref{eq:sahaideal}). In the high temperature regime $x\ll 1$, leading to $\alpha^{\text{id}} \sim 1 $ (fully ionized) and the r.h.s of the number density Eq.~(\ref{eq:sahaidealzerot}) is quadratic in $n_{\eta} $. At late times, roughly when temperature becomes of the order the binding energy, the contribution of the bound states becomes strongly enhanced due to Boltzmann factor $\propto e^{\beta|E_B|}$ in $K^{\text{id}}(T)$. For this low temperature regime $ x \gg 1 $, leading to $\alpha^{\text{id}} \simeq 1/\sqrt{n_{\eta} K^{\text{id}}(T)}$ which is much smaller than unity.
This behavior is expected since bound states with energy $2 M - |E_B^i|$ are thermodynamically more favoured compared to the scattering states with energy $2M$ for $T < |E_B^i|$.
Therefore, the equilibrium limit for low temperature is that most of the particles are contained in the lowest bound state (ground state). Inserting the low temperature behavior of $\alpha^{\text{id}} \simeq 1/\sqrt{n_{\eta} K^{\text{id}}(T)}$ into Eq.~(\ref{eq:sahaidealzerot}) results in the fact that at late times the r.h.s the of number density equation is \emph{linear} in $n_{\eta}$ and the proportionality factor is effectively the bound-state decay rate. Since typically the decay rates are much larger compared to the Hubble rate and the number density equation is linear in  $n_{\eta}$, the total number density gets depleted exponentially fast in time.
This means at late times, if ionization equilibrium is assumed, also the free DM particle number decreases simply because it must follow the exponential fast decay of the bound states in order to maintain the imposed ionization equilibrium. In summary, if bound-state solutions are present and one would integrate numerically Eq.~(\ref{eq:sahaidealzerot}) until today, effectively no DM would remain as a consequence of the imposed ionization equilibrium. However, we know from the full coupled set of classical Boltzmann equations \cite{vonHarling:2014kha} that ionization equilibrium is \emph{not} maintained for all times during the DM depletion phase caused by bound-state decays.
This is because, once the decay rate exceeds the ionization rate, the system leaves the ionization equilibrium. The temperature dependence of the bound state formation (BSF) and ionization determines by how much the stable components are depleted during this critical epoch. Once the BSF rate drops below the cosmic expansion rate $H$ the depletion stops.

\item  If we instead would have treated bound and scattering states separately, with independent chemical potentials, we would have obtained Eq.~(\ref{eq:decoupled}).
In this equation one can see that the bound states are independent of the scattering states. Since the differential equation of the former is linear in the number density on the r.h.s., at some point all bound states start to decay away. The coupling between bound and scattering states via radiative processes are not included in our theory and therefore do not appear. As we have learned, in our theory treating bound states as composite particles there is only one chemical potential. Therefore, by naively just giving bound and scattering states different chemical potentials would lead to the fact that we describe the system not with a grand canonical ensemble. Moreover, the KMS condition does not hold for this case. In order to be able to introduce different chemical potentials might require to rewrite our theory in terms of effective operators, creating only scattering or only bound states respectively. Then, there might be for every operator a conserved charge and one can associate individually different chemical potentials. We will see later that this more 'phenomenological' procedure is definitely not applicable for finite temperature case. There it becomes impossible to introduce such operators since the eigenvalues might be not well defined when non-hermitian thermal corrections are included.

\item For going beyond the ionization equilibrium it is required to include ultra-soft terms from the beginning and re-derive the DM correlator EoM including those corrections. This we leave for future work and we restrict our equations to be valid as long as ionization equilibrium can be maintained. If no bound-state solution exist, our equations presented here only assume kinetic equilibrium \cite{vandenAarssen:2012ag, Binder:2017rgn}. If bound states exist, the validity of Eq.~(\ref{eq:sahaidealzerot}) can be estimated. Once the decay rate exceeds the ionization rate at late times, the regime of out-of-ionization equilibrium starts (see also \cite{vonHarling:2014kha}). Until then, our description is valid.
In the next Section, we will generalize this equation for the finite temperature case. We will assume that the thermal width in the effective potential is small.
Already in this case, the number of bound states is dynamical since the screening as well as the constant real part term in the effective potential are temperature dependent. They lead for decreasing temperature to an abrupt occurrence of bound-state poles in the spectral function (see also Section \ref{sec:finitetemp}). Therefore, at finite temperature the description via Eq.~(\ref{eq:sahaidealzerot}) is insufficient. The more general description can be obtained from our formalism by going back to our Master equation, expressing the annihilation or decays in terms of the two-particle spectral function. The spectral function automatically decides for a given temperature which part of the spectrum contributes to the continuum and which part is bounded, as we will see later. Moreover, as we discuss in the next Section, it is required at finite temperature to include non-ideal contributions to the idealized Saha equation, leading to finite temperature corrections of the chemical potential. Thus, for a consistent treatment we need to compute both: long-range modified annihilation and the chemical potential to the same order of approximation.
\end{itemize}

\subsection{Chemical potential in narrow thermal width approximation}
\label{sec:ionizatineq}
In the previous Section, we established what is the outcome of our Master equation in the limit of vanishing thermal corrections. 
In this Section, we compute the chemical potential for finite $g_{\chi}$ and $g_{\psi}$.
The only assumption we make is that the thermal width in the mediator correlator $D$ is subleading. Only the real-part correction in Eq.~\eqref{eq:inmediumpot} is kept. This is called the narrow width (or quasi particle) approximation. It leads to the fact that the correlator takes the simple form
\begin{align}
D(x,y)=\delta_\mathcal{C}(t_x,t_y) (-i)V(\mathbf{x}-\mathbf{y}),
\end{align}
where $V$ is the screened Yukawa potential.
To now evaluate the pressure explicitly, we make use of the structure of the Hamiltonian for this potential:
\begin{align}
H&= H_0 + M (N_{\eta} + N_{\xi}) + \frac{g_{\chi}^2}{2}\int_{\mathbf{x},\mathbf{y}}  V(\mathbf{x}-\mathbf{y}) \left[\eta^{\dagger}(\mathbf{x})\eta^{\dagger}(\mathbf{y})\eta(\mathbf{y})\eta(\mathbf{x})
+  \xi(\mathbf{x})\xi(\mathbf{y})\xi^{\dagger}(\mathbf{y})\xi^{\dagger}(\mathbf{x})
- 2 \eta^{\dagger}(\mathbf{x})\xi(\mathbf{y})\xi^{\dagger}(\mathbf{y})\eta(\mathbf{x})\right].
\end{align}
Substituting $g_{\chi}^2 \rightarrow \lambda g_{\chi}^2$ and taking the partial derivative of the partition function with respect to $\lambda$ we arrive at the convenient form:
\begin{align}
p \Omega - p_0 \Omega &=  T \int_0^1 \text{d} \lambda\frac{1}{Z_{\text{gr}}} \partial_{\lambda}    Z_{\text{gr}}(\Omega,T,\mu) \\&=  T \int_0^1 \text{d} \lambda \int_{\mathbf{x},\mathbf{y}} \frac{ g_{\chi}^2}{2} V(\mathbf{x}-\mathbf{y}) \left[2 G^{++--}_{\eta \xi}(\mathbf{x},\mathbf{y},\mathbf{x},\mathbf{y};t-t^{\prime},\lambda)|_{t=t^{\prime}} - G^{++--}_{\eta \eta}- G^{++--}_{\xi \xi}  \right].
\end{align}
Spin indices are summed over equal arguments and $G^{++--}_{\eta \eta}$, $G^{++--}_{\xi \xi}$ carry the same arguments as $G^{++--}_{\eta \xi}$. Applying KMS condition, and expressing $G^{++--}$ in terms of spectral function we arrive, after introducing Wigner coordinates and Fourier transformation, at:
\begin{align}
\beta p = \beta p_0 + e^{-\beta 2M} \int \frac{\text{d}^3\mathbf{P}}{(2 \pi)^3} e^{-\beta \mathbf{P}^2/4M} \int \frac{\text{d} E}{(2 \pi)} e^{-\beta E }\int_0^1 \text{d} \lambda \int_{\mathbf{r}} \frac{ g_{\chi}^2}{2} V(\mathbf{r}) \left[2 e^{\beta (\mu_{\eta} + \mu_{\xi})}  G^{\rho}_{\eta \xi}(\mathbf{r},\mathbf{r};E,\lambda) - e^{\beta 2\mu_{\eta} } G^{\rho}_{\eta \eta}- e^{\beta 2 \mu_{\xi}} G^{\rho}_{\xi \xi}  \right]. \label{eq:pressuredensity}
\end{align}
This equation tells us that the total pressure is equal to the ideal pressure, defined as 
\begin{align}
p_0 \Omega = T \ln \Tr[e^{-\beta(H_0-\mu_{\eta} N_\eta-\mu_{\xi} N_\xi)} ],
\end{align}
plus non-ideal contributions arising from DM long-range self-interactions.
It is possible to eliminate the $\lambda$ integration in Eq.~(\ref{eq:pressuredensity}) by partial integration and using the BS equation backwards. The final result can be expressed in terms of bound state contributions and the change of the scattering phase with respect to the energy for the scattering states. The equation is then known as Beth-Uhlenbeck formula. For the following discussion, it is however not required to explicitly give those expressions and therefore we just refer to the result in standard text books for non-ideal plasmas, see \cite{wernerebeling}\footnote{In non-ideal plasma literature, the non-ideal contribution in Eq.~(\ref{eq:pressuredensity}) is referred as the second-viral coefficient. It might be also interesting to note that the number of bound states are related to scattering phases according to the Levinson theorem.}. Let us remark that one can also directly solve Eq.~(\ref{eq:pressuredensity}) numerically via the solution of the two-particle spectral function.
This is the power of Eq.~(\ref{eq:pressuredensity}). One can solve self-consistently for $G_{\eta \xi,s}^{++--} (x,x,x,x)\big|_{\text{eq}}$ and the chemical potential (see below) entering our Master equation just by evaluating the two-particle spectral function \emph{without} specifying what is a bound or a scattering state. The two-particle spectral function automatically takes into account everything. This is because Eq.~(\ref{eq:pressuredensity}) is \emph{exact} in the narrow thermal width limit. We have just reduced the problem of evaluating the total pressure to the evaluation of a four-point correlation function.

For the moment, only the structure of Eq.~(\ref{eq:pressuredensity}) is important. Namely, when differentiating the total pressure equation with respect to chemical potential we obtain, according to Eq.~(\ref{eq:thermonumber}), for the total DM number densities:
\begin{align}
n_{\eta} &= n_{\eta,0} +  K(T) n_{\eta,0} n_{\xi,0}, \label{eq:totalnumber}\\
n_{\xi} &= n_{\xi,0} +  K(T)n_{\eta,0} n_{\xi,0},\label{eq:totalnumberanti}\\
n_{\eta,0}&= n_{\eta,\text{0}}^{\text{eq}} e^{\beta \mu_{\eta}},\label{eq:idealnumber_eta}\\
n_{\xi,0}&= n_{\xi,\text{0}}^{\text{eq}} e^{\beta \mu_{\xi}},\label{eq:idealnumber_xi}\\
n_{\eta,\text{0}}^{\text{eq}} &= n_{\xi,\text{0}}^{\text{eq}} =
2 \left( \frac{M T}{2 \pi} \right)^{3/2} e^{- \beta M}.
\label{eq:idealnumber}
\end{align}
The subscript '$0$' labels ideal number densities, i.e. $n_{\eta,0}$ is a number density for a free DM without gauge interactions as in the previous Section. 
Superscript 'eq' stands for chemical equilibrium and a symmetric plasma is assumed $\mu_{\eta} = \mu_{\xi}$. We stress that $n_{\eta,0}$ should not be confused with number densities of one quasi-particle excitation of DM in thermal plasma. 
This equation just tells us that the exact total DM number density, $n_\eta$ and $n_\xi$, including corrections from thermal plasma and bound states can be simply expressed as Eqs.~\eqref{eq:totalnumber} and \eqref{eq:totalnumberanti} by means of the ideal number densities given in Eqs.~\eqref{eq:idealnumber_eta}, \eqref{eq:idealnumber_xi}, and \eqref{eq:idealnumber}.
All the effects from interactions are encoded in $K(T)$. An explicit expression for $K(T)$ can be obtained by comparing Eq.~(\ref{eq:totalnumber}) to (\ref{eq:pressuredensity}). 
Again, it is only important to know that $K(T)$ includes bound state contributions as well as scattering parts which can be seen from the integration of spectral function in the whole energy range. One can see that in the limit of $K(T) \to 0$ the number densities of free DM are recovered.
If we would send finite temperature corrections to zero and only include the bound state contribution, then $K(T)$ just reproduces the ideal gas case from the previous section, see Eq.~\eqref{eq:sahaideal} and \eqref{eq:ionization_eq_ideal}. In this sense, \emph{standard Boltzmann equations (see also previous Section) typically used in numerical codes solve for the DM number density non-self-consistently}. This is because they miss the (small) non-ideal corrections coming from the scattering contributions in $K(T)$ and evaluate the chemical potential in the ideal gas approximation.

For a symmetric plasma $n_{\eta,0} = n_{\xi,0}$, Eq.~(\ref{eq:totalnumber}) is a quadratic equation in $n_{\eta,0}$ and the solution is given by:
\begin{align}
\frac{n_{\eta,0}}{n_{\eta}} = \alpha(n_{\eta} K(T)),\quad \alpha(x) =  \frac{1}{2x} \left( \sqrt{1+4x} -1 \right).\label{eq:saha}
\end{align}
where $\alpha$ gives the ratio between the ideal number density of free particles (no bound states and no interactions) and the total number density $ n_{\eta}$, including bound states, non-ideal and thermal corrections. This is the main difference compared to the ideal definition in the previous Section. The generalized Saha Eq.~(\ref{eq:saha}) accounts for non-ideal contributions like self-interactions as well as for finite temperature corrections. Important to note is that we do not have to define what is a bound or a scattering contribution to the total number density. The spectral function in Eq.~(\ref{eq:pressuredensity}) does the job automatically and takes all contributions into account when integrating over the whole energy range.

Since $n_{\eta,0}=n_{\eta,\text{0}}^{\text{eq}} e^{\beta \mu_{\eta}}$, we can finally determine the chemical potential $\mu_{\eta}$ as a function of total number density from Eq.~(\ref{eq:saha}).
\begin{claim}
\emph{Assuming a grand canonical state for a system having bound-state solutions in the spectrum, automatically implies the Saha ionization equilibrium.} 
Under this assumption, the chemical potential is set by
\begin{align}
\beta \mu_{\eta} = \ln\bigg[ \frac{ \alpha n_{\eta}}{n_{\eta,\text{0}}^{\text{eq}}} \bigg],
\end{align}
and our Master formula for the total number density Eq.~(\ref{eq:numbergrandcanonicalgeneral}) can be written in a fully closed form as:
\begin{align}
\dot{n}_{\eta} + 3 H  n_{\eta} 
&= - \frac{2 (\sigma v_{\text{rel}})G_{\eta \xi,s}^{++--} (x,x,x,x)\big|_{\text{eq}}}{n_{\eta,\text{0}}^{\text{eq}}(T) n_{\eta,\text{0}}^{\text{eq}}(T)} \bigg[\alpha^2 (n_\eta K(T)) n_{\eta}n_{\eta}- n_{\eta,\text{0}}^{\text{eq}} (T) n_{\eta,\text{0}}^{\text{eq}} (T) \bigg],\label{eq:numberdensitysaha}
\end{align}
where
\begin{align}
G^{++--}_{\eta \xi,s} (x,x,x,x)\big|_{\text{eq}} =  e^{-  \beta 2 M } \int_{-\infty}^{\infty} \frac{\text{d}^3\bm{P}}{(2 \pi)^3} e^{- \beta \bm{P}^2 / 4 M}
	\int_{-\infty}^{\infty} \frac{\text{d}E}{(2\pi)} e^{- \beta E} G^\rho_{\eta \xi} (\mathbf{0},\mathbf{0}; E)|_{l=0}.
\end{align}
\end{claim}
\noindent
Importantly, the obtained number density Eq.~(\ref{eq:numberdensitysaha}) is in general \textit{not quadratic in $n_\eta$} contrary to the conventional Lee-Weinberg equation because the generalized ionization fraction, $\alpha (n_\eta K(T))$, exhibits a non-trivial dependence on $n_\eta$.
As we have discussed in the previous section, this equation contains all the number violating processes of DM, i.e.~annihilations and bound-state decay.
The process which dominates the decrease of DM number density is determined by how $\alpha$ evolves in time as we have discussed at the end of the previous section.

The insights we gained in this Section are important for the understanding of our work. Let us put below these results more into context.
\begin{itemize}
\item Eq.~(\ref{eq:numberdensitysaha}) describes the out-of-chemical equilibrium (finite $\mu$) evolution of the total number density under the constraint of ionization equilibrium and is one of our main results. It is a generalization of the idealized vacuum Eq.~(\ref{eq:sahaidealzerot}), accounting for \emph{i}) finite temperature corrections to the annihilation/decay rates and \emph{ii}) non-ideal corrections to the chemical potential. The non-ideal corrections to the chemical potential consist of finite temperature corrections as well as scattering contributions. The main advantage of the Eq.~(\ref{eq:numberdensitysaha}) is that we do not have to define what is a bound or scattering state contribution since at finite temperature this is meaningless to do. All expressions needed in order to solve our generalized number density equation numerically can be obtained by evaluating the two-particle spectral function. Since one has to integrate over the whole energy spectrum, the result manifestly takes all contributions into account, without the need of differentiating between bound and scattering states. Later in Section \ref{sec:finitetemp}, we show the results for the two-particle spectral function $G^\rho_{\eta \xi} (\mathbf{0},\mathbf{0}; E)|_{l=0}$ in Eq.~(\ref{eq:numberdensitysaha}) including finite temperature corrections.

\item Let us discuss a bit more in detail the generalized Saha Eq.~(\ref{eq:saha}).
For vanishing finite temperature corrections and the ideal gas limit, we have emphasized that it reduces to the standard expression Eq.~(\ref{eq:iodegreeideal}). Since \eqref{eq:saha} looks the same as the one in the ideal gas limit aside from how $K(T)$ depends on $T$, one may understand how $\alpha$ evolves from the discussion at the end of the previous Section. However, at finite temperature its is more complicated to precisely estimate since the number of bound states is dynamical. Furthermore, it is also more complicated to discuss the case when ionization equilibrium is broken at finite temperature due to this reason.

Finite temperature effects might extend the period where the ionization rate is much larger compared to the decay rate, since $\psi$ particles can efficiently destroy the bound state. Thus it might be true that the validity of our equations holds longer compared to the vacuum case.
Another difficulty at finite temperature is, only in the limit of narrow thermal width it might be possible to estimate the validity of ~(\ref{eq:numberdensitysaha}). This is because in order to estimate the rates one has to define what is the decay width of the bound state, which becomes hard to define beyond the narrow thermal width limit. 
At late times of DM freeze-out, however, we naively expect that the thermal width becomes less important and one may estimate the rates by just including the real-part corrections (but still in this case the number of highly excited bound states can be dynamical).

\item According to previous discussions, we would like to emphasize again that our number density Eq.~(\ref{eq:numberdensitysaha}) is not applicable to the regime where bound-state decay rates exceed the ionization rates at late times causing an out-of-ionization equilibrium state (now disregarding the issue of how we can precisely estimate those rates at finite temperature). However, our more general equations Eqs.~(\ref{eq:twotimeother})-(\ref{eq:twotimeret}) do not assume ionization equilibrium and can be applied to any out-of-equilibrium state. The point is, since we have dropped for simplicity soft emissions from the beginning, there are no processes like BSF via the emission of a mediator relating the number of bound and free particles. Consequently, if one would solve the general Eqs.~(\ref{eq:twotimeother})-(\ref{eq:twotimeret}) numerically, with an initial out-of-ionization equilibrium state, the system would remain for all times in out-of-ionization equilibrium. It is required for future work to include soft emissions via e.g.\ an electric dipole operator in thermal plasma, to account for a correct description of the DM thermal history at late times. We will see, however, if ionization equilibrium can be guaranteed, our description accounts for sizable finite temperature corrections during the early phase of the freeze-out which can not be captured by the classical on-shell Boltzmann equation treatment as in \cite{vonHarling:2014kha}. Thus these different approaches are in some sense complementary. Furthermore, the approach in \cite{vonHarling:2014kha} uses only the main contribution of the ground state 1S, while via Eq.~(\ref{eq:numberdensitysaha}) and (\ref{eq:solutionnumerical}) it is very elegant and efficient to include \emph{all} (here only s-wave) bound state contributions (but under the assumption of ionization equilibrium).

\item From this Section we learned that the standard Boltzmann equations at zero temperature are a non-self-consistent set of equations. They miss scattering contributions in $K(T)$ which can now be fully accounted for. Bearing in mind that these contributions might be small, it might also be acceptable to adopt a non-self consistent solution of Eq.~(\ref{eq:numberdensitysaha}). By this we mean one can in principle compute the chemical potential in the narrow width approximation, however in the computation of $G^{++--}_{\eta \xi,s} (x,x,x,x)\big|_{\text{eq}} $, the two-particle spectral function can be solved in its most general form including finite temperature width (as we present in Section \ref{sec:finitetemp}). We emphasize again that a non-self-consistent solution might cause some troubles and care should be taken. In Appendix \ref{app:salpeter}, we discuss other non-self consistent computations of the chemical potential and point out their failure, especially for late times if bound-state solutions exist.

\end{itemize}

\subsection{Comparison to linear-response-theory method}
\label{sec:com2linear}
Our dark matter system is similar to heavy-quark pair annihilation in a thermal quark gluon plasma produced in heavy ion collisions. In literature, the annihilation rate of the heavy-quark pair into dileptons is estimated from linear response theory. Let us just quote their results in the following without deeply diving into the details. For comparison we translate the expressions for $SU(3)$ to our U$(1)$ by adjusting color and flavour factors and call the heavy (anti) quark dark matter ($\xi$) $\chi$ . The linearised Boltzmann equation around chemical equilibrium is given by
\begin{align}
\dot{n}_{\eta}+ 3H n_{\eta} = -\Gamma_\text{chem}(n_{\eta}-n_{\eta,\text{eq}})+ \mathcal{O}(n_{\eta}-n_{\eta,\text{eq}})^2,
\label{eq:linear_boltz}
\end{align}
where $\Gamma_\text{chem}$ is called the chemical equilibration rate. $\Gamma_\text{chem}$ can be extracted from linear response theory, assuming thermal equilibrium and a perturbation linear around chemical equilibrium. It is defined as \cite{Burnier:2007qm, Bodeker:2012gs, Bodeker:2012zm, Kim:2016zyy}:
\begin{align}
\Gamma_\text{chem} \equiv \frac{\Omega_{\text{chem.}}}{ (4\chi_\eta/\beta) M^2},
\end{align}
where $\Omega_{\text{chem.}}$ is a transport coefficient and $\chi_\eta$ is the heavy DM number susceptibility. The transport coefficient is quoted as
\begin{align}
\Omega_{\text{chem.}}/M^2 \simeq 16 (\sigma v_{\text{rel}}) \int_{\omega, \bm{P}} e^{-\beta(2M+\omega)} \rho_{\eta \xi}(\omega, \bm{P}),\label{eq:omegachem}
\end{align}
where we consider always only s-wave contributions and the tree level annihilation cross section is defined as in our case, while $\rho_{\eta \xi}$ is also called a spectral function but might be defined slightly differently than ours. This expression looks similar to our term in chemical equilibrium
\begin{align}
2 (\sigma v_{\text{rel}})G_{\eta \xi,s}^{++--} (x,x,x,x)|_{\text{eq}} = 2 (\sigma v_{\text{rel}})  \int_{\omega, \bm{P}} f_B (2 M +\omega ) G^\rho_{\eta \xi} (\bm{0}, \bm{0}; \omega, \bm{P}).\label{eq:compare}
\end{align}
Indeed, we find both spectral functions in Eqs.~(\ref{eq:omegachem}) and (\ref{eq:compare}) give identical results if we take the DM dilute limit of our equation. 

The number susceptibility is defined as the response of the total number density with respect to infinitesimal variation of the chemical potential:
\begin{align}
\chi_\eta \equiv  \left.  \frac{\partial n_{\eta} }{\partial \mu_{\eta} }\right|_{\mu_{\eta}=0}.
\end{align}
In the dilute limit, using Eq.~(\ref{eq:totalnumber}), we have $\chi_\eta = \beta n_{\eta}^{\text{eq}}$. So we find in total for the linearised equation using linear response theory:
\begin{align}
\dot{n}_{\eta}+ 3H n_{\eta} = - \frac{4 (\sigma v_{\text{rel}}) \int_{\omega, \bm{P}} e^{-\beta(2M+\omega)} \rho_{\eta \xi}(\omega, \bm{P})}{ n_{\eta}^{\text{eq}}n_{\eta}^{\text{eq}}} \left[ n_{\eta} n_{\eta}^{\text{eq}} -  n_{\eta}^{\text{eq}} n_{\eta}^{\text{eq}} \right].\label{eq:mikkonumber}
\end{align}

In the following, we compare this equation based on linear response theory with our Eq.~\eqref{eq:numberdensitysaha}.
We start from our more general expression and reproduce Eq.~\eqref{eq:mikkonumber} in the linear regime around chemical equilibrium while clarifying the underlying approximations. For this purpose, we have to linearise our equation around $n_\eta \sim  n_\eta^\text{eq} $, where 'eq' labels chemical equilibrium. 
On the one hand, as we have shown in Sec.~\ref{sec:solutions}, the spectral function, $G^\rho_{\eta\xi}$, does not depend on $n_\eta$ as long as the DM number densities are dilute.
On the other hand, the generalized ionization fraction, $\alpha$, does depend on $n_\eta$ and hence it must also be approximated as to be close to chemical equilibrium. One may easily see this by looking at the definition of $\alpha$ given in Eq.~\eqref{eq:saha}. From this expression it is clear that if the DM number densities are close to chemical equilibrium, the ionization fraction is always close to one, i.e.~$\alpha_\text{eq} \simeq 1$, because $n_{\eta,0}^\text{eq} \propto e^{- \beta M}$ and $K(T) \propto e^{\beta |E_B|}$. Hence, in the linear regime, the total DM number densities are fully ionized and can be approximated as the free scattering contributions: $n_\eta^\text{eq} \simeq n_{\eta,0}^\text{eq}$.
From these observations, we reproduce  Eq.~\eqref{eq:mikkonumber} from our Eq.~\eqref{eq:numberdensitysaha} \emph{in the linear regime around chemical equilibrium}.

The linear response theory method applies, by definition, only to the linear regime around chemical equilibrium where $n_\eta \sim  n_\eta^\text{eq} $. It is not possible via this method to see what is the correct form of the underlying Lee-Weinberg equation describing the non-linear regime where $n_\eta \gg n_\eta^\text{eq}$. Nevertheless, one might be tempted to use it by replacing the right-hand-side of Eq.~\eqref{eq:linear_boltz} with $\frac{\Gamma_\text{chem}}{2 n_\eta^\text{eq}} (n_\eta^2 - (n_\eta^\text{eq})^2)$ as done in~\cite{Kim:2016zyy, Kim:2016kxt, Biondini:2017ufr, Biondini:2018pwp, Biondini:2018xor}. Most important to note is that the collision term obtained from this replacement matches our equation only if the generalized ionization fraction is close to \emph{chemical equilibrium}, i.e.~$\alpha_{\text{eq}} \simeq 1$, which is no longer true at late times.
We can see how $\alpha$ depends on time from Eq.~\eqref{eq:saha}. 
As we have already discussed in detail at the end of Secs.~\ref{sec:ionizationeq_ideal} and \ref{sec:ionizatineq}, at late times of DM freeze-out the total comoving number density approaches a constant value while $K(T)$ starts to grow rapidly for $T < |E_B|$.
As a result, we find the generalized ionization fraction to be $\alpha \simeq 1 / \sqrt{n_\eta K(T)} \ll 1$ at late times, which invalidates the naive replacement above.
One can understand this behavior intuitively because the bound states of energy $2 M - |E_B|$ are thermodynamically favored compared to the scattering states of energy $2M$ at $T < |E_B|$. For a fixed total number of DM, the bound states dominate over the scattering states at some point as in the case of the recombination. There is also another issue when correcting $n_\eta^\text{eq}$ in $\frac{\Gamma_\text{chem}}{2 n_\eta^\text{eq}} (n_\eta^2 - (n_\eta^\text{eq})^2)$ only by the Salpeter term. Since this discussion requires some detailed knowledge about thermal corrections, we share it in Appendix \ref{app:salpeter}. In summary, we would like to emphasize that care must be taken if one matches the equation obtained by linear response theory to non-linear differential equations. 

Instead, one may  
match the equation obtained by linear response theory to our corrected form of the Lee-Weinberg equation in the non-linear regime. By this procedure it is now also clear what the limitation exactly is. As we have discussed in detail in the previous section, our master formula for the number density equation is valid as long as ionization equilibrium can be maintained. Ionization equilibrium is broken if the decay rate exceeds the ionization rate. The temperature where this happens can be estimated for finite temperature systems, at least in the narrow width case.

\section{Numerical results for two-particle spectral function at finite temperature}
\label{sec:finitetemp}
We turn to the numerical solution of the two-particle spectral function for the full in-medium potential. The effects of the finite temperature corrections can be simplest understood for the case of the Coulomb potential as given in Eq.~(\ref{eq:inmediumpot}). The first correction is a real constant term that shifts effectively only the energy by $\alpha_{\chi} m_D$. When only taking this correction into account one would thus expect that the infinite number of bound states in the spectral function of the Coulomb case just move to lower binding energies and similar shift to the threshold as well as to the positive energy spectrum. The second real-part correction is an exponential screening of the Coulomb potential with radius $m_D$. This introduces another effect. It leads to a disappearance of the bound states closest to the threshold since Yukawa potentials have only a finite number of bound states. The disappearance of bound states wins against the move of the poles towards lower energies at the Mott transition, where all bound states disappear and the spectrum is exclusively continuous. Additionally, we have imaginary-part corrections coming from the soft DM-$\psi$ scatterings, leading to a finite thermal width of the bound states. Once the thermal width is comparable to the binding energy, the bound-state poles are strongly broadened.

\begin{figure}[h]
\includegraphics[scale=0.42]{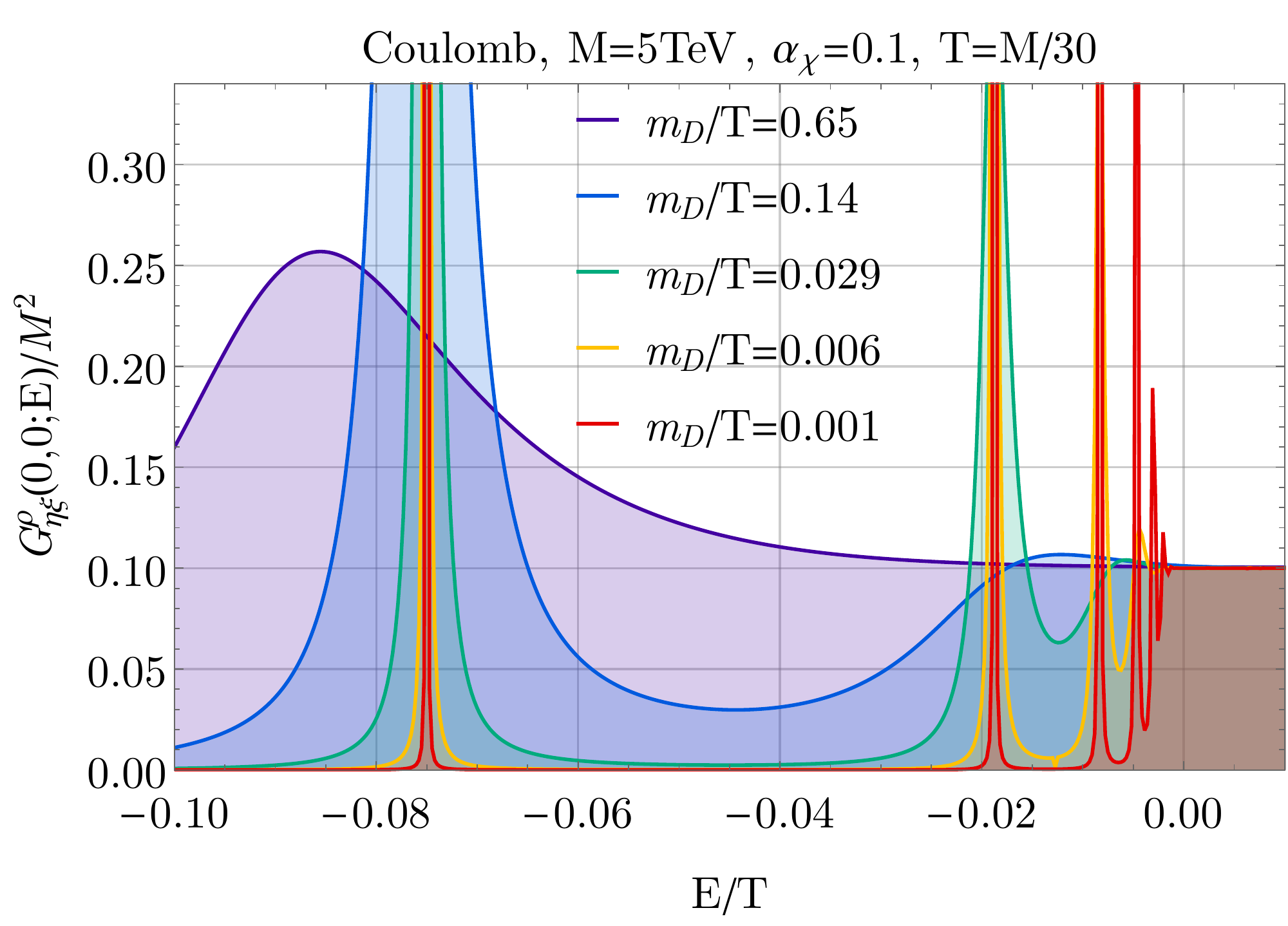}
\includegraphics[scale=0.42]{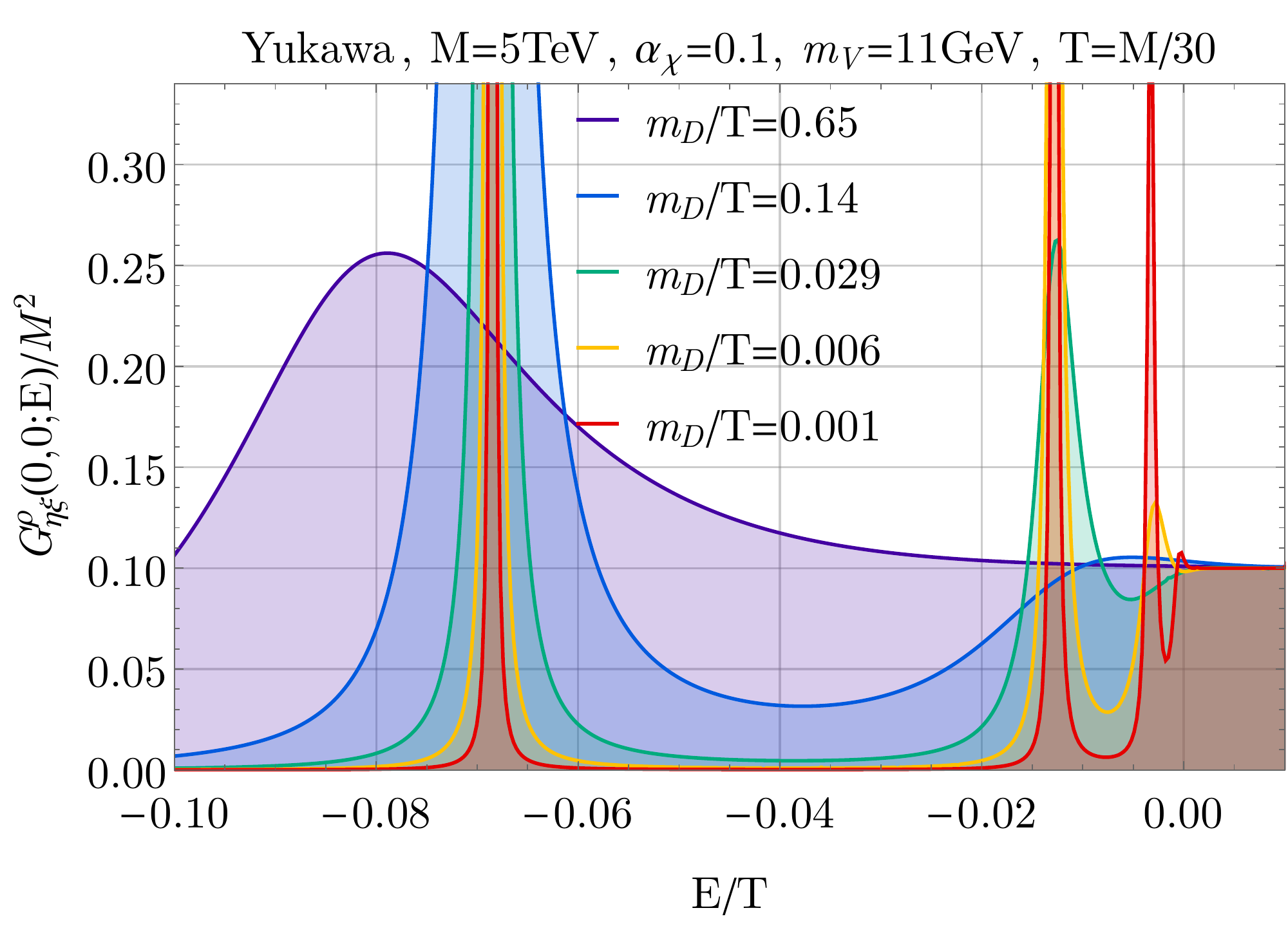}
\caption{Two-particle spectral function at finite temperature shown vs. the energy in units of typical freeze-out temperature. The violet line corresponds to the equal charge case $g_\psi=g_\chi$ of our minimal model and hence $m_D^2 = g_{\chi}^2T^2/3$.}
\label{fig:spectral}
\end{figure}

The combination of all effects are shown in Fig.~\ref{fig:spectral}, where we present the numerical solution of the two-particle spectral function for the full in-medium potential according to Eq.~(\ref{eq:solutionnumerical}).
We show the case of the Coulomb and Yukawa potential. In this figure, we have fixed the temperature to $T= M/30$ (slightly below the typical DM freeze-out temperature) and varied the Debye mass where the maximal value shown corresponds to the equal charge case $g_\psi=g_\chi$ of our minimal model: $m_D^2 = g_{\chi}^2T^2/3$.
The mass of the DM is fixed to 5 TeV and the coupling $\alpha_{\chi}=0.1$ very roughly chosen to account for the correct order of the abundance.

All finite temperature effects together lead to a continuous \emph{melting} of the bound-state poles. As can be seen, the melting of the bound states leads to the fact that even at negative energies, the spectrum is continuous at finite temperature.
The reshuffling of the spectrum towards lower energies affects the rates \emph{exponentially} according to Eq.~(\ref{eq:coll_th}). This is because the integrand (the spectral function) has more support at negative energies which is, due to Boltzmann factor, exponentially preferred in kinetic equilibrium. The whole integral stays convergent since at very low energies the spectral function starts to decrease faster than the Boltzmann factor increases. 
It now becomes clear that the notion of spectral function is more general compared to the vacuum case where one could separate the spectrum for bound and scattering states. Here, it is evident that such a distinction is impossible. It is also not necessary to do so since the integration of spectral function times Boltzmann factor takes already all contributions into account.

\begin{figure}[h]
\includegraphics[scale=0.42]{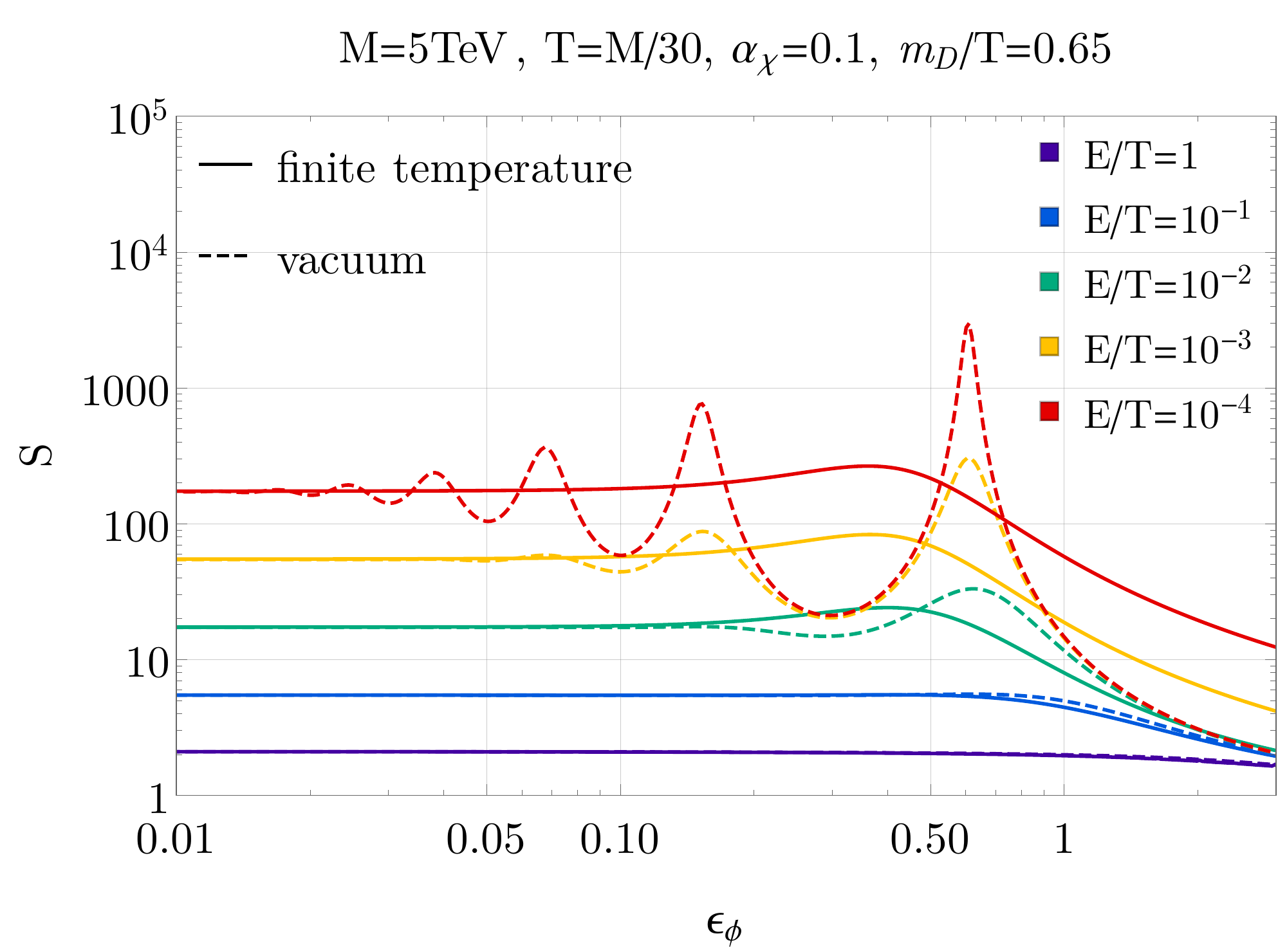}
\includegraphics[scale=0.42]{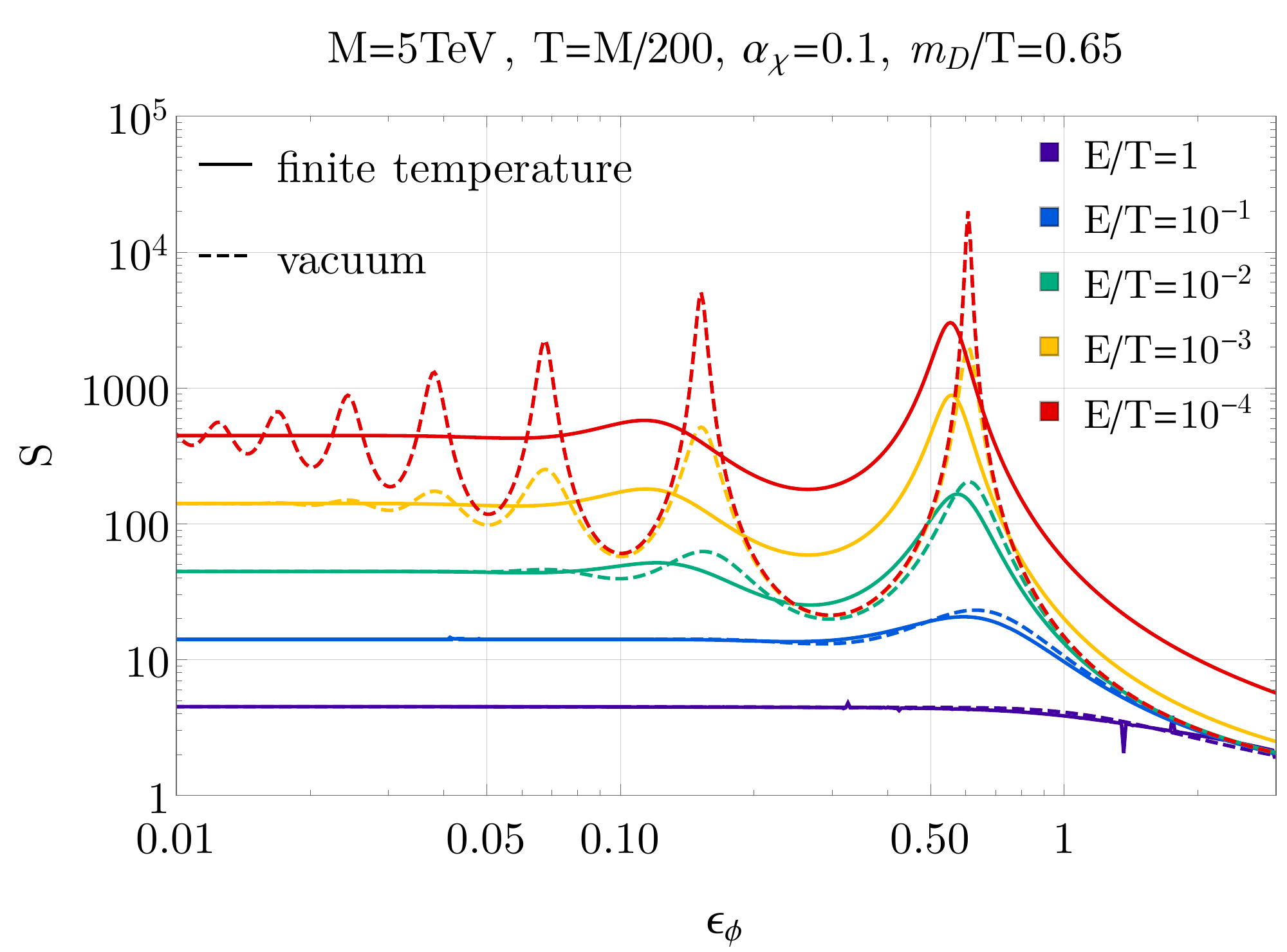}
\caption{Sommerfeld enhancement factor for a Yukawa potential shown vs. the mediator mass in units of the Bohr radius $\epsilon_\phi \equiv m_V/(\alpha_{\chi} M)$. Dashed and solid lines correspond to the vacuum limit and the full in-medium potential, respectively. 
Effects at different temperatures are compared. Left side for typical freeze-out temperature and right plot at a typical temperature where the annihilation rate would be much smaller compared to the Hubble rate.}
\label{fig:scattering}
\end{figure}

Nevertheless, let us discuss how only the positive energy spectrum is affected.
According to the theorem of Levinson, the scattering phases (and hence the wave function at the origin) depend on the amount or properties of the bound states. This means that thermal modifications of the bound states automatically affect also the positive energy spectrum. The impact on the positive energy spectrum depends on the melting status of the bound states. In general, there can be both, a suppression or further enhancement of the positive energy spectrum as can be seen by carefully looking at the value around $E=0$ in Fig. \ref{fig:spectral}. We would like to stress that a suppression of the positive spectrum does not imply that the total rate is less. For the computation of the rate one has to integrate the spectral function over the whole energy range, where $G^{++--}_{\eta \xi,s} (x,x,x,x)\big|_{\text{eq}} $ becomes due to the reshuffling towards lower energies exponentially enhanced. 

While for the Coulomb case, the impact of the melting on the positive energy spectrum is only very little (which does not mean that the overall effect is small), the impact for the Yukawa potential case can be much larger. In Fig.~\ref{fig:scattering}, we compare the positive energy solution of the Yukawa spectrum at zero and finite temperature, as a function of the mediator mass $m_V$.
The vacuum line (dashed) is obtained by solving the spectral function in the limit of vanishing finite temperature corrections. The Sommerfeld factor for this case can be obtained from Eq.~(\ref{eq:possol}) according to:
\begin{align}
S(v_{\text{rel}}) &= \frac{2\pi}{M^2 v_\text{rel}} G_{\eta \xi}^{\rho}(\mathbf{0},\mathbf{0};E)|_{E>0,l=0} \\
\Leftrightarrow S(E) &= \frac{\pi}{M \sqrt{ME}} G_{\eta \xi}^{\rho}(\mathbf{0},\mathbf{0};E)|_{E>0,l=0}. \label{eq:def_SEfinite}
\end{align}
In the second line we used $E=M v_{\text{rel}}^2/4$ for on-shell particles. At finite temperature this kinetic energy relation does not hold. We therefore use the second Eq.~(\ref{eq:def_SEfinite}) to define the Sommerfeld enhancement factor at finite temperature as shown in Fig.~\ref{fig:scattering}. The enhancement or suppression of the Sommerfeld enhancement factor due to the thermal effects is largest if the ground state is close to the threshold of $E=0$ (around the first peak from the right, here $\epsilon_{\phi} \sim 0.6$). For our minimal model, it is also shown in the right plot of Fig.~\ref{fig:scattering} that the whole temperature range of a typical freeze-out process can be affected. In the limit $m_V\rightarrow0$ the Coulomb limit is recovered. Again, this does not mean that it is sufficient to just take the standard expression of the Sommerfeld enhancement factor of the Coulomb potential to describe the DM freeze-out. There is also a contribution from the negative energy spectrum.
\begin{figure}[h]
\includegraphics[scale=0.42]{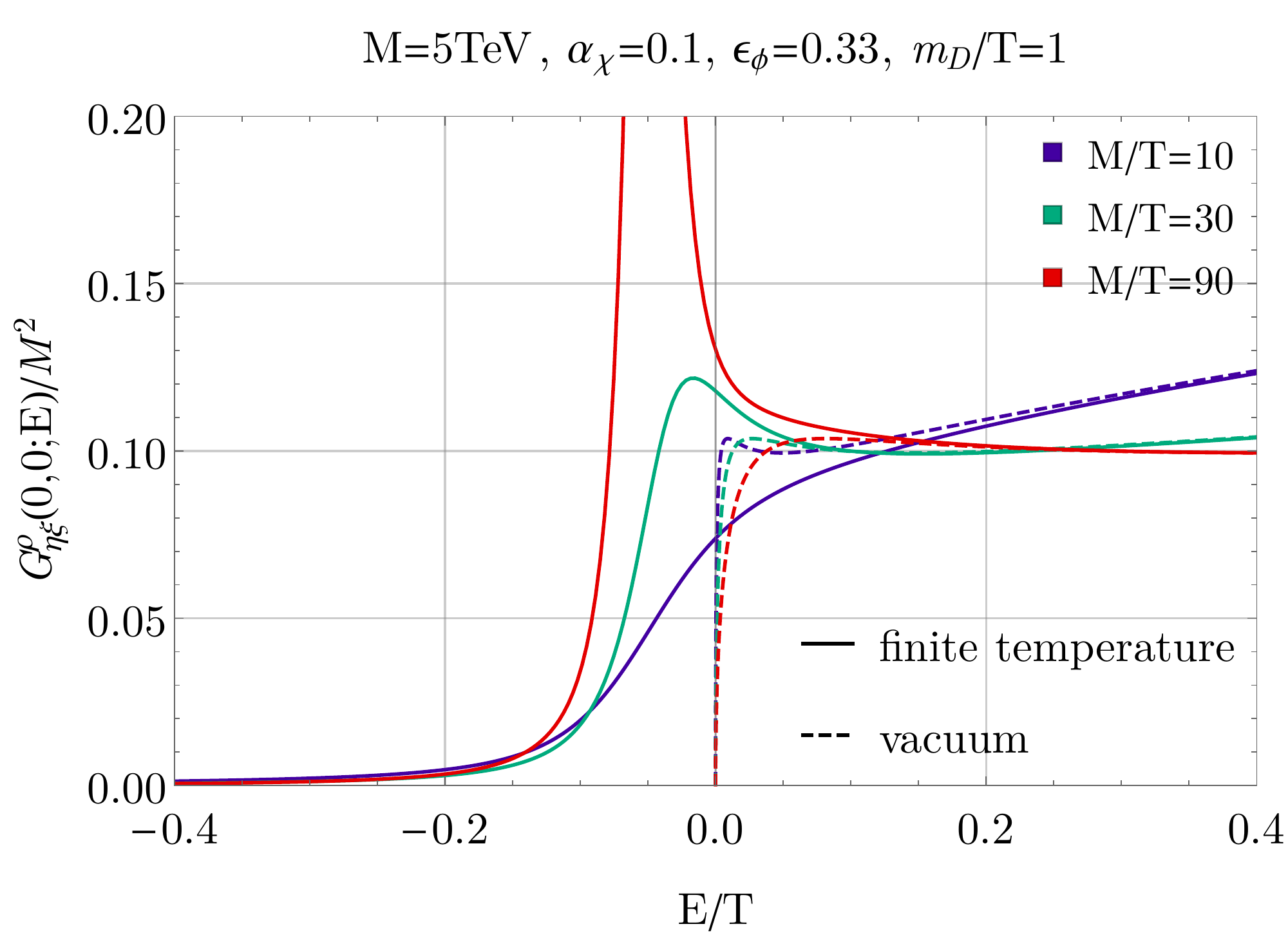}
\includegraphics[scale=0.42]{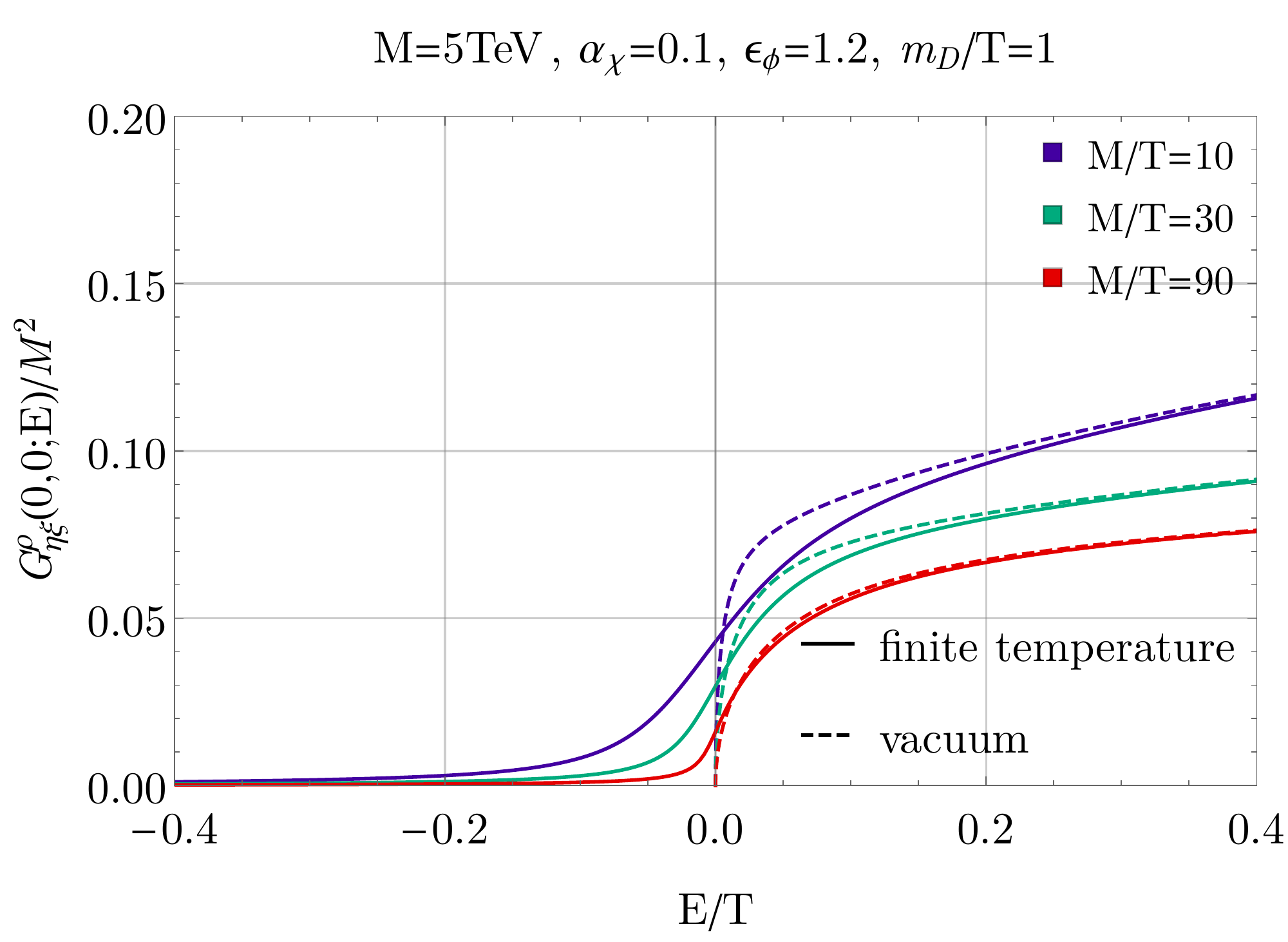}
\caption{Comparison between extreme examples, where only the ground state exists and is close to the threshold (left) and where no bound states exist (right). Here, the vacuum curve is defined as not including delta-peaked bound state contributions.}
\label{fig:poles}
\end{figure}
Therefore, one has to be careful in interpreting Fig.~\ref{fig:scattering}. On the one hand side the positive energy solution (Sommerfeld enhancement factor) can be suppressed or equal compared to the vacuum case, but on the other side the total $G^{++--}_{\eta \xi,s} (x,x,x,x)\big|_{\text{eq}} $ entering our master formula Eq.~(\ref{eq:numbergrandcanonicalgeneral}), which requires the integration over the whole energy spectrum, can be enhanced.

As an extreme example of this situation, let us discuss the case where there are no bound states (e.g.\ Yukawa potential in the Born regime $\epsilon_{\phi} \gg1$). The finite temperature spectral function for this case is shown in the right plot of Fig.~\ref{fig:poles}. Indeed, the positive energy spectrum is dominantly suppressed compared to the vacuum case but when integrating the spectral function over the Boltzmann factor in the whole energy range there is still an enhancement of order 1\% compared to the vacuum case. Another extreme example, where the corrections to the positive spectrum are strongest, is the case where only the ground state exists and is close to threshold \cite{Kim:2016kxt, Biondini:2017ufr, Biondini:2018pwp}. This example is shown in the left plot of Fig.~\ref{fig:poles}. We find in this case, the value of the integration of the spectral function times the Boltzmann factor is by up to 10\% (30\%) larger compared to the vacuum case \emph{without} bound-state peak at the typical freeze-out temperature $T=M/30$ $(T=M/90)$. The correction increases for lower temperature due to the Boltzmann factor. 

The total rate is only proportional to the integration of the spectral function times the Boltzmann factor. There are additional finite temperature corrections to the chemical potential (see previous Section) which can be also obtained from the spectral function. We have not explicitly computed those non-ideal corrections yet but leave it for future work once we have included ultra-soft emissions in our system description.

\section{Discussion}
\label{sec:discussion}

A self-interacting DM system, where long-range forces and bound-state solutions exist, is in general a complex ensemble where many processes with different rates are taking place at the same time during the DM thermal history. Essentially, there are three quite different approaches in the literature with distinct motivation to describe the evolution of the abundance of the stable components for such systems:
\begin{enumerate}
\item The first approach is based on a coupled set of classical on-shell Boltzmann equations.
If bound-state solutions are absent, the description of the DM freeze-out acquires dominantly corrections from the Sommerfeld-enhanced annihilation of free DM particles \cite{Hisano:2003ec, Hisano:2006nn}. If the two-particle spectrum has support at negative energies, the free DM particles can form a bound state via radiative processes \cite{vonHarling:2014kha, Oncala:2018bvl}. The reverse process can also happen, called ionization. If there are several bound-state solutions present, further processes like excitation or de-excitation can happen \cite{Petraki:2015hla, Petraki:2016cnz}. All those processes are in general coupled, and as we see, the list of Boltzmann equations needed to describe such systems can be long. When relying on those classical Boltzmann equation computations, treating  e.g.\ the number density of free particles and bound states separately and as idealized, potential strong modifications arising from higher-order plasma interactions might be missed. In this approach, however, it is always guaranteed that the non-linearity of out-of-chemical equilibrium reactions are accurately described. And there can be in general many such out-of-equilibrium reactions as listed above.

\item The second approach starts from the EoM of correlation functions on the Keldysh contour and takes into account some finite temperature corrections. The major difference to our work is that in \cite{Beneke:2014gla} it is assumed that the correlator hierarchy can by truncated at the lowest order, resulting in closed equations for the two-point functions in terms of the one-particle self-energy only. One of the equations are the so called kinetic equations, being the differential equations for describing the evolution of observables in terms of the macroscopic Wigner coordinates. In the one-particle self-energy approximation they are also known as Kadanoff-Baym equations. Expanding the self-energy in terms of the coupling to NLO results in the standard Boltzmann equation. At NNLO first finite temperature corrections enter. The advantage of a fixed order calculation is that infrared divergences, arising at NNLO cancel \cite{Beneke:2014gla}. At NNLO in the self-energy expansion of the kinetic equations, the thermal corrections turn out to be strongly suppressed, i.e.\ to high power in $T/M$, compared to the NLO result. One should, however, keep in mind that there are next to the kinetic equations also the equations for the microscopic Wigner coordinates, called mass-shell equations accounting for, e.g.,\ thermal corrections to the dispersion relation. 
Kinetic and mass-shell equations are in general coupled. Therefore, a self-consistent solution in principle requires to take 
account of corrections also from the mass-shell equation. In any case, the problem within this systematic approach is that a fixed order calculation can never account for correctly describing the Sommerfeld enhancement beyond the Born regime and also bound-state solutions will never appear.

\item The third approach addresses the description of long-range force systems at finite temperature in a non-perturbative sense, i.e.\ by resummation of the Coulomb divergent ladder diagrams including thermal corrections. Clearly, first attempts were made in the literature of heavy quark pair annihilation in a quark gluon plasma \cite{Burnier:2007qm}, produced in heavy-ion collisions at the LHC. More recently, 
some of these authors have applied the same techniques also to the DM freeze-out \cite{Kim:2016kxt}. The method is based on \emph{linear response theory} \cite{Bodeker:2012gs, Bodeker:2012zm, Kim:2016zyy}, estimating the DM Sommerfeld-enhanced annihilation and bound-state decay from a spectral function including finite temperature corrections. It has been shown in Ref.~\cite{Biondini:2017ufr, Biondini:2018pwp, Biondini:2018xor} that the DM overclosure bound, computed by this method, can be strongly affected by finite temperature effects if bound-state solutions exist. Compared to a fixed order calculation as in approach 2, the finite temperature corrections are larger. The reason is because the mass-shell equations are solved by resummation of the Hard thermal loop contribution. Albeit there are potentially strong effects, the linear response theory is strictly speaking valid only for systems close to thermal equilibrium, e.g.\ $n \sim n_\text{eq}$. At finite temperature the spectral function can in general depend on the DM density. Therefore, it is a priory not clear if the transport coefficients extracted from linear response theory can be inserted into a non-linear Boltzmann equation describing the DM freeze-out in a non-linear regime where $n \gg n^\text{eq}$. From vacuum computations it is known that the Sommerfeld effect can still be efficient in such a regime. This is because the transition $n \sim n_\text{eq}$ to $n \gg n^\text{eq}$ happens in a short time, since $n_\text{eq} \propto e^{-M/T}$ decreases rapidly.
To the best of our knowledge this method, inserting transport coefficients obtained from linear response theory into a non-linear Boltzmann equation,
has not been tested so far by using other treatments applying for generic out-of equilibrium situations.
\end{enumerate} 

Our formalism, presented in this work, aims towards a first step in unifying the approach 1 and 3, by generalizing the approach 2 for long-range force systems. In other words, we derived from the EoM of Keldysh correlation functions the number density equation for DM including finite temperature corrections and accounting for the full resummation of Coulomb divergent ladder diagrams. This allows to study the finite temperature corrected Sommerfeld-enhanced annihilation as well as bound-state decay. Moreover, our Master Eq.~(\ref{eq:numbergrandcanonicalgeneral}) is able to describe the correct non-linear transition to out-of-chemical equilibrium, i.e.\ the freeze-out process.
\bigskip

Although we have derived all equations on the Keldysh contour, and therefore they should be valid for any out-of-equilibrium situation of the system, the reader should be reminded what precisely our system is. While it remains true that we can describe correctly Sommerfeld-enhanced annihilation and bound-state decay in the presence of a relativistic plasma background for out-of equilibrium situations, we have dropped from the beginning, when deriving our nonrelativistic effective action, ultra-soft contributions of the fully relativistic action. Hence, Eqs.~(\ref{eq:twotimeother})-(\ref{eq:twotimeret}) are missing ultra-soft contributions leading to bound state formation and ionization processes via the emission or absorption of a mediator, as well as contributions to excitation or de-excitation processes if multiple bound states exist. Once ultra-soft terms are included in the system of equations, we expect the final equations, if finite temperature effects are neglected, to coincide with the full set of equations of approach 1. Moreover, the inclusion of emission and absorption in the Keldysh formalism might lead to new insights in the production rate of dileptons or photons, produced from heavy-quark pair-annihilation in a quark gluon plasma.

In the second half of the work, we have indirectly included all bound-state formation, ionization, excitation and de-excitation processes. This was achieved by assuming our system is in a grand canonical state with one single time dependent chemical potential as in our Master Eq.~(\ref{eq:numbergrandcanonicalgeneral}). The important observation that adopting a grand canonical picture automatically implies ionization equilibrium if bound states are present was by far not obvious to us. This key observation brought us to the conclusion that our equations in the limit of vanishing thermal corrections are equivalent to the coupled system of classical Boltzmann equations in the limit of ionization equilibrium. 
Thus we have shown that under certain assumptions our approach and approach 1 consistently fall together. Another important point based on this observation was that, since the ionization fraction at chemical equilibrium is close to unity, our and approach 3 are equivalent to approach 1 in the regime linear near chemical equilibrium.

Important to recognize was that our approach and approach 3 give different results if the transport coefficients extracted from linear response theory is inserted into a non-linear Boltzmann equation just by replacing $\Gamma_\text{chem} (n_\eta - n_\eta^\text{eq})$ with $\frac{\Gamma_\text{chem}}{2 n_\eta^\text{eq}} (n_\eta^2 - (n_\eta^\text{eq})^2)$ \cite{Biondini:2017ufr, Biondini:2018pwp, Biondini:2018xor}. 
This is because the ionization fraction depends on $n_\eta$ where another non-linearity comes in, and in particular, the ionization fraction will be much smaller than unity at late times.
This is intuitively because the bound states are exponentially favored compared to the scattering states for $T < |E_{B}|$. Furthermore, the ionization fraction counteracts against the exponential grow of $\Gamma_\text{chem}$ or of our $G^{++--}_{\eta \xi,s} (x,x,x,x)\big|_{\text{eq}} $ for late limes if bound-state solutions exist. In a word, while the spectral function is identical between the linear response and ours in the DM dilute limit, the ionization fraction makes the difference. This effect is non-negligible when at late times the DM gets depleted by bound state formation effects.

Our Master Eq.~(\ref{eq:numbergrandcanonicalgeneral}) can not be used at very late times where ionization equilibrium is not maintained.
Therefore, one has to be careful in relying on our so far simplified treatment 
for all times during the DM thermal history. More generally, when using our equations, 
it has to be ensured that the rates driving the system to kinetic and ionization equilibrium are much faster than any other rates leading to a potential out-of-kinetic or -ionization equilibrium state.
In the case of no $\psi$-particles (no finite temperature corrections), it was shown in Refs.~\cite{vonHarling:2014kha, Petraki:2015hla, Petraki:2016cnz} that the decay of the bound state becomes faster than the ionization via emission and absorption processes by an electric dipole operator at some point, which breaks the ionization equilibrium at late times. 
Later, when the bound state formation becomes inefficient compared to the cosmic expansion, the dark matter number freezes out completely. Estimating the valid regime of our approach in the presence of finite temperature corrections is a more complicated task.
To draw a definitive conclusion in our case, one has to estimate these processes, including emission and absorption of ultra-soft gauge bosons, in the presence of the thermal plasma.
As we have discussed already in detail, this might only be realizable if the thermal width is negligible compared to the real-part corrections. Furthermore, one has to keep in mind that the number of existing bound-state solutions is temperature dependent when already only real-part corrections are taken into account.

\bigskip

After this warning, we now would like to discuss the case where a grand canonical description with one single chemical potential is justified. As we have in detail presented in this work, all finite temperature corrections are then totally encoded in the solution of the DM two-particle spectral function. In the presence of ultra-relativistic fermionic particles in the background, the hard thermal loop resummed corrections to the DM system can be classified into three contributions [see effective in-medium potential Eq.~(\ref{eq:inmediumpot})]. The first two contributions are real-part corrections to the DM effective in-medium potential. The first one leads to an energy shift (Salpeter correction) in the DM two-particle spectral function by an amount of $\alpha_{\chi} m_D$ towards lower energies compared to the vacuum case. Second, the Debye mass $m_D$ leads to a screening of the Coulomb potential, resulting in a temperature dependent Yukawa potential with screening radius $m_D$. The third contribution to the effective potential is purely imaginary and originates from soft DM scattering processes with the ultra-relativistic fermions in the hot and dense background.

Let us first discuss the two real-part corrections and their implication on the two-particle spectral function for the case of a Coulomb potential. In the limit of vanishing real-part corrections, it is well known that a Coulomb system has infinitely many bound-state solutions. Furthermore, the bound-state solutions and the scattering states can clearly be separated
sharply at the energy $E=0$ in the two-particle spectral function (see Fig.~\ref{fig:spectralvacuum}). If the real-part finite-temperature corrections are included, bound states close to the threshold $E=0$ disappear in the spectrum. This is simple to understand. First of all, by the real-part corrections the Coulomb potential transforms into a Yukawa potential. Yukawa potentials have only a finite number of bound states. Secondly, due to the energy shift caused by the other real-part correction, the threshold is lowered. The combination of these two effects causes the disappearance of highly excited bound states close to the threshold and the spectrum is continuous instead of discrete. The effect gets stronger with increasing Debye mass leading at the Mott transition to a total disappearance of all bound states.

Already when only real-part corrections to the in-medium potential are included,
the number of bound states as well as their binding energies are temperature dependent according to the discussion above. It implies that a sharp definition of bound and scattering states can not be made for all times during the DM thermal history. However, for our total number density equation, it is NOT required at all to distinguish between bound and scattering states. For example, the computation of the ionization fraction via the generalized Saha Eq.~(\ref{eq:saha}) can always be performed without specifying what is a bound or scattering contribution. Another example is the spectral function entering the total number density in the production term. Also here, the integration of the spectral function automatically takes into account all contributions from the spectrum. Only in the absolute vacuum limit, it is possible to separate contributions. At finite temperature everything is mixed into one single object. A separation would only cause problems like unphysical jumps in projected thermodynamical quantities when bound states abruptly disappear (if only real-part corrections are included).

The mixing between scattering and bound-state solutions becomes even stronger once the imaginary contributions to the effective in-medium potential are included. As we have seen in Section \ref{sec:finitetemp}, these corrections lead to a thermal width of the peaks of the bound states and a continuous melting of the poles for increasing Debye mass, as illustrated in Fig.~\ref{fig:spectral}. 
Note that instead of changing the independent coupling in the Debye mass, we could have also increased the number of generations of ultra-relativistic particles in the plasma. The Debye mass is proportional to the square root of number of generations, and thus, if most of the particle content of the Standard Model would run in the thermal loop, we can have a large Debye mass although the coupling is still
small. The broadening of the peak and the shift towards lower binding energies increases the annihilation or decay rate exponentially (again its not necessary to distinguish between these two) due to the integration of the product of two-particle spectral function and the Boltzmann factor, as in Eq.~(\ref{eq:coll_th}).
\bigskip

Although, we focused on a simple U$(1)$ like theory and s-wave contributions in the present work, most of the equations for higher gauge theories, scalar mediators, or higher partial waves will change in the expected way. Let us already mention some major changes. The mediator self-energy would acquire further contributions from self-interactions as well as different colour or flavour pre-factors. The definition of the effective in-medium potential in Eq.~(\ref{eq:effectivepot}) remains the same and is computed from the specific dressed mediator correlator. The r.h.s of the Schr\"odinger-like Eq.~(\ref{eq:BSeq_st_dil}) will be proportional to the number of colours. Our number density Eq.~(\ref{eq:numberp}) in the case of velocity dependent tree-level annihilation cross sections (like in p-wave case) will have a space derivative on the r.h.s.. As expected, our formalism breaks down for temperatures around the confining scale of confining theories.

\section{Summary and conclusion}
\label{sec:conclusion}
Traditional computations of DM Sommerfeld-enhanced annihilation and bound-state decay rates rely on the assumption that reactions of such processes are taking place under perfect vacuum conditions. In this work we developed a comprehensive derivation of a more general description, taking into account non-ideal contributions arising from simultaneous interactions with the hot and dense plasma environment in the early Universe. We have derived the evolution equation for the DM number density which is applicable to the case where scattering and bound states get strongly mixed due to the influence of the thermal plasma surrounding. Our master Eq.~(\ref{eq:numbergrandcanonicalgeneral}) for the total DM density simultaneously accounts for annihilation and bound-state decay and hence its collision term is in general not quadratic in the DM number density. We showed that finite temperature effects can lead to strong modifications of the shape of the two-particle spectrum, which in turn modifies the DM annihilation or decay rates.

The Keldysh formalism we adopted throughout this work applies for the description of the dynamics of generic out-of-equilibrium states. Within this mathematical framework, we derived in the first part of this work directly from our nonrelativistic effective action the exact equation of motion of the DM two-point correlation functions. We extracted for the first time from those EoM the differential equation for the DM number density [see Eq.~(\ref{eq:numberp}) and (\ref{eq:numberantip})], which turns out to only depend on a special component of the DM four-point function on the Keldysh contour, namely $G^{++--}_{\eta \xi}$. Let us emphasize again that this equation for the number density is exact within our nonrelativistic effective action, however not closed since it depends on the solution of this four-point correlation function. The long-range force enhanced annihilations, the decay of bounded particles as well as the finite temperature corrections are all contained in the solution of this one single four-point correlator. 

In the second part of this work, we derived the EoM for the DM four-point function on the Keldysh contour. We developed the approximations needed in order to close the hierarchy of correlators but at the same time keep the resummation of Coulomb divergent ladder diagrams as well as the finite temperature corrections. Based on our approximation and resummation scheme, the final form of the equation for our target component $G^{++--}_{\eta \xi}$ is physically sound and maintains important relations like the KMS condition in equilibrium. The coupled system of equations is general enough to apply for the description of DM out-of-chemical equilibrium states.

In the third part, we explored further approximations needed in order to obtain a simple solution to our target component and to reproduce from our general equations the results in the literature, based on different assumptions. So far existing literature has estimated transport coefficients from linear response theory and entered those into a non-linear Boltzmann equation by classical rate arguments \cite{Burnier:2007qm, Bodeker:2012gs,Biondini:2017ufr, Biondini:2018pwp, Biondini:2018xor, Bodeker:2012zm, Kim:2016zyy, Kim:2016kxt}. 
We have proven that our master Eq.~(\ref{eq:numbergrandcanonicalgeneral}) is equivalent to the method of linear response only in the linear regime close to chemical equilibrium.
Finally, we must point out that the Lee-Weinberg equation, adopted in \cite{Kim:2016zyy, Kim:2016kxt, Biondini:2017ufr, Biondini:2018pwp, Biondini:2018xor} to re-derive the DM overclosure bound in the non-linear regime, is not the correct form of the number density equation to use if bound-state solutions exist in the spectrum. The ionisation fraction causes the difference as discussed in great detail in our work.

When taking the vacuum limit, our master equation reduces correctly to the coupled system of classical Boltzmann equations for ideal number densities of bound and scattering states \emph{in the limit of ionization equilibrium}. In our method, it came out as a consequence of assuming the system is in a grand canonical state. Namely, we have proven that the assumption of a grand canonical state automatically implies the Saha ionization equilibrium if bound-state solution exist. One has to take the assumption of ionization equilibrium to be fulfilled for all times with a grain of salt for the following reason. From the vacuum treatment it is known that the duration of Saha ionization equilibrium is limited. Therefore, when using our Master equation one has to carefully check that this condition is satisfied for a sufficiently long period. 
And especially when the assumption of ionization equilibrium is not justified, one has to make sure that at least the abundance of the stable scattering states are not affected by out-of-ionization equilibrium effects which might be model dependent.

The reason why in our Keldysh formalism we can not resolve this issue at the moment lies in one particular approximation, made from the beginning. Ultra-soft emissions and absorptions were dropped for simplicity. We leave the inclusion of those quantities for future work, but expect once they are included we can fully recover the general set of coupled classical Boltzmann equations in the vacuum limit of our (future) updated equations. Moreover, this would allow us to describe Sommerfeld-enhanced annihilation and bound-state decay at finite temperature for the first time beyond the ionization equilibrium.

In the regime where ionization equilibrium is maintained, we have shown that finite temperature effects strongly mix bound and scattering states and the effects are all encoded in the solution of the two-particle spectral function. Let us remark that the numerical results for the spectral function obtained in Section \ref{sec:finitetemp} are compatible with the linear response theory approach \cite{Burnier:2007qm, Bodeker:2012gs,Biondini:2017ufr, Biondini:2018pwp, Biondini:2018xor, Bodeker:2012zm, Kim:2016zyy, Kim:2016kxt}, although we started from a completely different method. The component $G_{\eta \xi,s}^{++--}\big|_{\text{eq}}$ in our master Eq.~(\ref{eq:numbergrandcanonicalgeneral}) can be enhanced by much more than $10 \%$. In addition, our master equation is applicable to the non-linear regime beyond the limitation of linear response if, at least, the ionization equilibrium is maintained. These results make it definitely worthwhile to 
further generalize our Keldysh description in order to correctly describe the out-of-ionization equilibrium transition at late times by including contributions from the ultra-soft scale.

\acknowledgments{ 
\vspace{-0.2cm}
We are thankful to Kalliopi Petraki, Andrzej Hryczuk, Mikko Laine, and Shigeki Matsumoto for very useful comments on our manuscript. 
T.B. thanks the DESY Theory Group and Kavli IPMU for hospitality during an early phase of this project.
This work is supported by Grant-in-Aid for Scientific Research from the Ministry of Education, Science, Sports, and Culture
(MEXT), Japan,  World Premier International Research Center Initiative (WPI Initiative), MEXT, Japan, 
and the JSPS Research Fellowships for Young Scientists (K.M.). 
T.B. and L.C. received funding from the German Research Foundation (Deutsche Forschungsgemeinschaft (DFG)) through the Institutional Strategy of the University of G\"ottingen (RTG 1493), the European Union's Horizon 2020 research and innovation programme InvisiblesPlus RISE under the Marie Sklodowska-Curie grant agreement No 690575, and from the European Union's Horizon 2020 research and innovation programme Elusives ITN under the Marie Sklodowska-Curie grant agreement No 674896.
}


\appendix

\section{Semigroup property of free correlators}
\label{app:semigr}
Free correlators $G_0$ fulfil \emph{semigroup properties}:
\begin{align}
G_0^{R}(x,y) &= +\int\text{d}^3z \; G_0^{R}(x,z) G_0^{R}(z,y),\;\text{for }t_x>t_z>t_y, \label{eq:first}\\
G_0^{R}(x,y) &= +\int\text{d}^3z \text{d}^3w \; G_0^{R}(x,w)  G_0^{R}(w,z) G_0^{R}(z,y),\;\text{for }t_x>t_w>t_z>t_y,\\
G_0^{A}(x,y) &= -\int\text{d}^3z \; G_0^{A}(x,z) G_0^{A}(z,y),\;\text{for }t_x<t_z<t_y, \\
G_0^{A}(x,y) &= +\int\text{d}^3z \text{d}^3w \;  G_0^{A}(x,w) G_0^{A}(w,z) G_0^{A}(z,y),\;\text{for }t_x<t_w< t_z<t_y, \\
G_0^{+-/-+}(x,y) &= +\int\text{d}^3z \; G_0^{R}(x,z) G_0^{+-/-+}(z,y),\;\text{for }t_x>t_z, \\
G_0^{+-/-+}(x,y) &= -\int\text{d}^3z \; G_0^{+-/-+}(x,z) G_0^{A}(z,y),\;\text{for }t_z<t_y,\\
G_0^{+-/-+}(x,y) &= -\int\text{d}^3z \text{d}^3w \; G_0^{R}(x,w) G_0^{+-/-+}(w,z)G_0^{A}(z,y),\;\text{for }t_x>t_w \text{ and }t_z<t_y .
\end{align}
Note, there is no time-integration here. All relations follow from  the first Eq.~(\ref{eq:first}) by using Eq.~(\ref{eq:complexrelations}). It might be helpful to prove the first equation from definition, where we have:
\begin{align}
G_0^{R}(x,y)=\theta(t_x-t_y)\left[G_0^{-+}(x,y)-G_0^{+-}(x,y) \right]= G_0^{\rho}(x,y),\;\text{for }t_x>t_y.
\end{align}
The free spectral function for non-relativistic particles is in Fourier space given by $G_0^{\rho}(\omega,\mathbf{p})=(2\pi)\delta\left(\omega -\frac{\mathbf{p}^2}{2m}\right)$. Then it follows:
\begin{align}
&\int\text{d}^3z \; G_0^{R}(x,z) G_0^{R}(z,y)\\ &= \int\text{d}^3z \; G_0^{\rho}(x,z) G_0^{\rho}(z,y),\;\text{for }t_x>t_z>t_y\\
&= \int\text{d}^3z \int \frac{\text{d}^3p}{(2\pi)^3} \frac{\text{d}\omega}{(2\pi)} (2\pi)\delta\left(\omega -\frac{\mathbf{p}^2}{2m}\right)e^{i\left(\omega(t_x-t_z)-\mathbf{p}\cdot(\mathbf{x}-\mathbf{z})\right)}\int \frac{\text{d}^3p^{\prime}}{(2\pi)^3} \frac{\text{d}\omega^{\prime}}{(2\pi)} (2\pi)\delta\left(\omega^{\prime} -\frac{(\mathbf{p}^{\prime})^2}{2m}\right)e^{i\left(\omega^{\prime}(t_z-t_y)-\mathbf{p}^{\prime}\cdot(\mathbf{z}-\mathbf{y})\right)}\\
&= \int \frac{\text{d}^3p}{(2\pi)^3} e^{i\left(\frac{p^2}{2m}(t_x-t_y)-\mathbf{p}\cdot(\mathbf{x}-\mathbf{y})\right)}\\
&=G_0^{R}(x,y)\;,\;\text{for }t_x>t_y.\;\; \Box
\end{align}

\section{Annihilation term}
\label{app:annihilation}
Although one can directly compute $\Gamma_s (x,y)$ defined on the closed-time-path contour, it is instructive to see how one can recover it from the more common computation of annihilations.
Usually, we compute the matrix element by means of Feynman correlators so as to evaluate annihilations, which means that both $x$ and $y$ are on the $\mathcal C_+$ contour. And thus, it yields the upper left component of $\Gamma_s$, i.e.\ $\Gamma_s^{++}$. The question is how to recover the remaining three components.
To answer it, let us go back one step further, namely before integrating out hard products of the annihilation. 
Suppose that the interaction with them takes the following form; $O_s (x) O_H [\chi (x)] e^{-2 i M x_0} + \text{H.c.}$,
with $\chi$ being the  products of the annihilation.
Here $O_s$ is defined as Eq.~\eqref{eq:ann_swave_u1} and $O_H$ represents hard degrees of freedom.
Then, $\Gamma_s$ is obtained from the cuttings of
$\langle T_{\mathcal C} O_H^\dag [\chi (x)] O_H [\chi (y)] \rangle e^{2 i M ( x_0 - y_0)}$. 
Since we assume that the background plasma is in thermal equilibrium and does not change by $ \eta$ and $\xi$ reactions, the two point correlator of $O_H$ only depends on the space-time difference $x-y$:
\begin{align}
 G_{O_H} (x-y) \equiv 
\langle T_{\mathcal C} O_H^\dag [\chi (x)] O_H [\chi (y)]  \rangle .
\end{align}
By definition, incoming energy/momentum from $O_s$ is much smaller than $M$, which justifies the following approximation:
\begin{align}
	\int_{x,y \in \mathcal C_+}  O_s^\dag(x) O_s(y) 
	i G_{O_H}^{++} (x - y) e^{2 i M ( x_0 - y_0)}
	\simeq 
	\int_{x,y \in \mathcal C_+} O_s^\dag(x) O_s(y) 
	i G_{O_H}^{++} (2 M , \bm{0} ) \delta (x-y).
\end{align}
Taking the imaginary part and
comparing it with $\Gamma_s^{++}$, one can see that the usual computation corresponds to $\Im i G_{O_H}^{++} (2 M , \bm{0})$.
As the background plasma is assumed to be close to equilibrium, we can use the Kubo-Martin-Schwinger (KMS) relation, which essentially connects all the other combinations, $G_{O_H}^{-+}$, $G_{O_H}^{+-}$, $G_{O_H}^{--}$, with this one.
Moreover, because of $M \gg T$, one can safely neglect $G_{O_H}^{+-}$.
As a result, we end up with
\begin{align}
	\Im i G_{O_H}^{++} (2 M, \bm{0}) &= \frac{\pi (\alpha_\chi^2 + \alpha_\chi \alpha_\psi)}{M^2}, \\
	\Im i G_{O_H}^{+-} (2 M, \bm{0}) &= 0, \\
	\Im i G_{O_H}^{-+} (2 M, \bm{0}) &=  
	\Im i G_{O_H}^\text{ret} (2 M, \bm{0}) - \Im i G_{O_H}^\text{adv} (2 M, \bm{0})
	= 2 \Im i G_{O_H}^{++} (2 M, \bm{0}), \\
	\Im i G_{O_H}^{--} (2 M, \bm{0}) 
	&= - \Im i G_{O_H}^\text{ret} (2 M, \bm{0}) 
	+ \Im i G_{O_H}^{-+} (2 M, \bm{0})
	=\Im i G_{O_H}^{++} (2 M, \bm{0}),
\end{align}
which results in Eq.~\eqref{eq:ann_swave_u1}.
Here, several properties of equilibrium correlators were used, as given in Section \ref{sec:rtf}.

\section{Hard thermal loop approximation}
\label{app:htl}
The Fourier transform of the ideal $\psi$ two-point function is in the Keldysh representation given by:
\begin{align}
S(P) =(\slashed{P}+m)\left[ \begin{pmatrix} \frac{i}{P^2-m^2+i\epsilon} & 0 \\ 0 & \frac{-i}{P^2-m^2-i\epsilon}\end{pmatrix} - 2\pi\delta(P^2-m^2)  \begin{pmatrix} n_F(|p^0|) & -\theta(-p^0)+ n_F(|p^0|)\\ -\theta(p^0)+ n_F(|p^0|) & n_F(|p^0|)\end{pmatrix}\right].
\end{align}
Combining different components, the following retarded, advanced and symmetric propagator can be obtained 
\begin{align}
S^{R/A}(P)&= \frac{i(\slashed{P}+m)}{P^2-m^2\pm i\text{sign}(p^0)\epsilon},\\
S^s(P) &\equiv S^{++}(P) + S^{--}(P) = 2 \pi(\slashed{P}+m)\left[1-2n_F(|p^0|) \right]\delta(P^2-m^2) .
\end{align}
The inverse relations are given by:
\begin{align}
S^{++} =\frac{1}{2} \left(S^{s}+S^{R}+S^{A} \right),\;S^{+-}=\frac{1}{2} \left( S^{s}-S^{R}+S^{A} \right),\; S^{-+}=\frac{1}{2} \left(S^{s}+S^{R}-S^{A}  \right).
\end{align}
By using these relations, the one-loop expression of the retarded mediator correlator can be simplified as
\begin{align}
\Tr[\gamma^{\mu} S^{++}(x,y) \gamma^{\nu} S^{++}(y,x)] - \Tr[\gamma^{\mu} S^{+-}(x,y) \gamma^{\nu} S^{-+}(y,x)]=\frac{1}{2}\Tr[\gamma^{\mu} S^{R}(x,y) \gamma^{\nu} S^{S}(y,x)] + \frac{1}{2}\Tr[\gamma^{\mu} S^{S}(x,y) \gamma^{\nu} S^{A}(y,x)].
\end{align}
Let us take the limit $ m \ll T$, where $m$ is the $\psi$ mass, leading in Fourier space to the following retarded self-energy of the mediator:
\begin{align}
\Pi^{00}_R(P)=-g_{\psi}^2 8 \pi \int \frac{\text{d}^4K}{(2\pi)^4} \left[(k^0-p^0)k^0 + (\mathbf{k}-\mathbf{p})\cdot \mathbf{k} \right] \left[1-2n_F(|k^0|) \right]\delta(K^2) \frac{1}{(K-P)^2 - i\text{sgn}(k^0-p^0)\epsilon}.
\end{align}
Dropping the vacuum part and integrating over $k^0$ one obtains:
\begin{align}
\Pi^{00}_R= g_{\psi}^2 8 \pi \int \frac{\text{d}^3k}{(2\pi)^4} \frac{n_F(|\mathbf{k}|)}{|\mathbf{k}|} \left[ \frac{(|\mathbf{k}|-p^0)|\mathbf{k}| + (\mathbf{k}-\mathbf{p})\cdot \mathbf{k} }{(|\mathbf{k}|-p^0)^2 -(\mathbf{k}-\mathbf{p})^2  - i\text{sgn}(|\mathbf{k}|-p^0)\epsilon} + \frac{(|\mathbf{k}|+p^0)|\mathbf{k}| + (\mathbf{k}-\mathbf{p})\cdot \mathbf{k} }{(|\mathbf{k}|+p^0)^2 -(\mathbf{k}-\mathbf{p})^2  + i\text{sgn}(|\mathbf{k}|+p^0)\epsilon} \right].
\end{align}
The result so far is exact up to the fact that we neglected the $\psi$ mass and ignored the vacuum contribution.
Now, the \emph{Hard-Thermal-Loop approximation} \cite{Braaten:1989mz} assumes that the external energy $p^0$ and momentum $|\mathbf{p}|$ are smaller compared to the typical loop momentum  $|\mathbf{k}|$ which is of the order temperature (hard), since the integrand contains $n_F$.
Expanding the term in the brackets to leading order in $p^0/|\mathbf{k}|$ and $|\mathbf{p}|/|\mathbf{k}|$, all remaining integrals can be performed analytically leading to the finite result:
\begin{align}
\Pi^{00}_{\text{R,A}}(P)\simeq -m^2_D \left[ 1- \frac{p^0}{2 |\mathbf{p}|} \ln{ \left(\frac{p^0 +|\mathbf{p}| \pm i \epsilon}{p^0 -|\mathbf{p}| \pm i \epsilon}\right)}  \right],
\end{align}
where the Debye mass is defined as $m_D^2 = g_\psi^2 T^2 /3$. This result coincides with the result obtained from the imaginary time formalism, where instead one has to perform a sum over Matsubara frequencies.

\section{Vacuum limit of two-particle spectral function}
\label{app:spec_at_vacuum}

Here we will derive Eq.~\eqref{eq:possol} starting from Eq.~\eqref{eq:solutionnumerical} in the limit of vanishing finite temperature corrections. As we have briefly mentioned in Section \ref{sec:explsol}, for the vacuum limit one has to carefully take into account the imaginary part $i\epsilon$ in the retarded equation, representing the small width which will be taken to be zero in the end. We will see that the result does not depend on this $i \epsilon$-prescription as long as $\epsilon$ is small enough.

Suppose that the potential almost vanishes for a large enough $r$.
For a Yukawa type potential, this is true for $m_{V} r \gg 1$.
The $\epsilon$ parameter should be much smaller than this mass parameter, namely $\epsilon \ll m_{V}$. In the case of the Coulomb potential one may introduce another small mass parameter to the gauge boson. In the end of the computation one can take it to be zero while keeping $m_V \gg \epsilon$.
For $m_{V}r \gg 1$, the homogeneous solutions, $g_{>,<}$, can be well approximated by the plane wave:
\begin{align}
	g_> (r) &\to C_> e^{i \sqrt{ME} r - \epsilon r}, \label{eq:plane_>}\\
	g_< (r) & \to \frac{1}{2} \left( C_< e^{i \sqrt{ME} r - \epsilon r} + C_<^\dag e^{-i \sqrt{ME}r  + \epsilon r} \right). \label{eq:plane_<}\
\end{align}
The Wronskian tells us that there exists a non-trivial relation between two coefficients $C_{>,<}$. Since the Wronskian, $W(r) = g_>(r) g_<' (r) - g_>'(r) g_< (r)$, does not depend on $r$, one may equate it at $r=0$ and $r \to \infty$, which yields
\begin{align}
	1 = W(0) = W(\infty) = - i \sqrt{ME} C_> C_<^\dag
	\quad \leftrightarrow \quad
	C_> = \frac{i}{\sqrt{ME}} \frac{1}{C_<^\dag}.
		\label{eq:coefs}
\end{align}
We have taken $\epsilon$ to be zero in the end of the computation.

Let us evaluate the integral given in Eq.~\eqref{eq:solutionnumerical} by means of Eq.~\eqref{eq:plane_<}. The first observation is that there is no imaginary part for $g_<$ if $\epsilon$ is zero from the beginning.
Thus, the integrand becomes relevant only after $r \gtrsim \epsilon^{-1} \gg m_{V}^{-1}$.
As a result one may evaluate the integral in Eq.~\eqref{eq:solutionnumerical} by substituting Eq.~\eqref{eq:plane_<}:
\begin{align}
	G_{\eta \xi}^{\rho}(\mathbf{0},\mathbf{0};E)|_{E>0,l=0}
	&= \frac{1}{2 \pi} \Tr[\mathbf{1}_{2\times2}] M \frac{1}{\abs{C_<}^2}
	\lim_{\epsilon \to 0} \Im \int_{m_{V}^{-1}}^\infty \dd r \frac{1}{\cos^2 [ ( \sqrt{ME} + i \epsilon ) r + \delta_{C_<} ]} \\
	&= \frac{1}{2 \pi} \Tr[\mathbf{1}_{2\times2}]  M \frac{1}{\abs{C_<}^2}\frac{1}{\sqrt{ME}}.
\end{align}
Finally, substituting Eq.~\eqref{eq:coefs} into this equation, we arrive at
\begin{align}
	G_{\eta \xi}^{\rho}(\mathbf{0},\mathbf{0};E)|_{E>0,l=0}
	= \frac{1}{4 \pi} \Tr[\mathbf{1}_{2\times2}] M^2 v_\text{rel} \abs{C_>}^2,
\end{align}
where we have used $E = M v_\text{rel}^2 / 4$. 

Now we are in a position to discuss its relation to the conventional definition of the Sommerfeld enhancement factor. In the limit of $\epsilon \to 0$, the wave function propagates to infinity. Then one may obtain the Sommerfeld enhancement factor by extracting the amplitude of the wave function at the infinity, which is nothing but the relation $S(v_\text{rel}) = \abs{C_>}^2$, see Ref.~\cite{Cirelli:2007xd} for instance. Utilizing this relation, we finally get Eq.~\eqref{eq:possol}. For conventional reason, let us give the s-wave Sommerfeld enhancement factor for the Coulomb case consistent with our equations:
\begin{align}
S(v_\text{rel}) =\frac{2\pi \alpha_{\chi}}{v_\text{rel}} \frac{1}{1 - e^{-\frac{2\pi \alpha_{\chi}}{v_\text{rel}}}}.
\end{align}

For the bound state, it is much easier to solve the equation directly rather starting from Eq.~\eqref{eq:solutionnumerical}.
One may express the spectral function by means of the wave functions for the bound states~\cite{Kim:2016kxt}:
\begin{align}
	G_{\eta \xi}^{\rho}(\mathbf{0},\mathbf{0};E)|_{E<0,l=0} 
	= 2 \pi \Tr[\mathbf{1}_{2\times2}] \sum_n \delta(E - E_B^n) \abs{\psi_{n}^{(B)}(0)}^2,
\end{align}
where $\psi_n^{(B)}$ represents the normalized wave function for the $n$-th bound state.
For instance, the wave function for the lowest energy state $n = 1$ is given by
\begin{align}
	\psi_1^{(B)} = \frac{1}{\sqrt{\pi}}\left(\frac{\alpha_\chi M}{2}\right)^{3/2} e^{-\alpha_\chi M r/2}.
	\label{eq:1s0}
\end{align}
The decay rate of the bound state is related to its wave function at the origin. For the lowest state, one can easily show this from Eq.~\eqref{eq:1s0}:
\begin{align}
	\left(\sigma v_\text{rel} \right) \abs{\psi_1^{(B)} (0)}^2
	= \frac{1}{4} \Gamma_{^1S_0},
\end{align}
where the decay rate of the lowest bound state is given by
\begin{align}
	\Gamma_{^1S_0} = (\alpha_\chi^2 + \alpha_\chi \alpha_\psi) \alpha_\chi^3 \frac{M}{2}.
\end{align}

Similar calculation holds in the limit of negligible thermal width but finite real-part corrections. For this case one should substitute the kinetic energy $E\rightarrow E - \Re V_\text{eff}(\infty) =  M v_\text{rel}^2 / 4$ and similar for the bound state energy. Also one has to take $\epsilon$ smaller than $m_D$ and $m_V$. 

\section{Number density and chemical potential in grand canonical ensemble}
\label{app:salpeter}
In this Section, we present an alternative way of how to derive the chemical potential as a function of the total number density directly from the EoM. It is convenient to write the EoM in integral form as
\begin{align}
G_\eta (x,y)\! &=\! G_{\eta,0}(x,y)-\!  g^2_{\chi} \!\int_{w,z \in \mathcal{C}}G_{\eta,0}(x,w) D (w,z)[G_{\eta\xi}(w,z,y,z)-G_{\eta\eta}(w,z,y,z)],\label{eq:massshells}\\
\bar{G}_\xi (x,y)\!	&=\! G_{\xi,0}(x,y)\!-\! g^2_{\chi}  \!\int_{w,z \in \mathcal{C}}  G_{\xi,0}(x,w)D (w,z) [G_{\eta\xi}(z,w,z,y)-G_{\xi\xi}(w,z,y,z)].\label{eq:massshells2}
\end{align}
The number density is given by the $+-$ component, namely in Wigner coordinates it reads $n_{\eta}(\mathbf{R},T) = \Tr[G_{\eta}^{+-}(x,x)]= \Tr[G_{\eta}^{+-}(\mathbf{0},0;\mathbf{R},T)]$.
Now one can already clearly see the structure that the total number density is given by the ideal number density plus interactions. These equations are exact within our nonrelativistic effective action. Expanding $D$ around the narrow width limit and assuming a grand canonical state, one should recover Eq.~(\ref{eq:totalnumber}) for the number density since the only assumption entering there is the narrow width approximation. Eq.~(\ref{eq:totalnumber}) is the correct thermodynamic definition of the number density which should coincide with the result obtained by solving for the number density from these integral EoM when assuming a thermodynamic picture like the grand canonical ensemble. It requires a rigorous proof of this claim, which we would like to give somewhere else.

Instead, we would like to give some approximations in order to obtain an analytic solution of Eq.~(\ref{eq:massshells}). We restrict the discussion by assuming the system is in a grand canonical state.
Utilizing the KMS condition for finite chemical potential, one can formally solve $G_{\eta}^{+-}$ in terms of spectral function:
\begin{align}
n_{\eta}=\Tr[G_{\eta}^{+-}(\mathbf{0},0)] = \int \frac{\text{d}^3 \mathbf{p}}{(2\pi)^3} \frac{\text{d} \omega}{(2\pi)} \frac{1}{e^{\beta(M + \omega - \mu_{\eta})}+1}  \Tr[G_{\eta}^{\rho}(\mathbf{p},\omega)] 
\simeq \int \frac{\text{d}^3 \mathbf{p}}{(2\pi)^3} \frac{\text{d} \omega}{(2\pi)}e^{-\beta(M + \omega - \mu_{\eta})} \Tr[G_{\eta}^{\rho}(\mathbf{p},\omega)]. \label{eq:gpmnumber}
\end{align}
In the last equality we assumed the DM gas to be dilute and approximated the Fermi-Dirac distribution as a Maxwellian. By KMS relation we have formally solved for the number density in terms of chemical potential and spectral function. The spectral function can be computed from the retarded component $G_{\eta}^R$, according to $G_{\eta,0}^{\rho}(\mathbf{p},\omega)= G_{\eta,0}^{R}(\mathbf{p},\omega)-G_{\eta,0}^{A}(\mathbf{p},\omega) $ (see Section \ref{sec:rtf}). The EoM of the retarded correlator can be obtained from Eq.~(\ref{eq:massshells}) by subtracting $+-$ from the $++$ component, e.g.\ $G_{\eta}^R = G_{\eta}^{++} -G_{\eta}^{+-} $. In subsequent sections we solve the retarded equation in various approximations, compute the spectral function and finally evaluate Eq.~(\ref{eq:gpmnumber}).

\subsection{Ideal gas approximation}
\label{app:idealgas}
The ideal gas approximation can be defined as the zeroth order contribution in Eq.~(\ref{eq:massshells}). This means we have to know the solution of the free retarded correlator.
It can be obtained from the differential form of the retarded equations.
In Fourier space of the microscopic Wigner coordinates it is given by $G_{\eta,0}^{R}(\mathbf{p},\omega)=i\delta_{ij}/(\omega-\mathbf{p}^2/(2M) + i \epsilon)$, leading to $G_{\eta,0}^{\rho}(\mathbf{p},\omega)=\delta_{ij} (2\pi) \delta(\omega-\mathbf{p}^2/2M)$.
Inserting this into Eq.~(\ref{eq:gpmnumber}) results in the ideal gas approximation of the free number density:
\begin{align}
n_{\eta,0}= 2 \int \frac{\text{d}^3 \mathbf{p}}{(2\pi)^3} e^{-\beta(M + \frac{\mathbf{p}^2}{2M} - \mu_{\eta})}.
\end{align}
We can invert this relation to obtain the chemical potential as a function of the number density in the ideal gas approximation:
\begin{align}
\beta \mu_{\eta}^{\text{id}} = \ln\left[ \frac{n_{\eta,0}}{n_{\eta,0}^{\text{eq}}} \right],\; \text{where } n_{\eta,0}^{\text{eq}} = 2 \int \frac{\text{d}^3 \mathbf{p}}{(2\pi)^3} e^{-\beta(M + \mathbf{p}^2/2M)}.
\end{align}
And similar expressions for anti-particle $\xi$. 

\subsection{Hartree-Fock approximation}
\label{app:hf}
The Hartree-Fock approximation is the zero order approximation of the four-point correlator.
Using Eqs.~(\ref{eq:hartree1})-(\ref{eq:hartree2}), one obtains:
\begin{align}
[G_{\eta\xi}(x,z,y,z)-G_{\eta\eta}(x,z,y,z)] &\simeq  G_{\eta,0}(x,z) G_{\eta,0}(z,y),\label{eq:hffree}\\
[G_{\eta\xi}(z,x,z,y) -G_{\xi\xi}(x,z,y,z) ] &\simeq  \bar{G}_{\xi,0}(x,z) \bar{G}_{\xi,0}(z,y),
\end{align}
valid for symmetric DM. Inserting this into Eq.~(\ref{eq:massshells})-(\ref{eq:massshells2}) and subtracting the components $G_{\eta}^{++} -G_{\eta}^{+-}$ , one obtains for the retarded equations:
\begin{align}
G^{R}_{\eta}(x,y) &= G^{R}_{\eta,0}(x,y) + \int \text{d}^4z \text{d}^4w \; G^{R}_{\eta,0}(x,z)(-i) \Sigma^{R}_{\eta}(z,w)G^{R}_{\eta,0}(w,y),\label{eq:retunresummedp}\\
\bar{G}^{R}_{\xi}(x,y) &= \bar{G}^{R}_{\xi,0}(x,y) + \int \text{d}^4z \text{d}^4w \; \bar{G}^{R}_{\xi,0}(x,z)(-i) \bar{\Sigma}^{R}_{\xi}(z,w)\bar{G}^{R}_{\xi,0}(w,y).\label{eq:retunresummedap}
\end{align}
The single particle self-energies are defined on the CTP contour as we have introduced in Eq.~(\ref{eq:self}). The retarded component is $\Sigma^{R} = \Sigma^{++}-\Sigma^{+-}$
and can be written as:
\begin{align}
\Sigma^{R}(x,y)/(-i g_{\chi}^2) &= D^{++}(x,y) G_0^{++}(x,y) - D^{+-}(x,y) G_0^{+-}(x,y)\\
 &=  D^{++}(x,y) G_0^{R}(x,y) - D^{R}(x,y) G_0^{+-}(x,y)\\
 &=  D^{-+}(x,y) G_0^{R}(x,y) - D^{R}(x,y) G_0^{+-}(x,y).
\end{align}
In the last step, the definition $ D^{++}(t,t^{\prime}) = \theta(t-t^{\prime}) D^{-+}(t,t^{\prime}) + \theta(t^{\prime}-t) D^{+-} (t,t^{\prime})$ was used and the fact that the retarded function $G_0^{R}$ projects out only the $D^{-+} (t,t^{\prime})$ contribution due to equal times in $D$ and $G_0^{R}$. A key observation is that the self-energy can depend on the DM number density and chemical potential due to the $G_0^{+-}$ contribution. This might lead to a non-linear dependence of the total number density on $e^{\beta \mu}$ when inserting the spectral function obtained from the retarded equation into Eq.~(\ref{eq:gpmnumber}). In the following we would like to perturbatively resum Eqs.~(\ref{eq:retunresummedp})-(\ref{eq:retunresummedap}) which brings us later to the widely used Salpeter correction. Let us therefore drop the dependence of the self-energy on $G_0^{+-}(x,y)$. Then the equation is closed in terms of the retarded correlators $G_0^{R}$.
Within this approximation, the self-energy can be written in Fourier space as (to leading order gradient expansion):
\begin{align}
\Sigma^{R}_{\eta}(\mathbf{p},\omega) &=  g_{\chi}^2\int \frac{\text{d}^3 \mathbf{p}_1}{(2\pi)^3} \frac{\text{d} \omega_1}{(2\pi)}\frac{\text{d} \omega_2}{(2\pi)}\frac{G^{\rho}_{\eta,0}(\omega_1,\mathbf{p}-\mathbf{p}_1) D^{-+}(\omega_2,\mathbf{p}_1)}{\omega - \omega_1 - \omega_2+i\epsilon} \\
&= g_{\chi}^2\int \frac{\text{d}^3 \mathbf{p}_1}{(2\pi)^3} \frac{\text{d} \omega_1}{(2\pi)}  \frac{D^{-+}(\omega_1,\mathbf{p}_1)}{\Omega-\omega_1 +i\epsilon}\\
&= g_{\chi}^2\frac{1}{2}\int \frac{\text{d}^3 \mathbf{p}_1}{(2\pi)^3} \frac{\text{d} \omega_1}{(2\pi)}  \frac{ D^{-+}(\omega_1,\mathbf{p}_1) +  D^{-+}(\omega_1,\mathbf{p}_1) }{\Omega-\omega_1  +i\epsilon} \\
&= g_{\chi}^2\frac{1}{2}\int \frac{\text{d}^3 \mathbf{p}_1}{(2\pi)^3} \frac{\text{d} \omega_1}{(2\pi)}  \frac{ D^{-+}(\omega_1,\mathbf{p}_1) +  D^{+-}(-\omega_1,\mathbf{p}_1) }{\Omega-\omega_1  +i\epsilon} \\
&=-i g_{\chi}^2\frac{1}{2}\int \frac{\text{d}^3 \mathbf{p}_1}{(2\pi)^3}  D^{++}(\Omega,\mathbf{p}_1).
\end{align}
where $\Omega= \omega - (\mathbf{p}-\mathbf{p}_1)^2/2M$. The final form is convenient for inserting the static HTL approximation of $ D^{++}$ as given in Eq.~(\ref{eq:staticHTL}). Performing the one-loop calculation results in 
\begin{align}
\lim_{\Omega\rightarrow0}\Sigma^{R}(\mathbf{p},\omega) = -i g_{\chi}^2\frac{1}{2}\int \frac{\text{d}^3 \mathbf{p}_1}{(2\pi)^3} \lim_{\Omega\rightarrow0} D^{++}(\Omega,\mathbf{p}_1)=-\frac{1}{2} \left( \alpha_{\chi} m_D + i \alpha_{\chi} T \right).\label{eq:selfenergyhfHTL}
\end{align}
One can recognize that this result is exactly half the effective in-medium potential for two particles at large distance, see Eq.~(\ref{eq:inmediumpot}).
Now, for perturbative resummation we replace the retarded correlators at the end of Eqs.~(\ref{eq:retunresummedp})-(\ref{eq:retunresummedap}) by the fully dressed one: $G^{R}_{\eta,0}(w,y) \rightarrow G^{R}_{\eta}(w,y)$, $\bar{G}^{R}_{\xi,0}(w,y) \rightarrow G^{R}_{\xi}(w,y)$. Performing Wigner and Fourier transformation of the equation leads at the leading order in gradient expansion to:
\begin{align}
G^{R}_{\eta}(\mathbf{p},\omega) &= G^{R}_{\eta,0}(\mathbf{p},\omega) +  \; G^{R}_{\eta,0}(\mathbf{p},\omega)(-i) \Sigma^{R}_{\eta}(\mathbf{p},\omega)G^{R}_{\eta}(\mathbf{p},\omega),
\end{align}
and similar equation for the anti-particle.
Then, by using geometric series one ends up with the HTL single particle correlators:
\begin{align}
G^{R}_{\eta}(\mathbf{p},\omega)= \frac{i \delta_{ij}}{\omega - \mathbf{p}^2/2M - \Sigma^{R}_{\eta}(\mathbf{p},\omega) + i \epsilon},\;\bar{G}^{R}_{\xi}(\mathbf{p},\omega)= \frac{i  \delta_{ij}}{\omega - p^2/2M - \Sigma^{R}_{\xi}(\mathbf{p},\omega) + i \epsilon}. 
\end{align}
Computing the spectral function from the difference of retarded and advanced correlators results in a Breit-Wigner shape:
\begin{align}
G_{\eta}^{\rho}(\mathbf{p},\omega) &=  \delta_{ij} \frac{\Gamma_{\eta}}{(\omega-\mathbf{p}^2/2M- \Re(\Sigma^R_{\eta}))^2 + (\Gamma_{\eta}/2)^2 } \;.\label{eq:spectralbreitw}
\end{align}
where the particle width is defined by $\Gamma \equiv  2\Im(\Sigma^R) $. In summary, we evaluated the four-point correlator in the Hartree-Fock (HF) approximation and formally solved for $\rho$ as a function of the self-energy in static HTL approximation, shifting the energy by $\alpha_{\chi} m_D/2$ and broadening the peak via imaginary contributions. 
Finally, let us quote the chemical potential in HF and static HTL approximation:
\begin{align}
\beta \mu_{\eta}^{\text{HF}} = \ln\left[ \frac{n_{\eta}}{n_{\eta,\text{HF}}^{\text{eq}}} \right],\; \text{where } n_{\eta,\text{HF}}^{\text{eq}} = 2 \int \frac{\text{d}^3 \mathbf{p}}{(2\pi)^3} \frac{\text{d} \omega}{(2\pi)} e^{-\beta(M + \omega)}\frac{\Gamma}{(\omega-\mathbf{p}^2/2M- \Re(\Sigma^R_{\eta}))^2 + (\Gamma/2)^2 } .
\end{align}
This approximation might be already good enough if there are no bound states but the two-particle spectral function has support at negative energies due to thermal width. We also see that one-particle spectral function can have spectral support at negative energies. 
\subsection{Salpeter correction}
\label{app:salpeter}
Taking the limit $\Gamma \rightarrow0$ in the spectral function Eq.~(\ref{eq:spectralbreitw}) results in
\begin{align}
\lim_{\Gamma \rightarrow 0}G_{\eta}^{\rho}(p,\omega) = \delta_{ij}(2\pi) \delta(\omega-p^2/2M-\Re(\Sigma^R_{\eta})).
\end{align}
This is called the \emph{narrow width} or \emph{quasi particle} approximation,
taking only the real-part correction into account. Inserting the spectral function into 
Eq.~(\ref{eq:gpmnumber}) leads to the chemical potential and equilibrium number density:
\begin{align}
\beta \mu_{\eta}^{\text{SP}} = \ln\left[ \frac{n_{\eta}}{n_{\eta,\text{SP}}^{\text{eq}}} \right],\; \text{where } n_{\eta,\text{SP}}^{\text{eq}} = 2 \int \frac{\text{d}^3 \mathbf{p}}{(2\pi)^3}  e^{-\beta(M + p^2/2M+\Re[\Sigma^R_{\eta}])}.
\end{align} 
For the self-energy in the static HTL approximation, the real-part correction is $\Re(\Sigma^R_{\eta})= -\alpha_{\chi} m_D/2$, according to Eq.~(\ref{eq:selfenergyhfHTL}). This is the well-known \emph{Salpeter correction} to the equilibrium distribution.

The Salpeter correction is a simple first order approximation for the description of quasi particles in a plasma. As we have seen in Section \ref{sec:ionizatineq}, it might be however required for our number density equation to compute both, the annihilation/decay rates AND the chemical potential, to the same level of approximation in the four-point correlator. Especially when bound-state solutions exist it is required to solve the four-point correlator non-perturbatively (by resummation).
The Salpeter correction was obtained by approximating the four-point correlator as a product of free particles without self-interactions, see Eq.~(\ref{eq:hffree}) and Eq.~(\ref{eq:retunresummedp}). Thus, in this approximation of the chemical potential, bound state contributions never appear. It might lead to inconsistencies in the number density equation, like exponentially growing terms for late times. 

Let us illustrate, why the Salpeter correction is not enough to correctly describe the freeze-out at late times. We plug the chemical potential $\mu_{\eta}^{\text{SP}}$ in Salpeter approximation into our master formula for the number density and obtain:
\begin{align}
\dot{n}_{\eta} + 3 H  n_{\eta} 
&= - 2 (\sigma v_{\text{rel}})G_{\eta \xi,s}^{++--} (x,x,x,x)\big|_{\text{eq}} \bigg[ e^{\beta 2 \mu_{\eta}^{\text{SP}}} - 1 \bigg]\\
&= - \frac{2 (\sigma v_{\text{rel}})G_{\eta \xi,s}^{++--} (x,x,x,x)\big|_{\text{eq}}}{n_{\eta,\text{SP}}^{\text{eq}}(T) n_{\eta,\text{SP}}^{\text{eq}}(T)} \bigg[ n_{\eta}n_{\eta}- n_{\eta,\text{SP}}^{\text{eq}} (T) n_{\eta,\text{SP}}^{\text{eq}} (T) \bigg].
\end{align}
There are two reasons why this description fails at late times if bound-state solutions exist. First of all, since no bound states are included in the computation of the chemical potential in Salpeter approximation, the ionization degree would always be approximated as 1. Second, if one would compute the spectral function in $G_{\eta \xi,s}^{++--} (x,x,x,x)\big|_{\text{eq}}$ non-perturbatively, this would cause an exponential growing term at late times, caused by the bound states:  $G_{\eta \xi,s}^{++--} (x,x,x,x)\big|_{\text{eq}} \propto e^{\beta|E_B|}$.
In the Salpeter approximation, the denominator $n_{\eta,\text{SP}}^2$ can not kill this unphysical behavior. Note however, if one computes the chemical potential to the same level as the annihilation rates, the degree of ionization exhibits a term leading to a cancellation of the exponential growing term in $ G_{\eta \xi,s}^{++--}$. 
Our main number density equation automatically incorporates this because it evaluates the chemical potential and number density to the same level of approximation.
The Salpeter correction together with solving the spectral function non-perturbatively was used in Ref.~\cite{Biondini:2017ufr, Biondini:2018pwp, Biondini:2018xor}.


\bibliography{seft}

\end{document}